\documentclass[entropy,article,accept,moreauthors,pdftex]{mdpi} 
\usepackage{arburst}
\usepackage[labelformat=simple]{subcaption}

\DeclareCaptionLabelFormat{subcaptionlabel}{\normalfont(\textbf{#2}\normalfont)}
\setitemize{parsep=6pt,itemsep=0pt,leftmargin=*,labelsep=5.5mm}
\setenumerate{parsep=6pt,itemsep=0pt,leftmargin=*,labelsep=5.5mm}
\setlist[description]{itemsep=0mm}

\newcommand{\grandO}[1]{\ensuremath{\mathop{}\mathopen{}O\mathopen{}\left(#1\right)}}

\firstpage{1} 
\pubvolume{xx}
\issuenum{1}
\articlenumber{5}
\pubyear{2020}
\copyrightyear{2020}

\history{Received: 12 May 2020; Accepted: 11 June 2020; Published: date}
\updates{yes}
\Title{An Upper Bound on the Error Induced by Saddlepoint Approximations---{A}pplications to Information Theory}

\Author{Dadja Anade $^{1}$\orcidA{}, Jean-Marie Gorce $^{1}$\orcidB{}, Philippe Mary $^{2}$\orcidC{}, and Samir M. Perlaza $^{3,4,}$*\orcidD{} }
\AuthorNames{Dadja Anade, Jean-Marie Gorce, Philippe Mary, and Samir M. Perlaza}

\address{%
	$^{1}$ \quad Laboratoire CITI, a Joint Laboratory between INRIA, the Universit\'{e} de Lyon and the Institut National de Sciences Appliqu\'{e}es (INSA) de Lyon. 6 Av. des Arts, 69621 Villeurbanne, France; dadja.anade-akpo@inria.fr~(D.A.); jean-marie.gorce@insa-lyon.fr (J.-M.G.)\\
	$^{2}$ \quad IETR and the Institut National de Sciences Appliqu\'{e}es (INSA) de Rennes, 20 Avenue des Buttes de Co\"esmes, CS 70839, 35708 Rennes, France; philippe.mary@insa-rennes.fr\\
	$^{3}$ \quad INRIA, Centre de Recherche de Sophia Antipolis---M\'{e}diterran\'{e}e, 2004  Route des Lucioles, 06902~Sophia~Antipolis,~France\\
	$^{4}$ \quad Princeton University, Electrical Engineering Department, Princeton, NJ 08544, USA}
	
	\corres{Correspondence: samir.perlaza@inria.fr}

\firstnote{Authors are listed in alphabetical order.} 

\secondnote{{Parts}  
of this paper appear in the proceedings of the IEEE international Conference on Communications (ICC), Dublin, Ireland, June, 2020; and in the INRIA technical report RR-9329.} 

\abstract{This paper introduces an upper bound on the absolute difference between: $(a)$~the cumulative distribution function (CDF) of the sum of a finite number of independent and  identically distributed random variables with finite absolute third moment; and $(b)$~a saddlepoint approximation of such  CDF.
	This upper bound, which is particularly precise in the regime of large deviations, is used to study the  dependence testing (DT) bound and the meta converse (MC) bound on the decoding error probability (DEP) in point-to-point  memoryless channels. Often, these bounds cannot be analytically calculated and thus lower and upper bounds become particularly useful. 
	Within~this context, the main results include, respectively, new upper and lower bounds on the DT and MC bounds.
	A~numerical experimentation of these bounds is presented in the case of the binary symmetric channel, the additive white Gaussian noise channel, and the additive symmetric $\alpha$-stable noise channel.}


\begin{document}
\DeclareGraphicsExtensions{.png, .pdf, .jpg}
	
	\section{Introduction}
	This paper focuses on approximating the cumulative distribution function (CDF) of sums of a finite number of real-valued independent and identically distributed (i.i.d.) random variables with finite absolute third moment. 
	More specifically, let $Y_1$, $Y_2$, $\dots$, $Y_n$, with $n$ an integer and $2 \leqslant n < \infty$, be~real-valued random variables with probability distribution $P_{Y}$. Denote by $F_Y$ the CDF associated with $P_Y$, and, if it exists,  denote by $f_Y$ the corresponding probability density function (PDF). 
	Let also 
	\begin{equation}\label{EqXn}
	X_n = \displaystyle\sum_{t = 1}^{n}Y_t
	\end{equation}
	be a random variable with distribution $P_{X_n}$. Denote by $F_{X_n}$ the CDF and if it exists,  denote by $f_{X_n}$ the PDF associated with $P_{X_n}$. 
	The objective is to provide a positive function that approximates $F_{X_n}$ and an upper bound on the resulting approximation error. 
	In the following, a positive  function $g: \mathbb{R} \rightarrow \mathbb{R}_{+}$ is said to approximate $F_{X_n}$ with an \emph{approximation error} that is upper bounded by a function $\epsilon: \mathbb{R} \rightarrow \mathbb{R}_{+}$, if, for all $x \in \mathbb{R}$, 
	\begin{equation}\label{EqApprox}
	\abs{F_{X_n}(x) - g(x)} \leqslant \epsilon(x).
	\end{equation}

	The case in which $Y_1$, $Y_2$, $\dots$, $Y_n$ in~\eqref{EqXn} are
%
stable random variables with $F_{Y}$ analytically expressible is trivial. This is essentially because the sum $X_n$ follows the same distribution of a random variable $a_n Y + b_n$, where $(a_n,b_n) \in \mathbb{R}^2$ and $Y$ is a random variable whose CDF is $F_Y$. Examples of this case are random variables following the Gaussian, Cauchy, or Levy distributions~\citep{Zolotarev-Book}.

	In general, the problem of calculating the CDF of $X_n$ boils down to calculating $n - 1$ convolutions. More specifically, it holds that
	\begin{equation}\label{EqConvolution}
	f_{X_{n}}(x) = \displaystyle\int_{-\infty}^{\infty} f_{X_{n-1}}\left(x-t \right) f_{Y}(t) \mathrm{d}t,
	\end{equation}
	where $f_{X_1} = f_{Y}$.
	Even for discrete random variables and small values of $n$, the integral in \eqref{EqConvolution}  often requires excessive computation resources~\citep{Jens1995}.

	When the PDF of the random variable $X_n$ cannot be conveniently obtained but only the $r$ first moments are known, with $r \in \mathbb{N}$, an approximation of the PDF can be obtained by using an Edgeworth expansion. 
	Nonetheless, the resulting relative error in the large deviation regime  makes these approximations inaccurate~\citep{Feller-book-1971}.

	When the cumulant generating function (CGF) associated with $F_Y$, denoted by  $K_{Y}: \mathbb{R} \rightarrow \mathbb{R}$, is~known, the PDF $f_{X_n}$ can be obtained via the Laplace inversion lemma~\citep{Jens1995}. 
	That is, given two reals $\alpha_{-} < 0 $ and $\alpha_{+} > 0$, if $K_Y$ is analytic for all $z \in \lbrace  a + i b \in \mathbb{C}: (a,b) \in \mathbb{R}^2 \text{ and } \alpha_{-} \leqslant a \leqslant \alpha_{+} \rbrace \subset \mathbb{C}$, then,
	\begin{equation}\label{EqPDFLaplace}
	f_{X_n}(x) = \frac{1}{2 \pi i} \displaystyle\int_{\gamma - i \infty}^{\gamma + i \infty} \exp\left(n K_{Y}(z) - z x\right) \mathrm{d}z,
	\end{equation}
	with $i = \sqrt{-1}$ and $\gamma \in (\alpha_{-}, \alpha_{+})$. 
	Note that the domain of $K_Y$ in~\eqref{EqPDFLaplace} has been extended to the complex plane and thus it is often referred to as the complex CGF. With an abuse of notation, both the CGF and the complex CGF are identically denoted.

	In the case in which  $n$ is sufficiently large, an approximation to the Bromwich integral in~\eqref{EqPDFLaplace} can be obtained by choosing the contour to include the unique saddlepoint of the integrand as suggested in~\citep{Daniels-AMS-1954}. 
	The intuition behind this lies on the following observations: 
	\begin{enumerate}[leftmargin=2.3em,labelsep=4mm]
	\item[$(i)$] the saddlepoint, denoted by $z_{0}$, is unique, real  and  $z_{0} \in (\alpha_{-}, \alpha_{+})$; 
	\item[$(ii)$] within a neighborhood around the saddlepoint of the form $\left|z - z_{0} \right| < \epsilon$, with $z \in \mathbb{C}$ and $\epsilon > 0$ sufficiently small,  
	$\mathrm{Im}\left[ n K_{Y}(z) - z x \right] = 0$ and $\mathrm{Re}\left[ n K_{Y}(z) -  z x \right]$ can be assumed constant; and 
	\item[$(iii)$] outside such neighborhood, the integrand is negligible. 
	\end{enumerate}
	
	From $(i)$, it follows that the derivative of $n K_Y(t) - t x$ with respect to $t$, with $t \in \mathbb{R}$, is equal to zero when it is evaluated at the saddlepoint $z_0$. More specifically, for all $t \in \mathbb{R}$,
	\begin{equation}
	\frac{\mathrm{d}}{\mathrm{d}t} K_Y(t) = \mathbb{E}_{P_{Y}}\left[Y \exp\left(  t  Y - K_{Y}(t)\right)  \right],
	\end{equation}
	and thus
	\begin{equation}\label{EqSaddlePoint}
	\mathbb{E}_{P_{Y}}\left[Y \exp\left(  z_{0}  Y - K_{Y}(z_{0})\right)  \right] = \frac{x}{n},
	\end{equation}
	which shows the dependence of $z_{0}$ on both $x$ and $n$.

	A Taylor series expansion of the exponent $nK_{Y}(z) -  z x$ in the neighborhood of $z_{0}$, leads to the following asymptotic expansion in powers of $\frac{1}{n}$ of the Bromwich integral in~\eqref{EqPDFLaplace}:
	\begin{IEEEeqnarray}{lcl}\label{EqPDFExpansion}
		f_{X_n}(x) &  = & \hat{f}_{X_n}(x) \left( 1 + \frac{1}{n} \left( \frac{1}{8} \frac{K_{Y}^{(4)}(z_{0})}{\left( K_{Y}^{(2)}(z_{0}) \right)^2} - \frac{5}{24} \frac{\left( K_{Y}^{(3)}(z_{0}) \right)^2}{\left( K_{Y}^{(2)}(z_{0}) \right)^{3}} \right) + \grandO{\frac{1}{n^2}}\right),
	\end{IEEEeqnarray}
	where $\hat{f}_{X_n} : \mathbb{R} \rightarrow \mathbb{R}_+$ is 
	\begin{IEEEeqnarray}{lcl}
		\label{Eqhatf}
		\hat{f}_{X_n}(x) &  = & \sqrt{\frac{1}{2 \pi n K_{Y}^{(2)}(z_{0})}}\exp\left( n K_Y(z_{0}) - z_{0} x \right),
	\end{IEEEeqnarray}
	and for all $k \in \mathbb{N}$ and $t \in \mathbb{R}$, the notation $K_Y^{(k)}(t)$ represents the $k$-th real derivative of the CGF $K_Y$ evaluated at $t$.
	The first two derivatives $K_Y^{(1)}$ and $K_Y^{(2)}$ play a central role, and thus it is worth providing explicit expressions. That is,
	\begin{IEEEeqnarray}{lcl}
		\label{EqTiltACMean}
		K_Y^{(1)} (t) & \triangleq &   \mathbb{E}_{P_{Y}}\left[Y \exp\left(  t  Y - K_{Y}(t)\right)  \right], \mbox{ and } \\
		\label{EqTiltACSecondMoment}
		K_Y^{(2)} (t) & \triangleq & \mathbb{E}_{P_{Y}}\left[ \left| Y - K_{Y}^{(1)}(t) \right|^2 \exp\left(  t  Y - K_{Y}(t)\right) \right].
	\end{IEEEeqnarray}

	The function $\hat{f}_{X_n}$ in~\eqref{Eqhatf} is referred to as the \emph{saddlepoint approximation} of the PDF $f_{X_n}$ and was first introduced in~\citep{Daniels-AMS-1954}. %
	Nonetheless,  $\hat{f}_{X_n}$ is not necessarily a PDF as often its integral on $\mathbb{R}$ is not  equal to one. 
	A particular exception is observed only in three cases~\citep{Daniels-Biometrica-1980}. 
	First, when $f_{Y}$ is the PDF of a Gaussian random variable,  the saddlepoint approximation $\hat{f}_{X_n}$ is identical to $f_{X_n}$, for all $n > 0$. 
	Second and third, when  $f_{Y}$ is the PDF associated with a Gamma distribution and an inverse normal  distribution, respectively, the saddlepoint approximation $\hat{f}_{X_n}$  is exact up to a normalization constant for all $n > 0$.

	An approximation to the CDF $F_{X_n}$ can be obtained by integrating the PDF in~\eqref{EqPDFLaplace}, cf.,~\citep{Daniels-ISR-1987, Temme-1982,Lugannani-Rice-AAP-1980}.
	In~particular, the result reported in~\citep{Daniels-ISR-1987} leads to an asymptotic expansion of the CDF of $X_{n}$, for all $x \in\mathbb{R}$, of the form:
	\begin{IEEEeqnarray}{lcl}
		\label{EqPDFapprox}
		F_{X_n}(x) & = & \hat{F}_{X_n}(x)  + \grandO{\frac{1}{\sqrt{n}}\exp{\left(n K_{Y}(z_0) - x z_0 \right)}},  
	\end{IEEEeqnarray}
	where the function $\hat{F}_{X_n}: \mathbb{R} \rightarrow \mathbb{R}$ is the \emph{saddlepoint approximation} of $F_{X_n}$. That is, for all $x  \in \mathbb{R}$, 
	\begin{IEEEeqnarray}{lcl}
		\label{EqHatPDF}
		\hat{F}_{X_n}(x) & = & \mathds{1}_{\left\lbrace z_{0} > 0 \right\rbrace} + (-1)^{\mathds{1}_{\left\lbrace z_{0} > 0 \right\rbrace }}  \exp\left( n K_{Y}(z_{0}) - z_{0} x + \frac{1}{2} z_{0}^2 n K^{(2)}_{Y}(z_{0}) \right) Q\left(|z_{0}| \sqrt{n K^{(2)}_{Y}(z_{0})}  \right), \qquad
	\end{IEEEeqnarray}
	where the function $Q: \mathbb{R} \rightarrow [0,1]$ is the complementary CDF of a Gaussian random variable with zero mean and unit variance. That is,
	for all $t \in \mathds{R}$,
	\begin{equation}\label{EqGSCCDF}
	Q(t) = \frac{1}{\sqrt{2 \pi}}\int_{t}^{\infty} \exp\left({-\frac{x^2}{2}}\right) \me{d}x.
	\end{equation}
		
	Finally, from the central limit theorem~\citep{Feller-book-1971}, for large values of $n$ and for all $x \in \mathbb{R}$, a reasonable approximation to $F_{X_n}(x)$ is $1 - Q(x)$. In the following, this approximation is referred to as the \emph{normal approximation}  of $F_{X_n}$. 
	
	\subsection{Contributions}
	
	The main contribution of this work is an upper bound on the error induced by the saddlepoint approximation  $\hat{F}_{X_n}$ in~\eqref{EqHatPDF} (Theorem \ref{TheoSaddlePointBeryUni} in Section~\ref{SecSumsII}).
	This result builds upon two observations. The~first observation is that the CDF $F_{X_n}$ can be written for all $x \in \mathbb{R}$ in the form,
	\begin{IEEEeqnarray}{lcl}
		\label{EqFXRW}
		F_{\hspace*{-0.2ex}X_n\hspace*{-0.2ex}}\hspace*{-0.2ex}(\hspace*{-0.2ex}x\hspace*{-0.2ex})\hspace*{-0.4ex} = \hspace*{-0.4ex} \mathds{1}_{\{\hspace*{-0.2ex}z_0 \leqslant 0\hspace*{-0.2ex}\}}\hspace*{-0.2ex}\mathbb{E}_{\hspace*{-0.1ex}P_{\hspace*{-0.1ex}S_n\hspace*{-0.1ex}}\hspace*{-0.1ex}}\hspace*{-0.4ex}\left[\hspace*{-0.3ex} \exp\hspace*{-0.2ex}\left(\hspace*{-0.2ex} n K_{\hspace*{-0.1ex}Y\hspace*{-0.1ex}}(\hspace*{-0.2ex}z_{0}\hspace*{-0.2ex}) \hspace*{-0.3ex}-\hspace*{-0.3ex}z_{0} S_n\hspace*{-0.2ex}\right)\hspace*{-0.3ex} \mathds{1}_{\{\hspace*{-0.2ex}S_n \leqslant x\hspace*{-0.2ex}\}}\hspace*{-0.4ex}\right]
		\hspace*{-0.5ex}+\hspace*{-0.5ex} \mathds{1}_{\{\hspace*{-0.2ex}z_0 > 0\hspace*{-0.2ex}\}}\hspace*{-0.6ex}\left(\hspace*{-0.4ex}1\hspace*{-0.3ex}-\hspace*{-0.3ex}\mathbb{E}_{P_{S_n}}\hspace*{-0.5ex}\left[\hspace*{-0.3ex} \exp\hspace*{-0.2ex}\left(\hspace*{-0.2ex} n K_{Y}(z_{0}) -z_{0} S_n\hspace*{-0.2ex}\right)\hspace*{-0.2ex} \mathds{1}_{\{S_n > x\}}\hspace*{-0.2ex}\right]\hspace*{-0.3ex}\right),\qquad
	\end{IEEEeqnarray} 
	where the random variable
	\begin{equation}\label{EqSnsum}
	S_n = \sum_{t=1}^{n} Y_t^{(z_0)}
	\end{equation}
	has a probability distribution denoted by $P_{S_n}$, and the random variables $Y_1^{(z_0)}$, $Y_2^{(z_0)}$, $\ldots$, $Y_n^{(z_0)}$ are independent with probability distribution $P_{Y^{(z_0)}}$.
	The distribution $P_{Y^{(z_0)}}$ is an exponentially tilted distribution \citep{Esscher-SAT-1932} with respect to the distribution $P_Y$ at the saddlepoint $z_0$. More specifically, the~Radon--Nikodym derivative of the distribution $P_{Y^{(z_0)}}$ with respect to the distribution $P_{Y}$ satisfies for all $y\in\mathrm{supp}P_{Y}$,
	\begin{IEEEeqnarray}{l}
		\frac{\mathrm{d}P_{Y^{(z_0)}}}{\mathrm{d} P_{Y}} (y) = \exp\left(- \left( K_Y(z_0) - z_0 y \right) \right).
	\end{IEEEeqnarray}

	The second observation is that the saddlepoint approximation $\hat{F}_{X_n}$ in~\eqref{EqHatPDF} can be written for all $x \in \mathbb{R}$ in the form, 
	\begin{IEEEeqnarray}{l}
		\label{EqSFXRW}
		\hat{F}_{\hspace*{-0.2ex}X_n\hspace*{-0.2ex}}\hspace*{-0.2ex}(\hspace*{-0.2ex}x\hspace*{-0.2ex})\hspace*{-0.4ex} = \hspace*{-0.4ex} \mathds{1}_{\{\hspace*{-0.2ex}z_0 \leqslant 0\hspace*{-0.2ex}\}}\hspace*{-0.2ex}\mathbb{E}_{\hspace*{-0.1ex}P_{\hspace*{-0.1ex}Z_n\hspace*{-0.1ex}}\hspace*{-0.1ex}}\hspace*{-0.4ex}\left[\hspace*{-0.3ex} \exp\hspace*{-0.2ex}\left(\hspace*{-0.2ex} n K_{\hspace*{-0.1ex}Y\hspace*{-0.1ex}}(\hspace*{-0.2ex}z_{0}\hspace*{-0.2ex}) \hspace*{-0.3ex}-\hspace*{-0.3ex}z_{0} Z_n\hspace*{-0.2ex}\right)\hspace*{-0.3ex} \mathds{1}_{\{\hspace*{-0.2ex}Z_n \leqslant x\hspace*{-0.2ex}\}}\hspace*{-0.4ex}\right]
		\hspace*{-0.5ex}+\hspace*{-0.5ex} \mathds{1}_{\{\hspace*{-0.2ex}z_0 > 0\hspace*{-0.2ex}\}}\hspace*{-0.6ex}\left(\hspace*{-0.4ex}1\hspace*{-0.3ex}-\hspace*{-0.3ex}\mathbb{E}_{P_{Z_n}}\hspace*{-0.5ex}\left[\hspace*{-0.3ex} \exp\hspace*{-0.2ex}\left(\hspace*{-0.2ex} n K_{Y}(z_{0}) -z_{0} Z_n\hspace*{-0.2ex}\right)\hspace*{-0.2ex} \mathds{1}_{\{Z_n > x\}}\hspace*{-0.4ex}\right]\hspace*{-0.3ex}\right)\hspace*{-0.4ex},\qquad
	\end{IEEEeqnarray}
	where $Z_n$ is a Gaussian random variable with mean $x$,  variance $n K_Y^{(2)} (z_0)$, and probability distribution $P_{Z_n}$. 
	Note that the means of the random variable $S_n$ in~\eqref{EqFXRW} and $Z_n$ in~\eqref{EqSFXRW} are equal to $n K_Y^{(1)}(z_0)$, whereas their variances  are equal to $n K_Y^{(2)} (z_0)$. 
	Note also that, from~\eqref{EqSaddlePoint}, it holds that $x = n K_Y^{(1)}(z_0)$.

	Using these observations, it holds that the absolute difference between $F_{X_n}$ in~\eqref{EqFXRW} and $\hat{F}_{X_n}$ in~\eqref{EqSFXRW} satisfies for all $x \in \mathbb{R}$,
	\begin{IEEEeqnarray}{l}
		\nonumber
		\left| F_{X_n}(x)  - \hat{F}_{X_n}(x) \right|  \\
		\nonumber
		=\mathds{1}_{\{z_0 \leqslant 0 \}}
		\left|\mathbb{E}_{P_{S_n}}\left[ \exp\left( n K_{Y}(z_{0}) -z_{0} S_n\right) \mathds{1}_{\{S_n \leqslant x\}}\right] - \mathbb{E}_{P_{Z_n}}\left[ \exp\left( n K_{Y}(z_{0})-z_{0} Z_n\right) \mathds{1}_{\{Z_n \leqslant x\}}\right]\right|  \\
		+  \mathds{1}_{\{z_0 > 0 \}}
		\left|\mathbb{E}_{P_{S_n}}\left[ \exp\left( n K_{Y}(z_{0}) -z_{0} S_n\right) \mathds{1}_{\{S_n > x\}}\right] - \mathbb{E}_{P_{Z_n}}\left[ \exp\left( n K_{Y}(z_{0})-z_{0} Z_n\right) \mathds{1}_{\{Z_n > x\}}\right]\right|.  \qquad
	\end{IEEEeqnarray}
	A step forward (Lemma~\ref{lem:1ForLem2} in Appendix~\ref{plem:saddlePointBeryUni}) is to note that, when $x$ is such that $z_0\leqslant 0$, then,
	\begin{IEEEeqnarray}{l}
		\left|\mathbb{E}_{P_{S_n}}\left[ \exp\left( n K_{Y}(z_{0}) -z_{0} S_n\right) \mathds{1}_{\{S_n \leqslant x\}}\right] - \mathbb{E}_{P_{Z_n}}\left[ \exp\left( n K_{Y}(z_{0})-z_{0} Z_n\right) \mathds{1}_{\{Z_n \leqslant x\}}\right]\right|\nonumber\\
		\leqslant \exp\left( n K_{Y}(z_{0}) -z_{0} x\right) \min\left\{ 1,  2 \sup_{a\in\mathbb{R}}\left|F_{S_n} (a) - F_{Z_n}(a) \right| \right\}, 
	\end{IEEEeqnarray}
	and when $x$ is such that $z_0 > 0$, it holds that
	\begin{IEEEeqnarray}{l}
		\left|\mathbb{E}_{P_{S_n}}\left[ \exp\left( n K_{Y}(z_{0}) -z_{0} S_n\right) \mathds{1}_{\{S_n > x\}}\right] - \mathbb{E}_{P_{Z_n}}\left[ \exp\left( n K_{Y}(z_{0})-z_{0} Z_n\right) \mathds{1}_{\{Z_n >  x\}}\right]\right|\nonumber\\
		\leqslant \exp\left( n K_{Y}(z_{0}) -z_{0} x\right)  \min\left\{ 1,  2 \sup_{a\in\mathbb{R}}\left|F_{S_n} (a) - F_{Z_n}(a) \right| \right\},
	\end{IEEEeqnarray}
	where  $F_{S_n}$ and $F_{Z_n}$ are the CDFs of the random variables $S_n$ and $Z_n$, respectively.
	The final result is obtained by observing that $\sup_{a\in\mathbb{R}}\left|F_{S_n} (a) - F_{Z_n}(a) \right|$ can be upper bounded using the Berry--Esseen Theorem (Theorem~\ref{TheoBerry} in Section~\ref{SecSumsI}). This is essentially due to the fact that the random variable $S_n$ is the sum of $n$ independent random variables, i.e.,~\eqref{EqSnsum}, and $Z_n$ is a Gaussian random variable, and both $S_n$ and $Z_n$ possess identical means and variances. 
	Thus, the main result (Theorem~\ref{TheoSaddlePointBeryUni} in Section~\ref{SecSumsII}) is that, for all $x \in \mathbb{R}$,
	\begin{IEEEeqnarray}{lcl}
		\label{EqTheEquation}
		\left| F_{X_n}(x)  -\hat{F}_{X_n}(x) \right| 
		\leqslant
		\frac{2 \xi_{Y}(z_0)}{\sqrt{n}}\exp\left( n K_Y(z_0) -z_0\, x\right),
	\end{IEEEeqnarray} 
	where
	\begin{IEEEeqnarray}{l}
		\label{EqTiltACThirdMomentIn}
		\xi_Y (z_0)  =  c_1\left(\frac{\mathbb{E}_{P_{Y}}\left[ \left| Y - K_{Y}^{(1)}(z_0) \right|^3 \exp\left(  z_0  Y - K_{Y}(z_0)\right)  \right] }{\left(K_Y^{(2)}(z_0)\right)^{3/2}}  + c_2\right),
	\end{IEEEeqnarray}
	with
		\begin{subequations}\label{EqConstants}
			\begin{IEEEeqnarray}{l}
				c_1 \df 0.33554, \mbox{ and}\\
				c_2 \df 0.415.
			\end{IEEEeqnarray}
		\end{subequations}

	Finally, 
	note that \eqref{EqTheEquation} holds for any finite value of $n$ and admits the asymptotic scaling law with respect to $n$ suggested in \eqref{EqPDFapprox}.

	\subsection{Applications}
	In the realm of information theory, the normal approximation has played a central role in the calculation of bounds on the minimum decoding error probability (DEP) in point-to-point memoryless channels, cf.,~\citep{Polyanskiy2010, MolavianJazi2015}. Thanks to the normal approximation, simple approximations for  the dependence testing (DT) bound, the random coding union bound (RCU) bound, and the meta converse (MC) bound have been obtained in~\citep{Polyanskiy2010, MolavianJaziPhD}.  The success of these approximations stems from the fact that they are easy to calculate. Nonetheless, easy  computation  comes  at  the  expense  of  loose upper and lower bounds and thus uncontrolled approximation errors.

	On the other hand,  saddlepoint techniques have been extensively used to approximate existing lower and upper bounds on the minimum DEP. See, for instance,~\citep{fJos2018,Martinez2014} in the case of the RCU bound and the MC bound. 
	Nonetheless, the errors induced by saddlepoint approximations are often neglected due to the fact that calculating them involves a large number of optimizations and numerical integrations. 
	Currently, the validation of saddlepoint approximations is carried through Monte Carlo simulations. Within this context, the main objectives of this paper are twofold: 
		$(a)$ to analytically assess the tightness of the approximation of DT and MC bounds based on the saddlepoint approximation of the CDFs of sums of i.i.d. random variables;
		$(b)$ to provide new lower and upper bounds on the minimum DEP by providing a lower bound on the MC bound and an upper bound on the DT bound.  
		Numerical experimentation of these bounds is presented for the binary symmetric channel (BSC), the additive white Gaussian noise (AWGN) channel, and the additive symmetric $\alpha$-stable noise (S$\alpha$S) channel, where the new bounds are  tight and obtained at low computational cost.

	\section{Sums of Independent and Identically Distributed Random Variables}\label{SecSums}

	In this section, upper bounds on the absolute error of approximating $F_{X_n}$ by the \emph{normal approximation} and the \emph{saddlepoint approximation} are presented.
	
	\subsection{Error Induced by the Normal Approximation} \label{SecSumsI}
	Given a random variable $Y$, let the function  $\xi_Y: \mathbb{R}\xrightarrow{} \mathbb{R}$ be for all $t\in\mathbb{R}$ :
	\begin{IEEEeqnarray}{lcl}
		\label{EqTiltACThirdMoment}
		\xi_Y (t) & \triangleq &  c_1\left(\frac{\mathbb{E}_{P_{Y}}\left[ \left| Y - K_{Y}^{(1)}(t) \right|^3 \exp\left(  t  Y - K_{Y}(t)\right)  \right] }{\left(K_Y^{(2)}(t)\right)^{3/2}}  + c_2\right),  
	\end{IEEEeqnarray}
	where $c_1$ and $c_2$ are defined in~\eqref{EqConstants}.

	The following theorem, known as the Berry--Esseen theorem~\citep{Feller-book-1971},  introduces an upper bound on the approximation error induced by the normal approximation. 
	\begin{Theorem}[Berry--Esseen~{\citep{Shevtsova-arxiv-2011}}]\label{TheoBerry} Let $Y_1$, $Y_2$, $\ldots$, $Y_n$ be i.i.d random variables with probability distribution $P_Y$. Let also $Z_n$ be a Gaussian random variable with mean $n \, K_Y^{(1)}(0)$,  variance  $n \, K_Y^{(2)}(0)$, and CDF denoted by $F_{Z_n}$. Then, the CDF  of the random variable $X_n$ $=$ $Y_1$ $+$ $Y_2$ $+$ $\ldots$ $+$ $Y_n$, denoted by $F_{X_n}$, satisfies
		\begin{eqnarray}\label{EqUppBoundBerryEsseen}
			\underset{a\in\mathbb{R}}{\sup}\left|F_{X_n}(a) - F_{Z_n}(a)\right| \leqslant \min\left\{1,\frac{\xi_Y(0)}{\sqrt{n } }\right\},
		\end{eqnarray} 
		 where the functions $K_Y^{(1)}$,  $K_Y^{(2)}$ and $\xi_Y$ are defined in~\eqref{EqTiltACMean},~\eqref{EqTiltACSecondMoment}, and~\eqref{EqTiltACThirdMoment}.
	\end{Theorem}

	An immediate result from Theorem~\ref{TheoBerry} gives the following upper and lower bounds on $F_{X_{n}}(a)$, for~all  $a \in \mathbb{R}$, 
	\begin{IEEEeqnarray}{lcl}
		\label{EqUpperBerry}
		F_{X_n}(a) \leqslant  F_{Z_n}(a)  + \min\left\{1,\frac{\xi_Y(0)}{\sqrt{n } }\right\} \triangleq \bar{\Sigma}(a,n), \mbox{ and }  \\
		\label{EqLowerBerry}
		F_{X_n}(a) \geqslant F_{Z_n}(a)  - \min\left\{1,\frac{\xi_Y(0)}{\sqrt{n } }\right\} \triangleq \underline{\Sigma}(a,n).
	\end{IEEEeqnarray}

	The main drawback of Theorem~\ref{TheoBerry} is that the upper bound on the approximation error does not depend on the exact value of  $a$.  
	More importantly, for some values of $a$ and $n$,  the upper bound on the approximation error  might be particularly big, which leads to irrelevant results. 
	
	\subsection{Error Induced by the Saddlepoint Approximation}\label{SecSumsII}
	
	The following theorem introduces an upper bound on the approximation error induced by approximating the CDF $F_{X_{n}}$ of $X_{n}$ in~\eqref{EqXn}  by  the function $\eta_{Y}:$ $\mathbb{R}^2$$\times$ $\mathbb{N}$  $\rightarrow$ $\mathbb{R}$ defined such that for all $(\theta$, $a$, $n)$ $\in$ $\mathbb{R}^2$ $\times$ $\mathbb{N}$, 
	\begin{IEEEeqnarray}{l}
		\nonumber
		\eta_{Y}(\hspace*{-0.2ex}\theta,\hspace*{-0.2ex}a,\hspace*{-0.2ex}n\hspace*{-0.2ex})  \hspace*{-0.5ex}  \\
		\triangleq\mathds{1}_{\{\hspace*{-0.2ex}\theta > 0 \hspace*{-0.2ex}\}}
		\hspace*{-0.4ex} + \hspace*{-0.2ex} (\hspace*{-0.2ex}-\hspace*{-0.2ex}1)^{\mathds{1}_{\{\hspace*{-0.2ex}\theta > 0\hspace*{-0.2ex}\}}}\hspace*{-0.4ex}\exp\hspace*{-0.5ex}\left(\hspace*{-0.5ex}\frac{1}{2} n \theta^2 K_Y^{(2)}(\hspace*{-0.2ex}\theta\hspace*{-0.2ex}) \hspace*{-0.4ex} + \hspace*{-0.3ex} n K_Y(\theta)\hspace*{-0.4ex}-\hspace*{-0.3ex}n\theta K_Y^{(1)}(\hspace*{-0.2ex}\theta\hspace*{-0.2ex}) \hspace*{-0.6ex}\right) 
		Q\hspace*{-0.6ex}\left(\hspace*{-0.8ex}(\hspace*{-0.2ex}-\hspace*{-0.2ex}1\hspace*{-0.2ex})^{\mathds{1}_{\{\hspace*{-0.2ex}\theta \leqslant 0\hspace*{-0.2ex}\}}}\frac{a\hspace*{-0.3ex}+\hspace*{-0.3ex}n \theta K_Y^{(2)}(\hspace*{-0.2ex}\theta\hspace*{-0.2ex})\hspace*{-0.4ex}-\hspace*{-0.4ex} n K_Y^{(1)}(\theta) }{\sqrt{n K_Y^{(2)}(\theta)}}\hspace*{-0.6ex}\right),\qquad
		\label{EqEtaApprox}
	\end{IEEEeqnarray}    
	where the function $Q: \mathbb{R} \rightarrow [0,1]$ is the complementary CDF of the standard Gaussian distribution defined in~\eqref{EqGSCCDF}.
	Note that $\eta_Y(\theta,n,a)$ is identical to $\hat{F}_{X_n}(a)$, when $\theta$ is chosen to satisfy the saddlepoint $K^{(1)}_{Y}(\theta) = \frac{a}{n}$. 
	Note also that $\eta_Y(0,n,a)$ is the CDF of a Gaussian random variable with mean $n K_{Y}^{(1)}(0)$ and variance $n K_{Y}^{(2)}(0)$, which are the mean and the variance of $X_n$ in \eqref{EqXn}, respectively.
	
	\begin{Theorem}\label{lem:saddlePointBeryUni}
		Let $Y_1$, $Y_2$, $\ldots$, $Y_n$ be i.i.d. random variables with probability distribution $P_{Y}$ and CGF $K_{Y}$. Let~also $F_{X_n}$ be the CDF of the random variable $X_n$ $=$ $Y_1$ $+$ $Y_2$ $+$ $\ldots$ $+$ $Y_n$. Hence,  for all  $a$ $\in$ $\mathbb{R}$ and for all $\theta\in\Theta_Y$, it~holds that
		\begin{eqnarray}
			\label{EqTilteapprox}
			\left|F_{X_{n}}(a) -\eta_Y\left(\theta,a,n\right) \right| \leqslant  \exp\left(nK_Y(\theta)-\theta \, a \right)\min\left\{1,\frac{2\;  \xi_Y(\theta)}{\sqrt{n}}\right\},
		\end{eqnarray}
		where
		\begin{eqnarray}\label{EqThetaY}
			\Theta_Y \triangleq \lbrace t\in\mathbb{R}: K_Y(t)< \infty \rbrace;
		\end{eqnarray} 
		and the functions $\xi_Y$ and $\eta_Y$ are defined in~\eqref{EqTiltACThirdMoment} and~\eqref{EqEtaApprox}, respectively.
	\end{Theorem}
	\begin{proof} 
		The proof of  Theorem~\ref{lem:saddlePointBeryUni} is presented in Appendix~\ref{plem:saddlePointBeryUni}. 
	\end{proof}

	This result leads to the following upper and lower bounds on $F_{X_n}(a)$, for all  $a \in \mathbb{R}$,
	\begin{IEEEeqnarray}{lcl}
		\label{EqUpperBoundFX}
		F_{X_n}(a)  & \leqslant &   \eta_Y\left(\theta,a,n\right) + \exp\left(nK_Y(\theta)-\theta \, a \right)\min\left\{1,\frac{2\;  \xi_Y(\theta)}{\sqrt{n}}\right\}, \mbox{ and } \\
		\label{EqLowerBoundFX}
		F_{X_n}(a)  & \geqslant &   \eta_Y\left(\theta,a,n\right) - \exp\left(nK_Y(\theta)-\theta \, a \right)\min\left\{1,\frac{2\;  \xi_Y(\theta)}{\sqrt{n}}\right\},
	\end{IEEEeqnarray}
	with $\theta$ $\in$ $\Theta_{Y}$.

	The advantages of  approximating $F_{X_{n}}$ by using  Theorem~\ref{lem:saddlePointBeryUni} instead of Theorem~\ref{TheoBerry} are twofold.
	First, both the approximation $\eta_Y$ and the corresponding approximation error depend on the exact value of $a$. In particular, the approximation can be optimized for each value of $a$ via the parameter $\theta$.
	Second, the~parameter $\theta$ in~\eqref{EqTilteapprox} can be optimized to improve either the upper bound in~\eqref{EqUpperBoundFX} or the lower bound in~\eqref{EqLowerBoundFX} for some $a \in \mathbb{R}$. Nonetheless, such optimizations are not necessarily simple.

	An alternative to the optimization on $\theta$ in~\eqref{EqUpperBoundFX} and~\eqref{EqLowerBoundFX} is to choose $\theta$ such that it minimizes $nK_Y(\theta)-\theta \, a$. This follows the intuition that, for some values of $a$ and $n$, the term  $\exp(nK_Y(\theta)-\theta \, a)$ is the one that influences the most the value of the right-hand side of~\eqref{EqTilteapprox}.  
	To build upon this idea, consider the following lemma.
	\begin{Lemma} \label{LemmaH}
		Consider a random variable $Y$ with probability distribution $P_Y$ and CGF $K_{Y}$. Given $n\in\mathbb{N}$, let the function $h: \mathbb{R} \rightarrow \mathbb{R}$ be defined for all $a\in\mathbb{R}$ satisfying $\frac{a}{n} \in \mathrm{int}\set{C}_{Y}$, with $\mathrm{int}\set{C}_{Y}$ denoting the interior of the convex hull of $\mathrm{supp} \,P_{X_n}$, as follows:
		\begin{eqnarray}
			\label{EqFunctionH}
			h(a) = \inf_{\theta \in \Theta_{Y}} n K_Y(\theta) - \theta\, a,
		\end{eqnarray}
		where $\Theta_{Y}$ is defined in~\eqref{EqThetaY}.
		Then, the function $h$ is  concave and for all $a \in \mathbb{R}$,
		\begin{eqnarray}
			h(a) \leqslant h(n\mathbb{E}_{P_Y}[Y]) = 0.
		\end{eqnarray} 
		Furthermore,  
		\begin{eqnarray}
			\label{EqHRedef}
			h(a) =  n K_Y(\theta^{\star}) - \theta^{\star}\, a,
		\end{eqnarray}
		where $\theta^{\star}$ is the unique solution in $\theta$ to
		\begin{eqnarray}\label{EqThetaStar}
			n K_Y^{(1)}(\theta) = a,
		\end{eqnarray}
		with $K_Y^{(1)}$ is defined in~\eqref{EqTiltACMean}.
	\end{Lemma}
	\begin{proof}
		The proof of  Lemma  \ref{LemmaH} is presented in Appendix  \ref{pLemmaH}.
	\end{proof}

	Given $(a,n) \in \mathbb{R} \times \mathbb{N}$, the value of $h(a)$ in~\eqref{EqFunctionH} is the argument that minimizes the exponential term in~\eqref{EqTilteapprox}. 
	An interesting observation from Lemma~\ref{LemmaH} is that the maximum of $h$ is zero, and it is reached when $a = n\mathbb{E}_{P_Y}[Y] = \mathbb{E}_{P_{X_n}}[X_n]$. 
	In this case, $\theta^{\star} = 0$, and thus, from~\eqref{EqUpperBoundFX} and~\eqref{EqLowerBoundFX}, it holds that
	\begin{IEEEeqnarray}{lcl}
		\nonumber
		F_{X_n}(a)  & \leqslant &   \eta_Y\left(0,a,n\right) + \min\left\{1,\frac{2\;  \xi_Y(0)}{\sqrt{n}}\right\} \\
		\label{EqUpperBoundFXBad}
		& = &  F_{Z_n}(a)   + \min\left\{1,\frac{2\;  \xi_Y(0)}{\sqrt{n}}\right\},  \mbox{ and } \\
		\nonumber
		F_{X_n}(a)  & \geqslant &   \eta_Y\left(0,a,n\right) - \min\left\{1,\frac{2\; \xi_Y(0)}{\sqrt{n}}\right\} \\
		\label{EqLowerBoundFXBad}
		& = &  F_{Z_n}(a)   - \min\left\{1,\frac{2\; \xi_Y(0)}{\sqrt{n}}\right\},
	\end{IEEEeqnarray}
	where $F_{Z_n}$ is the CDF defined in Theorem~\ref{TheoBerry}.
	Hence, the upper bound in~\eqref{EqUpperBoundFXBad} and the lower bound in~\eqref{EqLowerBoundFXBad} obtained from Theorem~\ref{lem:saddlePointBeryUni} 
	are worse than those in~\eqref{EqUpperBerry} and~\eqref{EqLowerBerry} obtained from Theorem~\ref{TheoBerry}. 
	In~a nutshell, for values of $a$ around the vicinity of $n\mathbb{E}_{P_Y}[Y] = \mathbb{E}_{P_{X_n}}[X_n]$, it is more interesting to use Theorem~\ref{TheoBerry} instead of Theorem~\ref{lem:saddlePointBeryUni}.

	Alternatively, given that $h$ is non-positive and concave, when $\left| a - n\mathbb{E}_{P_Y}[Y] \right|$  $=$ $\big| a -$ $ \mathbb{E}_{P_{X_{n}}}[X_{n}] \big|$ $>$ $\gamma$, with $\gamma$ sufficiently large, it follows that
	\begin{IEEEeqnarray}{lcl}
		\exp\left(nK_Y(\theta^{\star})-\theta^{\star} \, a \right) < \min\left\{1,\frac{\xi_Y(0)}{\sqrt{n} }\right\}, 
	\end{IEEEeqnarray}
	with $\theta^{\star}$ defined in~\eqref{EqThetaStar}.
	Hence, in this case, the right-hand side of~\eqref{EqTilteapprox} is always smaller than the right-hand side of~\eqref{EqUppBoundBerryEsseen}.
	That is, for such values of $a$ and $n$, the  upper and lower bounds in~\eqref{EqUpperBoundFX} and~\eqref{EqLowerBoundFX} are better than those in~\eqref{EqUpperBerry} and~\eqref{EqLowerBerry}, respectively. 
	The following theorem leverages this observation. 
	
	\begin{Theorem}\label{TheoSaddlePointBeryUni}
		Let $Y_1$, $Y_2$, $\ldots$, $Y_n$ be i.i.d. random variables with probability distribution $P_{Y}$ and CGF $K_{Y}$. Let~also $F_{X_n}$ be the CDF of the random variable $X_n = Y_1 + Y_2 + \ldots + Y_n$. Hence,  for all  $a$ $\in$ $\mathrm{int} \, \set{C}_{X_n}$, with $\mathrm{int} \, \set{C}_{X_n}$ the interior of the convex hull of $\mathrm{supp}{P_{X_n}}$, it holds that
		\begin{eqnarray}
			\label{EqSPAprox}
			\left| F_{X_n}(a )  -\hat{F}_{X_n}(a) \right| 
			\leqslant
			\exp\left( n K_Y(\theta^{\star}) - \theta^{\star} \, a\right) \min\left\{1,\frac{ 2 \,\xi_Y(\theta^{\star})}{\sqrt{n}}\right\},
		\end{eqnarray} 
		where $\theta^{\star}$ is defined in~\eqref{EqThetaStar},  and the functions $\hat{F}_{X_n}$ and $\xi_Y$ are defined in~\eqref{EqHatPDF}, and~\eqref{EqTiltACThirdMoment},  respectively.
	\end{Theorem}
	
	\begin{proof}
		The proof of  Theorem~\ref{TheoSaddlePointBeryUni} is presented in Appendix~\ref{pTheoSaddlePointBeryUni}.
	\end{proof}

	An immediate result from Theorem \ref{TheoSaddlePointBeryUni} gives the following upper and lower bounds on $F_X(a)$, for~all  $a \in \mathbb{R}$ , 
	\begin{IEEEeqnarray}{lcl}
		\label{EqUpperAnade}
		F_{X_n}(a )  & \leqslant & \hat{F}_{X_n}(a) + 
		\exp\left( n K_Y(\theta^{\star}) - \theta^{\star} \, a\right) \min\left\{1,\frac{ 2 \,\xi_Y(\theta^{\star})}{\sqrt{n}}\right\} \triangleq \bar{\Omega}(a,n), \mbox{ and } \quad \\
		\label{EqLowerAnade}
		F_{X_n}(a )  & \geqslant & \hat{F}_{X_n}(a) - 
		\exp\left( n K_Y(\theta^{\star}) - \theta^{\star} \, a\right) \min\left\{1,\frac{ 2  \,\xi_Y(\theta^{\star})}{\sqrt{n}}\right\} \triangleq \underline{\Omega}(a,n).
	\end{IEEEeqnarray}
	The following section presents two examples that highlight the observations mentioned above.
	\subsection{Examples}

\begin{figure}[H]
	\centering
			\includegraphics[width=.6\linewidth]{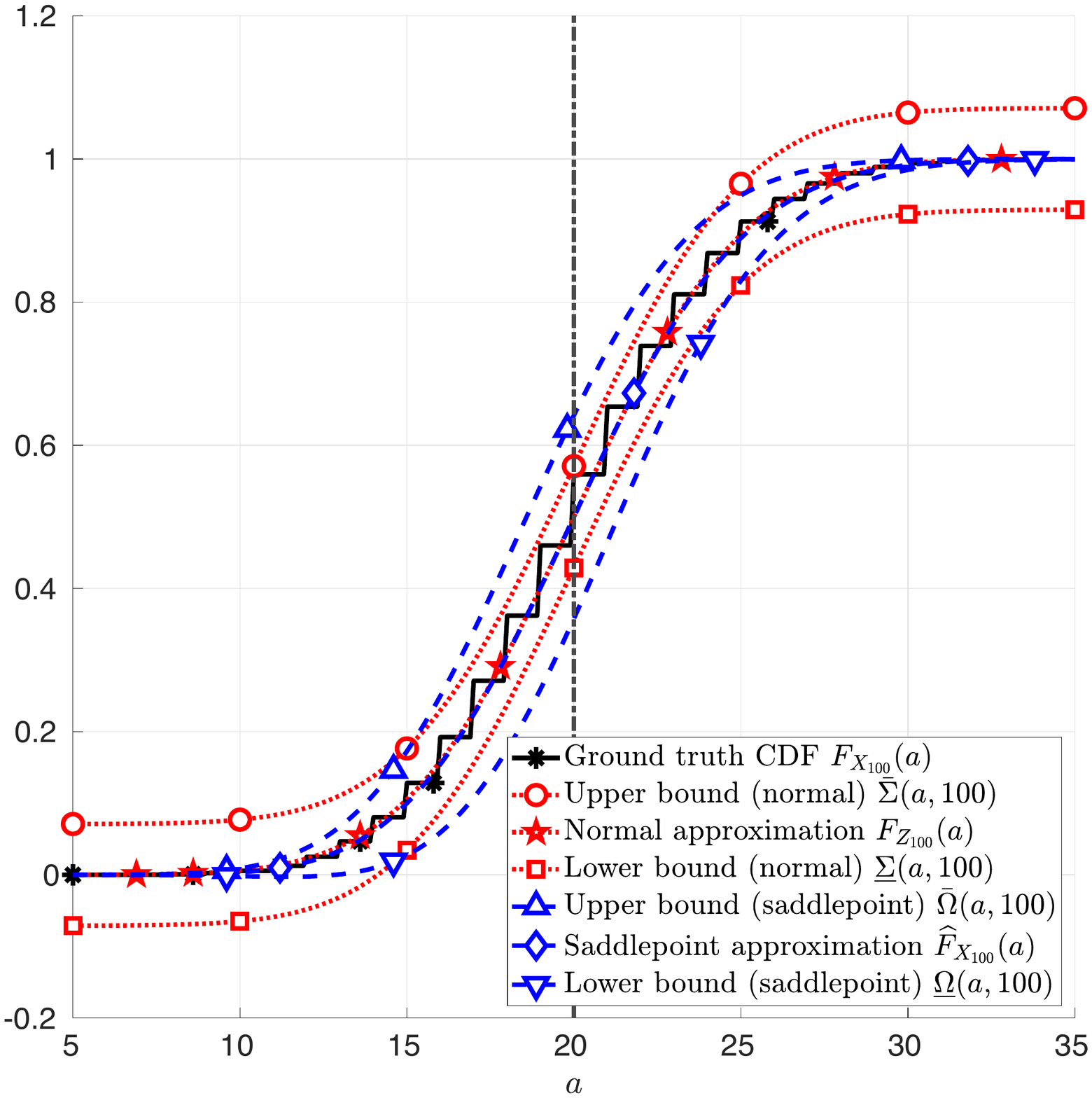}
			\caption{
				Sum of $100$ Bernoulli random variables with parameter $p = 0.2$. 
					The function $F_{X_{100}}(a)$ (asterisk markers $\boldsymbol{*}$) in Example \ref{ExampleDiscret}; 
					the function $F_{Z_{100}}(a)$ (star markers $\color{red} \star$) in \eqref{EqUppBoundBerryEsseen}; 
					the function $\hat{F}_{X_{100}}(a)$ (diamond markers $\color{blue} \diamond$) in \eqref{EqHatPDF}; 
					the function $\bar{\Sigma}(a,100)$ (circle marker $\color{red} \circ$) in \eqref{EqUpperBerry}; 
					the function $\underline{\Sigma}(a, 100)$ (square marker~$\color{red} \square$) in \eqref{EqLowerBerry}; 
					the function $\bar{\Omega}(a, 100)$ (upward-pointing triangle marker $\color{blue} \triangle$) in \eqref{EqUpperAnade};  
					and the function $\underline{\Omega}(a,100)$ (downward-pointing triangle marker $\color{blue} \triangledown$) in \eqref{EqLowerAnade} 
					are plotted as functions of $a$, with $a \in [5,35]$.
			}
			\label{FigBin}
	\end{figure}

	\begin{Example}[Discrete random variable]\label{ExampleDiscret} Let the random variables $Y_1$, $Y_2$, $\ldots$, $Y_n$ in \eqref{EqXn} be  i.i.d. Bernoulli random variables with parameter $p=0.2$ and $n=100$. In this case, $\mathds{E}_{P_{X_n}}\left[ X_n \right] =  n \mathds{E}_{P_Y}\left[ Y \right] = 20$.
		Figure \ref{FigBin} depicts the CDF $F_{X_{100}}$ of $X_{100}$ in \eqref{EqXn}; the normal approximation $F_{Z_{100}}$ in \eqref{EqUppBoundBerryEsseen}; and the saddlepoint approximation $\hat{F}_{X_{100}}$  in \eqref{EqHatPDF}. Therein, it is also depicted  the upper and lower bounds due to the normal approximation $\bar{\Sigma}$ in \eqref{EqUpperBerry} and $\underline{\Sigma}$ in \eqref{EqLowerBerry}, respectively; and the upper and lower bounds due to the saddlepoint approximation $\bar{\Omega}$   in \eqref{EqUpperAnade} and $\underline{\Omega}$  in \eqref{EqLowerAnade}, respectively. These functions are plotted as a function of $a$, with $a$ $\in$ $[5,35]$.
	\end{Example}
	
\begin{figure}[H]
	\centering
			\includegraphics[width=.6\linewidth]{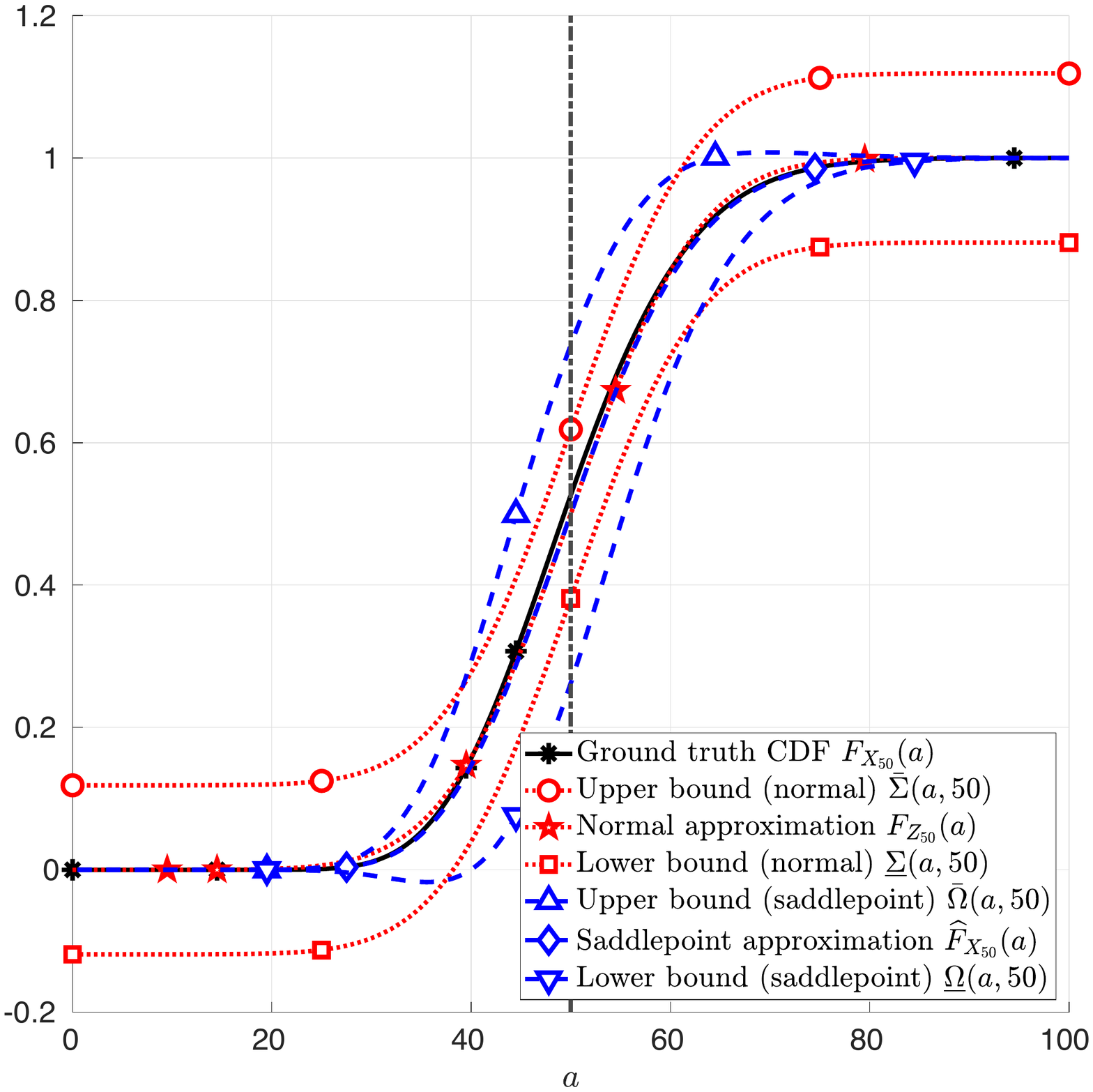}
			\caption{
				Sum of $50$ Chi-squared  random variables with parameter $k = 1$.
					The function $F_{X_{50}}(a)$ (asterisk markers $\boldsymbol{*}$) in Example \ref{ExampleContinous}; 
					the function $F_{Z_{50}}(a)$ (star markers $\color{red} \star$) in \eqref{EqUppBoundBerryEsseen}; 
					the function $\hat{F}_{X_{50}}(a)$ (diamond markers $\color{blue} \diamond$) in \eqref{EqHatPDF}; 
					the function $\bar{\Sigma}(a,50)$ (circle marker $\color{red} \circ$) in \eqref{EqUpperBerry}; 
					the function $\underline{\Sigma}(a, 50)$ (square marker $\color{red} \square$) in \eqref{EqLowerBerry}; 
					the function $\bar{\Omega}(a, 50)$ (upward-pointing triangle marker $\color{blue} \triangle$) in \eqref{EqUpperAnade};  
					and the function $\underline{\Omega}(a,50)$ (downward-pointing triangle marker $\color{blue} \triangledown$) in \eqref{EqLowerAnade} 
					are plotted as functions of $a$,  with $a \in [0,100]$.
			}
			\label{FigChi}
	\end{figure}

	\begin{Example}[Continuous random variable]\label{ExampleContinous}
		Let the random variables $Y_1$, $Y_2$, $\ldots$, $Y_n$ in \eqref{EqXn} be  i.i.d. chi-squared random variables with parameter $k = 1$ and $n=50$. In this case, $\mathds{E}_{P_{X_n}}\left[ X_n \right] =  n \mathds{E}_{P_Y}\left[ Y \right] = 50$.
		Figure \ref{FigChi} depicts the CDF $F_{X_{50}}$ of $X_{50}$ in \eqref{EqXn};
		the normal approximation $F_{Z_{50}}$ in \eqref{EqUppBoundBerryEsseen}; 
		and the saddlepoint approximation $\hat{F}_{X_{50}}$  in~\eqref{EqHatPDF}.
		Therein, it is also depicted  the upper and lower bounds due to the normal approximation $\bar{\Sigma}$ in \eqref{EqUpperBerry} and $\underline{\Sigma}$  in \eqref{EqLowerBerry}, respectively; and the upper and lower bounds due to the saddlepoint approximation $\bar{\Omega}$   in \eqref{EqUpperAnade} and $\underline{\Omega}$  in \eqref{EqLowerAnade}, respectively. These functions are plotted as a function of $a$, with $a$ $\in$ $[0,100]$.
	\end{Example}
	
	\unskip

	\section{Application to Information Theory: Channel Coding}\label{secApplication}
	
	This section focuses on the study of the DEP in point-to-point memoryless channels. 
	The problem is formulated in Section~\ref{SecSystemModel}. 
	The main results presented in this section consist of lower and upper bounds on the DEP. The former, which are obtained building upon  the existing DT  bound~\citep{Polyanskiy2010}, are~presented in Section~\ref{SecDTBound}. The latter, which are obtained from the MC bound~\citep{Polyanskiy2010},  are~presented in Section~\ref{SecMCBound}.  
	
	\subsection{System Model} \label{SecSystemModel}
	
	Consider a point-to-point communication in which a transmitter aims at sending information to one receiver through a noisy memoryless channel. Such a channel can be modeled by a random transformation
	\begin{IEEEeqnarray}{lcl}\label{EqGeneralRamdonTransfomation}
		(\set{X}^n, \set{Y}^n, P_{\vec{Y}|\vec{X}}),
	\end{IEEEeqnarray}
	where $n \in$ $\mathbb{N}$ is the blocklength and $\mathcal{X}$ and $\mathcal{Y}$ are the channel input and channel output sets. Given~the channel inputs $\vec{x}$ $=$ $(x_{1}$, $x_{2}$, $\ldots$, $x_{n})$ $\in$ $\set{X}^n$, the outputs $\vec{y}$ $=$ $(y_1$, $y_2$, $\ldots$, $y_n)$ $\in$ $\set{Y}^n$ are observed at the receiver  with probability 
	\begin{IEEEeqnarray}{lcl}
		\label{EqMemChan}
		P_{\vec{Y}|\vec{X}}(\vec{y}|\vec{x}) = \prod_{t=1}^n P_{Y|X}(y_t|x_t),
	\end{IEEEeqnarray}
	where, for all $x \in \mathcal{X}$, $P_{Y|X = x} \in \triangle\left( \mathcal{Y}\right)$, with  $\triangle\left( \mathcal{Y} \right)$, the set of all possible probability distributions whose support is a subset of $\mathcal{Y}$.
	The objective of the communication is to transmit a message index $i$, which is a realization of a random variable  $W$ that is uniformly distributed  over the set 
	\begin{IEEEeqnarray}{lcl}\label{EqWset}
		\mathcal{W} \triangleq  \lbrace 1,2, \ldots, M\rbrace, 
	\end{IEEEeqnarray}
	with 
	$1$ $<$ $M$ $<$ $\infty$. 
	To achieve this objective, the transmitter uses an $(n$, $M$, $\lambda)$-code, where $\lambda$ $\in$ $[0,1]$.
	\begin{Definition}[$(n$, $M$,$\lambda)$-code]\label{defP2PCode}
		Given a tuple $(M$, $n$, $\lambda)$ $\in$ $\mathbb{N}^2$$\times$ $[0,1]$, an $(n$, $M$, $\lambda)$-code for the random transformation in~\eqref{EqGeneralRamdonTransfomation} is a system
		\begin{eqnarray}
			\label{EqP2PCode}
			\Bigg\{\bigg(\vec{u}(1), \set{D}(1)\bigg), \bigg(\vec{u}(2), \set{D}(2)\bigg),  \ldots,\bigg(\vec{u}(M), \set{D}(M)\bigg)\Bigg\},
		\end{eqnarray}
		where for all $(j, \ell) \in \set{W}^2$, with  $j \neq \ell$:
		\begin{subequations} \label{eqDefP2PCode}
			\begin{eqnarray}
				\label{eqDefP2PCode1}
				\vec{u}(j)  & = & (u_{1}(j), u_{2}(j), \ldots, u_{n}(j)) \in \set{X}^{n},\\
				\label{eqDefP2PCode2} 
				\set{D}(j) \cap \set{D}(\ell) &  = & \emptyset,\\
				\label{eqDefP2PCode3}
				\bigcup_{j \in \set{W}}\set{D}(j) & \subseteq & \set{Y}^n,\mbox{ and }  \\
				\label{eqDefAvgProbEr}
				\frac{1}{M}\sum_{i=1}^{M} \mathbb{E}_{P_{\vec{Y}|\vec{X} = \vec{u}(i)} }\left[\mathds{1}_{\left\{Y\notin \set{D}(i) \right\}}\right] & \leqslant & \lambda.  
			\end{eqnarray}
		\end{subequations}
	\end{Definition}

	To transmit message index $i$ $\in$ $\mathcal{W}$, the transmitter uses the codeword $\vec{u}(i)$. For all  $t$ $\in$ $\lbrace$ $1$,$2$,$\ldots$, $n \rbrace$, at channel use $t$, the transmitter inputs  the symbol $u_{t}(i)$ into the channel. 
	Assume that, at the end of channel use $t$, the receiver observes the output $y_{t}$. After $n$ channel uses, the receiver uses the vector \mbox{$\vec{y}$ $=$ $(y_{1}$,$y_{2}$,$\ldots$, $y_{n})$} and determines that the symbol $j$ was transmitted if $\vec{y}$ $\in$ $\mathcal{D}(j)$, with $j$ $\in$ $\mathcal{W}$.

	Given the $(n$,$M$,$\lambda)$-code described by the system in~\eqref{EqP2PCode}, the DEP of the message index $i$ can be computed as $\mathbb{E}_{P_{\vec{Y}|\vec{X} = \vec{u}(i)} }\Big[$ $\mathds{1}_{\lbrace Y\notin\set{D}(i) \rbrace}\Big]$. As a consequence, the average DEP is \begin{equation}
	\frac{1}{M} \sum_{i=1}^{M} \mathbb{E}_{P_{\vec{Y}|\vec{X} = \vec{u}(i)} }\Big[ \mathds{1}_{\lbrace Y\notin \set{D}(i) \rbrace}\Big].
	\end{equation}
	Note that, from~\eqref{eqDefAvgProbEr}, the average DEP of such an $(n,M,\lambda)$-code is upper bounded by $\lambda$. 
	Given a fixed pair $(n$,$M)$ $\in$ $\mathbb{N}^2$, the minimum $\lambda$ for which an $(n$,$M$,$\lambda)$-code exists is defined hereunder. 
	
	\begin{Definition}\label{DefminEr} Given a pair $(n$,$M)$ $\in$ $\mathbb{N}^2$, the minimum average DEP for the random transformation in~\eqref{EqGeneralRamdonTransfomation}, denoted by $\lambda^{*}(n,M)$, is given by
		\begin{eqnarray}
			\label{EqMinEr}
			\lambda^{*}(n,M) = \min\left\{\lambda \in [0,1]: \exists (n,M,\lambda)\text{-code}\right\}.
		\end{eqnarray}
	\end{Definition}

	When $\lambda$ is chosen accordingly with the reliability constraints, an $(n,M,\lambda)$-code is said to transmit at an information rate  $R = \frac{\mathrm{log}_2(M)}{n}$ bits per channel use.

	The remainder of this section introduces the DT and MC bounds. 
	The DT bound is one of the tightest existing upper bounds on $\lambda^{*}(n,M)$ in~\eqref{EqMinEr}, whereas the MC bound is one of the tightest lower~bounds.
	
	\subsection{Dependence Testing Bound} \label{SecDTBound}
	This section describes an upper bound on  $\lambda^{*}(n,M)$, for a fixed pair $(n,M) \in \mathbb{N}^2$.
	Given a probability distribution $P_{\vec{X}} \in \triangle\left( \mathcal{X}^n \right)$, let the random variable $\iota\left( \vec{X}; \vec{Y}\right)$ satisfy
	\begin{equation}\label{EqDefMI}
	\iota\left( \vec{X}; \vec{Y}\right) \triangleq \mathrm{ln}\left(\rndder{ P_{\vec{X} \vec{Y}}}{P_{\vec{X}}P_{\vec{Y}}}(\vec{X},\vec{Y} )\right),
	\end{equation}
	where the function $\rndder{P_{\vec{X} \vec{Y}}}{P_{\vec{X}}P_{\vec{Y}}}: \mathcal{X}^n \times \mathcal{Y}^n \rightarrow \mathbb{R}$ denotes the Radon--Nikodym derivative of the joint probability measure $P_{\vec{X} \vec{Y}}$ with respect to the product of probability measures $P_{\vec{X}} P_{\vec{Y}}$, with  $P_{\vec{X} \vec{Y}} = P_{\vec{X}}P_{\vec{Y}| \vec{X}}$ and $P_{\vec{Y}}$ the corresponding marginal. 
	Let the function $T: \mathbb{N}^2 \times \triangle\left( \mathcal{X}^n \right) \rightarrow \mathbb{R}_+$ be for all $(n$,$M)$ $\in$ $\mathbb{N}^2$ and for all probability distributions $P_{\vec{X}} \in \triangle\left( \mathcal{X}^n \right)$, 
	\begin{IEEEeqnarray}{lcl}
		\label{EqT} 
		T(n,M,P_{\vec{X}}) & =  & \mathbb{E}_{P_{\vec{X}}P_{\vec{Y}|\vec{X}}}\left[\mathds{1}_{\left\{\iota (\vec{X};\vec{Y}) \leqslant\me{ln}\left( \frac{M-1}{2} \right) \right\}}\right] 
		+ \frac{M-1}{2} \mathbb{E}_{P_{\vec{X}}P_{\vec{Y}}}\left[\mathds{1}_{\left\{\iota (\vec{X};\vec{Y}) > \ln{\frac{M-1}{2}}\right\}}\right]. \quad
	\end{IEEEeqnarray}
	
	Using this notation, the following lemma states the DT bound. 
	
	\begin{Lemma}[Dependence testing bound~\citep{Polyanskiy2010}]\label{LemmaDTBound}
		Given a pair $(n$,$M)$ $\in$ $\mathbb{N}^2$, the following holds for all $P_{\vec{X}} \in  \triangle\left( \mathcal{X}^n \right)$,  with respect to the random transformation in~\eqref{EqGeneralRamdonTransfomation}:
		\begin{eqnarray}
			\label{EqDTBound} 
			\lambda^{*}(n,M) \leqslant  T(n,M,P_{\vec{X}}),
		\end{eqnarray} 
		with the function $T$ defined in~\eqref{EqT}.
	\end{Lemma} 
	
	Note that the input probability distribution $P_{\vec{X}}$ in Lemma~\ref{LemmaDTBound} can be chosen among all possible probability distributions $P_{\vec{X}} \in  \triangle\left( \mathcal{X}^n \right)$ to minimize the right-hand side of~\eqref{EqDTBound}, which improves the bound. 
	Note also that with some loss of optimality, the optimization domain can be restricted to the set of  product probability distributions for which for all $\vec{x}$ $\in$ $\set{X}^n$,
	\begin{IEEEeqnarray}{lcl}
		\label{EqiidRandCod}
		P_{\vec{X}}(\vec{x}) = \prod_{t=1}^n P_X(x_t),
	\end{IEEEeqnarray} 
	with $P_X \in  \triangle\left( \mathcal{X} \right)$. 
	Hence, subject to~\eqref{EqMemChan}, the random variable $\iota (\vec{X};\vec{Y})$ in~\eqref{EqDefMI} can be written as the sum of  i.i.d. random variables, i.e., 
	\begin{IEEEeqnarray}{l}
		\label{EqIndDensMut}
		\iota(\vec{X};\vec{Y}) = \sum_{t=1}^n \iota (X_t;Y_t).
	\end{IEEEeqnarray}
	This observation motivates the application of the results of Section~\ref{SecSums} to provide upper and lower bounds on the function $T$ in~\eqref{EqT}, for some given values $(n,M) \in \mathbb{N}^2$ and a given distribution $P_{\vec{X}} \in \triangle\left( \mathcal{X}^n\right)$ for the random transformation in~\eqref{EqGeneralRamdonTransfomation} subject to~\eqref{EqMemChan}. 
	These bounds become significantly relevant when the exact value of $T(n,M,P_{\vec{X}})$ cannot be calculated with respect to the random transformation in~\eqref{EqGeneralRamdonTransfomation}. In such a case, providing upper and lower bounds on  $T(n,M,P_{\vec{X}})$ helps in approximating its exact value subject to an error sufficiently small such that the approximation is relevant.
	
	\subsubsection{Normal Approximation} \label{SecDTNA}
	This section describes the normal  approximation of the function $T$ in~\eqref{EqT}. That is, the random variable $\iota (\vec{X};\vec{Y})$ is assumed to satisfy \eqref{EqIndDensMut} and to follow a Gaussian distribution.
	More specifically, for~all $P_{X} \in \triangle\left( \mathcal{X} \right)$, let 
	\begin{IEEEeqnarray}{l}
		\mu(P_X) \df \mathbb{E}_{P_{X}  P_{Y|X}}\left[\iota(X;Y)\right], \\
		\sigma(P_X) \df \mathbb{E}_{P_{X}  P_{Y|X}} \Big[\big( \iota(X;Y) -\mu(P_X) \big)^2 \Big],
		\mbox{ and }\\
		\xi(P_X) \df c_1\left(\frac{\mathbb{E}_{P_{X}  P_{Y|X}}\Big[ \big|\iota(X;Y) -\mu(P_X)\big|^3\Big]}{\sigma(P_X)^{\frac{3}{2}}}+c_2\right), 
	\end{IEEEeqnarray}
	with $c_1$ and $c_2$ defined in~\eqref{EqConstants}, be functions of the input distribution $P_X$. In particular, $\mu(P_X)$ and $\sigma(P_X)$ are respectively the first moment and the second central moment of the random variables  $\iota (X_1;Y_1)$, $\iota (X_2;Y_2)$ $\ldots$ $\iota (X_n;Y_n)$.
	Using this notation, consider the functions $D: \mathbb{N}^2 \times \triangle\left( \mathcal{X} \right) \rightarrow \mathbb{R}_{+}$ and $N: \mathbb{N}^2 \times \triangle\left( \mathcal{X} \right) \rightarrow \mathbb{R}_{+}$ such that for all $(n,M) \in \mathbb{N}^2$ and for all $P_X \in \triangle\left( \mathcal{X} \right)$, 
	\begin{IEEEeqnarray}{l}
		\label{EqD}
		D(n,M,P_X) = \max \hspace{-0.1ex} \left\{ 0,  \alpha\hspace{-0.1ex}\left(\hspace{-0.1ex}n,M, P_X\right) \hspace{-0.1ex} - \hspace{-0.1ex} \frac{\xi(P_X)}{\sqrt{n}} \hspace{-0.1ex} \right\},
		\mbox{ and }\\
		N(n,M,P_X) = \label{EqN}
		\min\hspace{-.2ex} \left\{  1,  \alpha\left(n,M, P_X\right) \hspace{-.5ex} + \hspace{-.5ex}  \frac{5\;\xi(P_X)}{\sqrt{n}} \hspace{-.5ex} + \hspace{-.5ex}  \frac{2 \ln{2}}{\sigma(P_X)^{\frac{1}{2}} \sqrt{2n\pi}}  \right\}, 
	\end{IEEEeqnarray}
	where 
	\begin{IEEEeqnarray}{lcl}
		\label{EqAlpha}
		\alpha\left(n,M, P_X\right) &\triangleq&  Q\left(\frac{n \mu(P_X) -\ln{\frac{M-1}{2}} }{\sqrt{n \sigma(P_X)}}\right). 
	\end{IEEEeqnarray}
	Using this notation, the following theorem introduces  lower and upper bounds on the function $T$  in~\eqref{EqT}.
	
	\begin{Theorem}\label{Theo3} 
		Given a pair $(n,M) \in \mathbb{N}^2$, for all input distributions $P_{\vec{X}} \in \triangle\left( \mathcal{X}^n \right)$ subject to~\eqref{EqiidRandCod},  the following holds with respect to the random transformation in~\eqref{EqGeneralRamdonTransfomation} subject to~\eqref{EqMemChan},
		\begin{eqnarray}
			\label{EqDTNA}
			D(n,M,P_X) \leqslant T(n,M,P_{\vec{X}}) \leqslant N(n,M,P_X),
		\end{eqnarray}
		where the functions $T$, $D$ and $N$ are defined in~\eqref{EqT},~\eqref{EqD} and~\eqref{EqN}, respectively.
	\end{Theorem}

	\begin{proof}
		The proof of Theorem~\ref{Theo3}  is presented in~\citep{MolavianJaziPhD}. Essentially, it relies on Theorem~\ref{TheoBerry} for upper and lower bounding the terms  $\mathbb{E}_{P_{\vec{X}}P_{\vec{Y}|\vec{X}}}\left[\mathds{1}_{\left\{\iota (\vec{X};\vec{Y}) \leqslant\me{ln}\left( \frac{M-1}{2} \right) \right\}}\right] $ in~\eqref{EqT}. The upper bound on $\mathbb{E}_{P_{\vec{X}}P_{\vec{Y}}}\left[\mathds{1}_{\left\{\iota (\vec{X};\vec{Y}) > \ln{\frac{M-1}{2}}\right\}}\right]$ in~\eqref{EqT} follows from {Lemma~$47$ in~\citep{Polyanskiy2010}}.
	\end{proof}
	
	In ~\citep{MolavianJaziPhD}, the function $\alpha(n,M,P_X)$ in~\eqref{EqAlpha} is often referred to as the \textit{normal approximation} of $T(n,M,P_{\vec{X}})$, which is indeed a language abuse. 
	In Section~\ref{SecSumsI}, a comment is given on the fact that the lower and upper bounds, i.e., the functions $D$ in~\eqref{EqD} and $N$ in~\eqref{EqN}, are often too far from the normal approximation $\alpha$ in~\eqref{EqAlpha}.  
	
	\subsubsection{Saddlepoint Approximation}\label{SecDTSD}
	This section describes an approximation of the function $T$ in~\eqref{EqT} by using the saddlepoint approximation of the CDF of the random variable $\iota (\vec{X};\vec{Y})$, as suggested in Section~\ref{SecSumsII}.
	Given a distribution $P_{X} \in \triangle\left( \mathcal{X} \right)$, the moment generating function of $\iota(X;Y)$ is
	\begin{IEEEeqnarray}{lcl}
		\label{EqApvarphiMC}
		\varphi(P_X,\theta)     \df \mathbb{E}_{P_X P_{Y|X}}\left[\exp\left(\theta\, \iota(X;Y)\right) \right],
	\end{IEEEeqnarray} 
	with $\theta \in \mathbb{R}$.
	For all $P_{X} \in \triangle\left( \mathcal{X} \right)$ and for all $\theta \in \mathbb{R}$, consider the following functions:
	\begin{IEEEeqnarray}{lcl}
		\label{EqApMuMC}
		\mu(P_X,\theta) \df \mathbb{E}_{P_{X}  P_{Y|X}}\left[\frac{\iota(X;Y) \exp\left(\theta\, \iota(X;Y)\right)}{\varphi(P_X,\theta)} \right],\\
		\label{EqApVMC}
		V(P_X,\theta) \df\mathbb{E}_{P_{X}  P_{Y|X}} \hspace{-.5ex} \Bigg[ \hspace{-.8ex} \left(\iota(X;Y) \hspace{-.5ex} - \hspace{-.5ex} \mu(P_X,\theta)\right)^2 
		\hspace{-.5ex} \frac{\exp\left(\theta \hspace{-.05ex} \iota(X;Y)\right)}{\varphi(P_X, \theta)} \hspace{-.5ex} \Bigg],
		\mbox{ and }\\
		\label{EqApTMC}
		\xi(P_X,\theta)  \df c_1\left(\frac{\mathbb{E}_{P_{X}  P_{Y|X}} \hspace{-.5ex}  \left[ \hspace{-.5ex}  \left|\iota(X;Y) \hspace{-.5ex}  - \hspace{-.5ex} \mu(P_X,\theta)\right|^3      \frac{\exp\left(\theta \iota(X;Y)\right)}{\varphi(P_X,\theta)} \right]}{\left(V(P_X,\theta)\right)^{3/2}} + c_2\right),
	\end{IEEEeqnarray} 
	where $c_1$ and $c_2$ are defined in~\eqref{EqConstants}.
	Using this notation, consider the functions $\beta_1: \mathbb{N}^2 \times \mathbb{R}  \times \triangle\left( \mathcal{X} \right) \rightarrow \mathbb{R}_{+}$ and  $\beta_2: \mathbb{N}^2 \times \mathbb{R}  \times \triangle\left( \mathcal{X} \right) \rightarrow \mathbb{R}_{+}$:  
	\begin{IEEEeqnarray}{l}
		\label{EqBeta1}
		\beta_1(n,M,\theta,P_X) \hspace{-.2ex} 
		\hspace*{-0.5ex}=\hspace*{-0.5ex} \hspace{-.2ex} \mathds{1}_{\{\hspace{-.25ex}\theta > 0\hspace{-.25ex}\}}\hspace{-.8ex} + \hspace{-.7ex} (\hspace{-.4ex}-\hspace{-.4ex}1\hspace{-.4ex})^{\mathds{1}_{\{\hspace{-.24ex}\theta > 0 \hspace{-.24ex}\}}}\exp\hspace{-.7ex}\Bigg(\hspace{-.8ex}n \ln{\varphi(\hspace{-.3ex}P_X,\theta\hspace{-.25ex})}\hspace{-.4ex} -\hspace{-.4ex}\theta \ln{\hspace{-.6ex}\frac{M\hspace{-.2ex}-\hspace{-.2ex}1}{2}\hspace{-.6ex}}\hspace{-.4ex}
		+\hspace{-.4ex}\frac{1}{2}\theta^2 n V(\hspace{-.4ex}P_X,\theta\hspace{-.25ex})\hspace{-.9ex}\Bigg)\hspace{-.4ex}
		Q\hspace{-.6ex}\left(\hspace{-.8ex}\sqrt{n V(\hspace{-.3ex}P_X,\theta\hspace{-.2ex})} |\theta|\hspace{-.4ex}\right), \hspace{-0.5ex}\qquad 
	\end{IEEEeqnarray}
		and
	\begin{IEEEeqnarray}{l}
		\label{EqBeta2}
		\beta_2(n,M,\theta,P_X) \hspace{-.2ex} \nonumber\\
		{= \mathds{1}_{\{\hspace{-.25ex}\theta  \leqslant \hspace{-.35ex} -1\hspace{-.25ex}  \}} \hspace{-.5ex} + \hspace{-.5ex} (\hspace{-.5ex}-\hspace{-.25ex}1\hspace{-.5ex})^{\mathds{1}_{\{\theta  \leqslant \hspace{-.5ex} -\hspace{-.25ex}1 \}}}\exp\hspace{-.5ex}\left(\hspace{-.5ex}n \ln{\hspace{-.3ex}\varphi(\hspace{-.2ex}P_X,\hspace{-.2ex}\theta\hspace{-.2ex})\hspace{-.3ex}}\hspace{-.55ex}-\hspace{-.55ex}\left(\hspace{-.25ex}\theta\hspace{-.5ex}+\hspace{-.5ex}1\hspace{-.25ex}\right)\hspace{-.25ex}\ln{\hspace{-.8ex}\frac{M\hspace{-.5ex}-\hspace{-.5ex}1}{2}\hspace{-.25ex}}\hspace{-.55ex} +\hspace{-.55ex}\frac{1}{2}(\hspace{-.25ex}\theta\hspace{-.55ex}+\hspace{-.55ex}1\hspace{-.25ex})^2 n V\hspace{-.2ex}(\hspace{-.25ex}P_X, \hspace{-.2ex}\theta\hspace{-.2ex})\hspace{-.85ex}\right) Q\hspace{-.5ex}\left(\hspace{-.85ex}\sqrt{n V(\hspace{-.25ex}P_X,\theta\hspace{-.25ex})} |\hspace{-.15ex}\theta\hspace{-.55ex}+\hspace{-.55ex}1\hspace{-.3ex}|\hspace{-.5ex}\right) \hspace{-0.5ex}.}\nonumber\\
	\end{IEEEeqnarray}
	Note that $\beta_{1}$ is the saddlepoint approximation of the CDF  of the random variable $\iota (\vec{X};\vec{Y})$ in~\eqref{EqIndDensMut} when $\boldsymbol{X}$ and $\boldsymbol{Y}$ follow the distribution $P_{\vec{X}}P_{\vec{Y}|\vec{X}}$. Note also that  $\beta_{2}$   is the saddlepoint approximation of the   complementary CDF of the random variable $\iota (\vec{X};\vec{Y})$ in~\eqref{EqIndDensMut} when $\boldsymbol{X}$ and $\boldsymbol{Y}$ follow the distribution $P_{\vec{X}}P_{\vec{Y}}$.

	Consider also the following functions:  
	\begin{IEEEeqnarray}{l}
		\label{EqG1}
		G_1(n,M, \theta, P_X) =  \beta_1(n,M, \theta, P_X) 
		- \frac{ 2 \xi(P_X,\theta)}{ \sqrt{n}} \exp\hspace{-.6ex}\Bigg(\hspace{-.6ex}n \ln{\varphi(P_X,\theta)} -\theta \ln{\hspace{-.6ex}\frac{M-1}{2}\hspace{-.55ex}}\hspace{-.8ex}\Bigg), \\
		\label{EqG2}
		G_2(n,M, \theta, P_X) =  \beta_2(n,M, \theta, P_X) 
		-\hspace{-.4ex} \frac{ 2\xi(P_X,\theta)}{\sqrt{n}} \exp\hspace{-.6ex}\Bigg(\hspace{-.8ex}n\ln{\varphi(P_X,\theta)}\hspace{-.25ex}-\hspace{-.5ex}(\hspace{-.25ex}\theta\hspace{-.5ex}+\hspace{-.5ex}1\hspace{-.5ex}) \ln{\hspace{-.6ex}\frac{M\hspace{-.3ex}-\hspace{-.3ex}1}{2}\hspace{-.55ex}}\hspace{-1.2ex}\Bigg),\quad \\
		G(n,M, \theta, P_X) =  \max\left\{ 0, G_1(n,M, \theta, P_X) \right\}
			\label{EqU}
			+  \frac{M-1}{2} \max\left\{ 0, G_2(n,M, \theta, P_X) \right\}, \quad
		\mbox{ and }\\
		\label{EqL} 
		S(n,M,\theta,P_X)\hspace{-.2ex}=\hspace{-.2ex} \min\left\{1, \beta\left(n,M,\theta,P_X\right)+\frac{ 4 \xi(P_X,\theta)}{\sqrt{n}}   \exp\left(n\ln{\varphi(P_X,\theta)}-\theta \ln{\frac{M-1}{2}}\right)\right\},\quad		\quad
	\end{IEEEeqnarray}
where, 
	\begin{IEEEeqnarray}{l}
		\label{EqBeta}
		\beta(n,M, \theta, P_X)  =  \beta_1(n, M,\theta,P_X) + \frac{M-1}{2} \beta_2(n, M, \theta, P_X), \qquad
	\end{IEEEeqnarray}
	with $\beta_1$ in~\eqref{EqBeta1} and $\beta_2$ in~\eqref{EqBeta2}. Often, the function $\beta$ in \eqref{EqBeta} is referred to as the \emph{saddlepoint  approximation} of  the function $T$ in~\eqref{EqT}, which is indeed a language abuse. 

	The following theorem introduces new lower and upper bounds on the function $T$  in~\eqref{EqT}.
	
	\begin{Theorem}\label{TheoNew} 
		Given a pair $(n,M) \in \mathbb{N}^2$,  for all input distributions $P_{\vec{X}} \in \triangle\left( \mathcal{X}^n \right)$ subject to~\eqref{EqiidRandCod}, the following holds with respect to the random transformation in~\eqref{EqGeneralRamdonTransfomation} subject to~\eqref{EqMemChan},
		\begin{eqnarray}
			G(n,M, \theta, P_X)   \leqslant  T(n,M,P_{\vec{X}}) \leqslant  S(n,M, \theta, P_X)
		\end{eqnarray}
		where  $\theta$ is the unique solution in $t$ to 
		\begin{eqnarray}\label{EqDTBoundThetaStarA}
			n \mu(P_X,t) = \ln{\frac{M-1}{2}},
		\end{eqnarray}
		and the functions $T$, $G$, and $S$ are defined in~\eqref{EqT},~\eqref{EqU}, {and~\eqref{EqL}.}
	\end{Theorem}
	\begin{proof}
		The proof of Theorem~\ref{TheoNew} is provided in Appendix~\ref{pTheoNew}.
		In a nutshell, the proof relies on Theorem~\ref{TheoSaddlePointBeryUni} for independently bounding the terms $\mathbb{E}_{P_{\vec{X}}P_{\vec{Y}|\vec{X}}}\left[\mathds{1}_{\left\{\iota (\vec{X};\vec{Y}) \leqslant \me{ln}\left( \frac{M-1}{2} \right) \right\}}\right]$  and $\mathbb{E}_{P_{\vec{X}}P_{\vec{Y}}}\left[\mathds{1}_{\left\{\iota (\vec{X};\vec{Y}) > \me{ln}\left( \frac{M-1}{2} \right) \right\}}\right]$ in~\eqref{EqT}.
	\end{proof}

	\subsection{ Meta Converse Bound}\label{SecMCBound}
	This section describes a lower bound on  $\lambda^{*}(n,M) $, for a fixed pair $(n,M) \in \mathbb{N}^2$.
	Given two probability distributions $P_{\vec{X}\vec{Y}} \in \triangle\left( \mathcal{X}^n\times\mathcal{Y}^n \right)$ and $Q_{\vec{Y}} \in \triangle\left( \mathcal{Y}^n \right)$, let the random variable $\tilde{\iota}\left( \vec{X}; \vec{Y}|Q_{\vec{Y}}\right)$~satisfy
	\begin{equation}\label{EqGenInfMesVec}
	\tilde{\iota}\left( \vec{X}; \vec{Y}|Q_{\vec{Y}}\right) \triangleq \mathrm{ln}\left(\rndder{ P_{\vec{X} \vec{Y}}}{P_{\vec{X}}Q_{\vec{Y}}}(\vec{X},\vec{Y} )\right).
	\end{equation}
	For all $(n$,$M$,$\gamma)$ $\in$ $\mathbb{N}^2 \times \mathbb{R}$ and for all probability distributions $P_{\vec{X}} \in \triangle\left( \mathcal{X}^n \right)$ and $Q_{\vec{Y}} \in \triangle\left( \mathcal{Y}^n \right)$, let the function $C: \mathbb{N}^2 \times \triangle\left( \mathcal{X}^n \right) \times \triangle\left( \mathcal{Y}^n \right) \times {\mathbb{R}_+} \rightarrow \mathbb{R}_+$ be  
	\begin{IEEEeqnarray}{l}
		\label{EqMetConSym}
		C(n,M,P_{\vec{X}},Q_{\vec{Y}},\gamma)\df \mathbb{E}_{P_{\vec{X}}P_{\vec{Y}|\vec{X}}}\left[\mathds{1}_{\left\{\tilde{\iota}\left( \vec{X}; \vec{Y}|Q_{\vec{Y}}\right) \leqslant \me{ln}\left( \gamma \right) \right\}}\right] + \gamma\left( \mathbb{E}_{P_{\vec{X}}Q_{\vec{Y}}}\left[\mathds{1}_{\left\{\tilde{\iota}\left( \vec{X}; \vec{Y}|Q_{\vec{Y}}\right) > \me{ln}\left( \gamma \right) \right\}}\right]-\frac{1}{M}\right).\quad 
	\end{IEEEeqnarray}
	Using this notation, the following lemma describes the MC bound. 
	\begin{Lemma}[MC Bound~\citep{Polyanskiy2010,fJos2018}]\label{LemMart2018}
		Given a pair $(n$,$M)$ $\in$ $\mathbb{N}^2$, the following holds for all $Q_{\vec{Y}} \in\Delta(\set{Y}^n)$, with~respect to the random transformation in~\eqref{EqGeneralRamdonTransfomation}:
		\begin{eqnarray}
			\label{EqMetCon} 
			\lambda^{*}(n,M)   \geqslant   \inf_{P_{\vec{X}}\in\Delta(\set{X}^n)} \max_{\gamma \geqslant 0}  C(n,M,P_{\vec{X}},Q_{\vec{Y}},\gamma),
		\end{eqnarray} 
		where the function $C$ is defined in~\eqref{EqMetConSym}.
	\end{Lemma}

	Note that the output probability distribution $Q_{\vec{Y}}$ in Lemma~\ref{LemMart2018} can be chosen among all possible probability distributions $Q_{\vec{Y}} \in  \triangle\left( \mathcal{Y}^n \right)$  to maximize the right-hand side of~\eqref{EqMetConSym}, which improves the bound. 
	Note also that, with some loss of optimality, the optimization domain can be restricted to the set of  probability distributions for which for all $\vec{y}$ $\in$ $\set{Y}^n$,
	\begin{IEEEeqnarray}{lcl}
		\label{EqoutRandCod}
		Q_{\vec{Y}}(\vec{y}) = \prod_{t=1}^n Q_Y(y_t),
	\end{IEEEeqnarray} 
	with $Q_Y \in  \triangle\left( \mathcal{Y} \right)$. Hence, subject to~\eqref{EqMemChan}, for all $\vec{x}\in\set{X}^n$, the random variable $\tilde{\iota} (\vec{x};\vec{Y}|Q_{\vec{Y}})$ in~\eqref{EqMetConSym} can be written as the sum of the independent random variables, i.e., 
	\begin{IEEEeqnarray}{l}
		\label{EqGenIndDensMutCond}
		\tilde{\iota} (\vec{x};\vec{Y}|Q_{\vec{Y}}) = \sum_{t=1}^n \tilde{\iota} (x_t;Y_t|Q_{Y}).
	\end{IEEEeqnarray}
	With some loss of generality, the focus is on a channel transformation of the form in~\eqref{EqGeneralRamdonTransfomation} for which the following condition holds:   
	The infimum in \eqref{EqMetCon} is achieved by a product distribution, i.e.,  $P_{\vec{X}}$ is of the form in~\eqref{EqiidRandCod}, when the probability distribution $Q_{\vec{Y}}$ satisfies~\eqref{EqoutRandCod}.
	Note that this condition is met by memoryless channels such as the BSC, the AWGN and S$\alpha$S channels with binary antipodal inputs, i.e., input alphabets are of the form $\mathcal{X} = \lbrace a, -a \rbrace$, with $a \in \mathbb{R}$. This follows from the fact that  the random variable $\tilde{\iota} (\vec{x};\vec{Y}|Q_{\vec{Y}})$ is invariant of the choice of $\vec{x}\in\set{X}^n$ when the probability distribution $Q_{\vec{Y}}$ satisfies~\eqref{EqoutRandCod} and  for all $y\in\set{Y}$,
	\begin{IEEEeqnarray}{l}\label{EqUIOD}
		Q_{Y}(y) = \frac{P_{Y|X}(y|-a)+ P_{Y|X}(y|a)}{2}.
	\end{IEEEeqnarray}
	Under these conditions, the random variable $\tilde{\iota} (\vec{X};\vec{Y}|Q_{\vec{Y}})$ in~\eqref{EqMetConSym} can be written as the sum of i.i.d. random variables, i.e., 
	\begin{IEEEeqnarray}{l}
		\label{EqGenIndDensMut}
		\tilde{\iota} (\vec{X};\vec{Y}|Q_{\vec{Y}}) = \sum_{t=1}^n \tilde{\iota} (X_t;Y_t|Q_{Y}).
	\end{IEEEeqnarray}
	This observation motivates the application of the results of Section~\ref{SecSums} to provide upper and lower bounds on the function $C$ in~\eqref{EqMetConSym}, for some given values $(n,M) \in \mathbb{N}^2$ and given distributions $P_{\vec{X}} \in \triangle\left( \mathcal{X}^n\right)$ and $Q_{\vec{Y}} \in \triangle\left( \mathcal{Y}^n\right)$. These bounds become significantly relevant when the exact value of $C(n,M,P_{\vec{X}}, Q_{\vec{Y}},\gamma)$ cannot be calculated with respect to the random transformation in~\eqref{EqGeneralRamdonTransfomation}. In such a case, providing upper and lower bounds on  $C(n,M,P_{\vec{X}}, Q_{\vec{Y}},\gamma)$ helps in approximating its exact value subject to an error sufficiently small such that the approximation is relevant.

	\subsubsection{Normal Approximation}\label{SecMCNA}
	This section describes the normal  approximation of the function $C$ in~\eqref{EqMetConSym}, that is to say, the~random variable  $\tilde{\iota} (\vec{X};\vec{Y}|Q_{\vec{Y}})$ is assumed to satisfy~\eqref{EqGenIndDensMut} and to follow a Gaussian distribution. More specifically, for all $(P_{X},Q_Y) \in \triangle\left( \mathcal{X} \right)\times \triangle\left( \mathcal{Y} \right)$, let 
	\begin{IEEEeqnarray}{l}
		\tilde{\mu}(P_X,Q_Y) \df \mathbb{E}_{P_{X}  P_{Y|X}} \left[\tilde{\iota}(X;Y|Q_Y)\right],\\
		\tilde{\sigma}(P_X,Q_Y) \df \mathbb{E}_{P_{X}  P_{Y|X}} \Big[ \big( \tilde{\iota}(X;Y|Q_Y) - \tilde{\mu}(P_X,Q_Y) \big)^2\Big],
		\mbox{ and }\\
		\tilde{\xi}(P_X,Q_Y) \df c_1\left(\frac{\mathbb{E}_{P_{X}  P_{Y|X}}\Big[\big|\tilde{\iota}(X;Y|Q_Y)-\tilde{\mu}(P_X,Q_Y)\big|^3\Big]}{\left(\tilde{\sigma}(P_X,Q_Y)\right)^{3/2}}+c_2\right)
	\end{IEEEeqnarray}
with $c_1$ and $c_2$ defined in~\eqref{EqConstants}, be functions of the input and output distributions $P_X$ and $Q_Y$, respectively. In particular, $\tilde{\mu}(P_X,Q_Y)$ and $\tilde{\sigma}(P_X,Q_Y)$ are respectively the first moment and the second central moment of the random variables $\tilde{\iota}(X_1;Y_1|Q_Y), \tilde{\iota}(X_2;Y_2|Q_Y),\ldots \tilde{\iota}(X_n;Y_n|Q_Y)$. 
	Using this notation, consider the functions $\tilde{D}: \mathbb{N}^2 \times \triangle\left( \mathcal{X} \right)\times \triangle\left( \mathcal{Y} \right)\times\mathbb{R}_{+} \rightarrow \mathbb{R}_{+}$ and $\tilde{N}: \mathbb{N}^2 \times \triangle\left( \mathcal{X} \right)\times \triangle\left( \mathcal{Y} \right)\times\mathbb{R}_{+} \rightarrow \mathbb{R}_{+}$ such that, for all $(n,M,\gamma) \in \mathbb{N}^2\times{\mathbb{R}_+}$ and for all $P_X \in \triangle\left( \mathcal{X} \right)$ and for all $Q_Y \in \triangle\left( \mathcal{Y} \right)$, 
	\begin{IEEEeqnarray}{l}
		\label{EqTD}
		\tilde{D}(n,M,P_X,Q_Y,\gamma) = \max \hspace{-0.1ex} \left\{\hspace{-0.1ex} 0,  \tilde{\alpha}\hspace{-0.1ex}\left(\hspace{-0.1ex}n,M, P_X,Q_Y,\gamma\right) \hspace{-0.1ex} - \hspace{-0.1ex} \frac{\tilde{\xi}(P_X,Q_Y)}{\sqrt{n}} \hspace{-0.1ex} \right\}, 
		\mbox{ and }\\
		\tilde{N}(n,M,P_X,Q_Y,\gamma) = \label{EqTN}
		\min\hspace{-.2ex} \left\{\hspace{-.2ex} 1, \hspace{-.2ex} \tilde{\alpha}\left(n,M, P_X, Q_Y,\gamma\right) \hspace{-.5ex} + \hspace{-.5ex}  \frac{5\;\tilde{\xi}(P_X,Q_Y)}{\sqrt{n}} \hspace{-.5ex} + \hspace{-.5ex}  \frac{2 \ln{2}}{\tilde{\sigma}(P_X,Q_Y)^{\frac{1}{2}} \sqrt{2n\pi}} \hspace{-0.8ex} \right\}, \quad
	\end{IEEEeqnarray}
	where
	\begin{IEEEeqnarray}{lcl}
		\label{EqTAlpha}
		\tilde{\alpha}\left(n,M, P_X, Q_Y, \gamma\right) &\triangleq&  Q\left(\frac{n \tilde{\mu}(P_X,Q_Y) -\ln{\gamma} }{\sqrt{n \tilde{\sigma}(P_X,Q_Y)}}\right) - \frac{\gamma}{M}. 
	\end{IEEEeqnarray}
	Using this notation, the following theorem introduces lower and upper bounds on the function $C$  in~\eqref{EqMetConSym}.
	
	\begin{Theorem}\label{Theo3MC} 
		Given a pair $(n,M) \in \mathbb{N}^2$, for all input distributions $P_{\vec{X}} \in \triangle\left( \mathcal{X}^n \right)$ subject to~\eqref{EqiidRandCod}, for all output distributions $Q_{\vec{Y}} \in \triangle\left( \mathcal{Y}^n \right)$ subject to~\eqref{EqoutRandCod}, and for all $\gamma \geqslant 0$, the following holds with respect to the random transformation in~\eqref{EqGeneralRamdonTransfomation} subject to~\eqref{EqMemChan},
		\begin{eqnarray}
			\label{EqMCNA}
			\tilde{D}(n,M,P_X,Q_Y,\gamma) \leqslant C(n,M,P_{\vec{X}}, Q_{\vec{Y}},\gamma) \leqslant \tilde{N}(n,M,P_X,Q_Y,\gamma),
		\end{eqnarray}
		where the functions $C$, $\tilde{D}$, and $\tilde{N}$ are defined in~\eqref{EqMetConSym},~\eqref{EqTD}, and~\eqref{EqTN}, respectively.
	\end{Theorem}

	\begin{proof}
		The proof of Theorem~\ref{Theo3MC} is partially presented in~\citep{Polyanskiy2010}. Essentially, {it relies on} Theorem~\ref{TheoBerry} for upper and lower bounding the term  $\mathbb{E}_{P_{\vec{X}}P_{\vec{Y}|\vec{X}}}\left[\mathds{1}_{\left\{\tilde{\iota} (\vec{X};\vec{Y}|Q_{\vec{Y}}) \leqslant\me{ln}\left( \gamma \right) \right\}}\right] $ in~\eqref{EqMetConSym}; and using {Lemma~$47$ in~\citep{Polyanskiy2010}} for upper bounding the term $\mathbb{E}_{P_{\vec{X}}Q_{\vec{Y}}}\left[\mathds{1}_{\left\{\tilde{\iota} (\vec{X};\vec{Y}|Q_{\vec{Y}}) >\me{ln}\left( \gamma \right)\right\}}\right]$ in~\eqref{EqMetConSym}.
	\end{proof}
	
	The function $\tilde{\alpha}\hspace{-0.1ex}\left(\hspace{-0.1ex}n,M, P_X,Q_Y,\gamma\right)$ in~\eqref{EqTAlpha} is often referred to as the \textit{normal approximation} of $C(n,M,P_{\vec{X}})$, which is indeed a language abuse. 
	In Section~\ref{SecSumsI}, a comment is given on the fact that the lower and upper bounds on the normal approximation, i.e., the functions $\tilde{D}$ in~\eqref{EqTD} and $\tilde{N}$ in~\eqref{EqTN}, are~often too far from the normal approximation $\tilde{\alpha}$ in~\eqref{EqTAlpha}.

	\subsubsection{Saddlepoint Approximation}\label{SecMCSD}
	This section describes an approximation of the function $C$ in~\eqref{EqMetConSym} by using the saddlepoint approximation of the CDF of the random variable $\tilde{\iota} (\vec{X};\vec{Y}|Q_{\vec{Y}})$, as suggested in Section~\ref{SecSumsII}.
	Given two distributions $P_{X} \in \triangle\left( \mathcal{X} \right)$ and $Q_{Y} \in \triangle\left( \mathcal{Y} \right)$, let the random variable $\tilde{\iota}(X;Y|Q_Y)$ satisfy 
	\begin{IEEEeqnarray}{lcl}
		\label{EqGenInfMes}
		\tilde{\iota}(X;Y|Q_Y) \df \ln{\rndder{P_{X}P_{Y|X}}{P_X Q_Y}(X,Y)},
	\end{IEEEeqnarray}
	where $P_{Y|X}$ is in~\eqref{EqMemChan}. The moment generating function of $\tilde{\iota}(X;Y|Q_Y)$ is
	\begin{IEEEeqnarray}{lcl}
		\label{EqTApvarphiMC}
		\tilde{\varphi}(P_{X},Q_Y,\theta)     \df \mathbb{E}_{P_X P_{Y|X}}\left[\exp\left(\theta\, \tilde{\iota}(X;Y|Q_Y)\right) \right],
	\end{IEEEeqnarray} 
	with $\theta \in \mathbb{R}$.
	For all $P_{X} \in \triangle\left( \mathcal{X} \right)$ and $Q_{Y} \in \triangle\left( \mathcal{Y} \right)$,  and for all $\theta \in \mathbb{R}$, consider the following functions:
	\begin{IEEEeqnarray}{lcl}
		\label{EqTApMuMC}
		\tilde{\mu}(P_{X},Q_Y,\theta) \df \mathbb{E}_{P_{X}  P_{Y|X}}\left[\frac{\tilde{\iota}(X;Y|Q_Y) \exp\left(\theta\, \tilde{\iota}(X;Y|Q_Y)\right)}{\tilde{\varphi}(P_{X},Q_Y,\theta)} \right] ,\\
		\label{EqTApVMC}
		\tilde{V}(P_{X},Q_Y,\theta) \df\mathbb{E}_{P_{X}  P_{Y|X}} \hspace{-.5ex} \Bigg[ \hspace{-.8ex} \left(\tilde{\iota}(X;Y|Q_Y) \hspace{-.5ex} - \hspace{-.5ex} \tilde{\mu}(P_{X},Q_Y,\theta)\right)^2 
		\hspace{-.5ex} \frac{\exp\left(\theta \hspace{-.05ex} \tilde{\iota}(X;Y|Q_Y)\right)}{\tilde{\varphi}(P_{X},Q_Y,\theta)} \hspace{-.5ex} \Bigg]\hspace{-.5ex} ,
		\mbox{ and }\\
		\label{EqTApTMC}
		\tilde{\xi}(P_{X},Q_Y,\theta)  \df c_1\left(\frac{\mathbb{E}_{P_{X}  P_{Y|X}} \hspace{-.5ex}  \left[ \hspace{-.5ex}  \left|\tilde{\iota}(X;Y|Q_Y) \hspace{-.5ex}  - \hspace{-.5ex} \tilde{\mu}(P_{X},Q_Y,\theta)\right|^3      \frac{\exp\left(\theta \tilde{\iota}(X;Y|Q_Y)\right)}{\tilde{\varphi}(P_{X},Q_Y,\theta)} \right]}{\left(\tilde{V}(P_{X},Q_Y,\theta)\right)^{3/2}}+c_2\right) \hspace{-.5ex},
	\end{IEEEeqnarray} 
	where $c_1$ and $c_2$ are defined in~\eqref{EqConstants}.
	Using this notation, consider the functions $\tilde{\beta}_1: \mathbb{N} \times{\mathbb{R}_+}\times \mathbb{R} \times \triangle\left( \mathcal{X} \right)\times \triangle\left( \mathcal{Y} \right) \rightarrow \mathbb{R}_{+}$ and  $\tilde{\beta}_2: \mathbb{N} \times {\mathbb{R}_+}\times\mathbb{R}  \times \triangle\left( \mathcal{X} \right)\times \triangle\left( \mathcal{Y} \right) \rightarrow \mathbb{R}_{+}$:
	\begin{IEEEeqnarray}{l}
		\tilde{\beta}_1(n,\gamma,\theta,P_{X}, Q_{Y})  \nonumber\\
		=\hspace*{-0.4ex} \mathds{1}_{\{\hspace*{-0.2ex}\theta > 0\hspace*{-0.2ex}\}}\hspace*{-0.5ex} + \hspace*{-0.5ex} (\hspace*{-0.3ex}-1\hspace*{-0.3ex})^{\mathds{1}_{\{\hspace*{-0.2ex}\theta > 0\hspace*{-0.2ex} \}}}\hspace*{-0.4ex}\exp\hspace*{-0.7ex}\left(\hspace*{-0.7ex}n \ln{\hspace*{-0.4ex}\tilde{\varphi}(P_{X},Q_Y,\theta)\hspace*{-0.4ex}}\hspace*{-0.4ex}-\hspace*{-0.4ex}\theta \ln{\gamma}
		\hspace*{-0.6ex}+\hspace*{-0.6ex} \frac{1}{2}\theta^2 n \tilde{V}(\hspace*{-0.4ex}P_{X},\hspace*{-0.4ex}Q_Y,\hspace*{-0.4ex}\theta)\hspace*{-0.8ex}\right)
		\hspace*{-0.4ex}Q\hspace*{-0.5ex}\left(\hspace*{-1.1ex}\sqrt{\hspace*{-0.4ex}n \tilde{V}(P_{X},Q_Y,\theta)\hspace*{-0.4ex}} |\theta|\hspace*{-0.8ex}\right)  \label{EqTBeta1}, 
		\mbox{ and }\qquad\\
		\label{EqTBeta2}
		\tilde{\beta}_2(n,\gamma,\theta,P_{X},Q_Y) \hspace{-.2ex} \nonumber\\
		= \hspace{-.2ex} \mathds{1}_{\{\hspace{-.25ex}\theta  \leqslant \hspace{-.35ex} -1\hspace{-.25ex}  \}} \hspace{-.5ex} + \hspace{-.5ex} (\hspace{-.5ex}-\hspace{-.25ex}1\hspace{-.5ex})^{\mathds{1}_{\{\theta  \leqslant \hspace{-.5ex} -\hspace{-.25ex}1 \}}}\exp\Big(\hspace{-.5ex}n \ln{\tilde{\varphi}(P_{X},Q_Y,\theta)}\hspace{-.5ex}-\hspace{-.5ex}\left(\hspace{-.25ex}\theta\hspace{-.5ex}+\hspace{-.5ex}1\hspace{-.25ex}\right)\hspace{-.5ex}\ln{\gamma}\hspace{-.5ex} + 
		\hspace{-.5ex}\frac{1}{2}(\hspace{-.25ex}\theta\hspace{-.5ex}+\hspace{-.5ex}1\hspace{-.25ex})^2 n \tilde{V}(\hspace{-.25ex}P_{X},Q_Y, \theta\hspace{-.2ex})\hspace{-.5ex}\Big) \hspace{-.25ex}\nonumber\\
		Q\hspace{-.5ex}\left(\hspace{-.85ex}\sqrt{n \tilde{V}(\hspace{-.25ex}P_{X},Q_Y,\theta\hspace{-.25ex})} |\hspace{-.15ex}\theta\hspace{-.5ex}+\hspace{-.5ex}1\hspace{-.25ex}|\hspace{-.5ex}\right) \hspace{-0.5ex}.
	\end{IEEEeqnarray}

	Note that $\tilde{\beta}_{1}$ and $\tilde{\beta}_{2}$ are the saddlepoint approximation of the CDF and the complementary CDF of the random variable $\tilde{\iota}(\vec{X};\vec{Y}|Q_{\vec{Y}})$ in~\eqref{EqGenIndDensMut} when $\left(\boldsymbol{X} , \boldsymbol{Y}\right)$ follows the distribution $P_{\vec{X}}P_{\vec{Y}|\vec{X}}$ and $P_{\vec{X}}Q_{\vec{Y}}$, respectively. Consider also the following functions:
	\begin{IEEEeqnarray}{l}
		\label{EqTG1}
		\tilde{G}_1(n,\gamma, \theta, P_{X}, Q_Y) 
		=  \tilde{\beta}_1(n,\gamma, \theta, P_{X},Q_Y) - \frac{ 2 \tilde{\xi}(P_{X},Q_Y,\theta)}{\sqrt{n}}
		\exp\left(n \ln{\tilde{\varphi}(P_{X},Q_Y,\theta)} -\theta \ln{\gamma}\right),\\
		\label{EqTG2}
		\tilde{G}_2(n,\gamma, \theta, P_{X},Q_Y) 
		= \tilde{\beta}_2(n,\gamma, \theta, P_{X},Q_Y) - \frac{ 2  \tilde{\xi}(P_{X},Q_Y,\theta)}{\sqrt{n}}
		\exp \left(n \ln{\tilde{\varphi}(P_{X},Q_Y,\theta)} - (\theta+1) \ln{\gamma}\right),\qquad\\
		\label{EqTU}
		\tilde{G}(n,\gamma,\theta,P_{X},Q_Y,M)
		= \max \left\{ 0, \tilde{G}_1( n, \gamma, \theta,  P_{X}, Q_Y) \right\} +  \gamma \max \left\{ 0, \tilde{G}_2( n, \gamma, \theta, P_{X}, Q_Y) \right\} - \frac{\gamma}{M},\\
		\label{EqTL} 
		\tilde{S}(n,\gamma,\theta,P_{X},Q_Y,M)
		\hspace*{-0.5ex}= \hspace*{-0.5ex}\min\hspace*{-0.5ex} \left\{\hspace*{-.5ex} 1, \tilde{\beta} \left( n, \gamma, \theta, P_{X}, Q_Y, M \right) \hspace*{-0.5ex} + \hspace*{-0.5ex} \frac{ 4  \tilde{\xi}(P_{X},Q_Y,\theta)}{\sqrt{n}} 
			\exp \hspace*{-0.2ex}\left(\hspace*{-0.2ex} n\ln{\hspace*{-0.4ex}\tilde{\varphi}(P_{X},Q_Y,\theta)\hspace*{-0.4ex}} \hspace*{-0.4ex}-\hspace*{-0.4ex}\theta \ln{\hspace*{-0.5ex}\gamma\hspace*{-0.2ex}} \hspace*{-0.4ex}\right) \hspace*{-0.5ex}\right\}\hspace*{-0.7ex},
	\end{IEEEeqnarray}
	 and
	\begin{equation} \label{EqTBeta}
	\tilde{\beta}(n,\gamma, \theta, P_{X}, Q_Y,M) =  \tilde{\beta}_1(n, \gamma,  \theta,P_{X}, Q_Y )  +  \gamma \tilde{\beta}_2(n, \hspace{-.3ex} \gamma,  \theta, P_{X}, Q_Y) - \frac{\gamma}{M}.
	\end{equation}

	The function $\tilde{\beta}(n,\gamma, \theta, P_{X}, Q_Y,M)$ in~\eqref{EqTBeta}  is referred to as  the \emph{saddlepoint  approximation} of  the function $C$ in~\eqref{EqMetConSym}, which is indeed a language abuse. 
	
	The following theorem introduces new lower and upper bounds on the function $C$ in \eqref{EqMetConSym}.
	
	\begin{Theorem}\label{TheoNewMC} 
		Given a pair $(n,M) \in \mathbb{N}^2$,  for all input distributions $P_{\vec{X}} \in \triangle\left( \mathcal{X}^n \right)$ subject to~\eqref{EqiidRandCod}, for all output distributions $Q_{\vec{Y}}\in \triangle\left( \mathcal{Y}^n \right)$ subject to~\eqref{EqGenIndDensMut} such that for all $x\in\set{X}$,
		$P_{Y|X=x}$ is absolutely continuous with respect to $Q_Y$, for all $\gamma\geqslant 0$, the following holds with respect to the random transformation in~\eqref{EqGeneralRamdonTransfomation} subject to~\eqref{EqMemChan},
		\begin{eqnarray}
			\label{EqCbounds}
			\tilde{G}(n,\gamma, \theta, P_{X}, Q_Y,M)   \leqslant  C(n,M, P_{\vec{X}}, Q_{\vec{Y}},\gamma) \leqslant  \tilde{S}(n,\gamma, \theta, P_{X},Q_Y,M) 
		\end{eqnarray}
		where  $\theta$ is the unique solution in $t$ to 
		\begin{eqnarray}\label{EqMCBoundThetaStarA}
			n \mu(P_{X},t) = \ln{\gamma},
		\end{eqnarray}
		and the functions $C$, $\tilde{G}$, and $\tilde{S}$ are defined in~\eqref{EqMetConSym},~\eqref{EqTU} and~\eqref{EqTL}.
	\end{Theorem}
	\begin{proof}
		The proof of Theorem \ref{TheoNewMC} is provided in Appendix~\ref{pTheoNewMC}.
	\end{proof}

	Note that, in \eqref{EqCbounds}, the parameter $\gamma$ can be optimized as in \eqref{EqMetCon}.
	
	\subsection{Numerical Experimentation} \label{sec:numresults}

	The normal and the saddlepoint approximations of the DT and MC bounds as well as their corresponding upper and lower bounds presented from Section~\ref{SecDTNA} to Section~\ref{SecMCSD}  are studied in the cases of the BSC, the AWGN channel, and the  S$\alpha$S channel. The latter is defined by the random transformation in~\eqref{EqGeneralRamdonTransfomation} subject to~\eqref{EqMemChan} and for all $(x,y)\in\set{X}\times \set{Y}$:
	\begin{IEEEeqnarray}{l}
		P_{Y|X}(y|x) = P_{Z}(y-x),
	\end{IEEEeqnarray}
	where $P_Z$ is a probability distribution satisfying for all $t \in \mathds{R}$, 
	\begin{IEEEeqnarray}{lcl}
		\label{EqZ}
		\mathbb{E}_{P_Z}\left[\exp\left(i t Z\right)\right] = \exp\left(-\left|\sigma t\right|^\alpha\right),
	\end{IEEEeqnarray}
	with $i = \sqrt{-1}$. The reals $\alpha \in (0,2]$ and $\sigma \in \mathbb{R}_{+}$ in~\eqref{EqZ} are parameters of the S$\alpha$S channel.

	In the following figures, Figures~\ref{FigBSCGB}--\ref{FigASGB}, the channel inputs are discrete $\mathcal{X} = \lbrace -1, 1 \rbrace$,  $P_X$ is the uniform distribution, and $\theta$ is chosen to be the unique solution to $t$ in \eqref{EqDTBoundThetaStarA} or \eqref{EqMCBoundThetaStarA} depending on whether the DT or MC bound is considered.  For the results relative to the MC bound, $Q_Y$ is chosen to be equal to the distribution $P_Y$, i.e., the marginal of $P_X P_{Y|X}$. The parameter $\gamma$ is chosen to maximize the function $C(n,2^{nR},P_X,Q_Y,\gamma)$ in~\eqref{EqMetConSym}. 
	The plots in Figures~\ref{FigBSCGB}a--\ref{FigASGB}a illustrate the function $T(n,2^{nR},P_X)$ in~\eqref{EqT} as well as the  bounds  in Theorems~\ref{Theo3} and \ref{TheoNew}. 
	Figures~\ref{FigBSCGB}b--\ref{FigASGB}b illustrate the function $C$ in~\eqref{EqMetConSym} and the bounds in Theorems~\ref{Theo3MC} and \ref{TheoNewMC}. 
	The normal approximations, i.e, $\alpha\left(n,2^{n R}, P_X\right)$ in~\eqref{EqAlpha} and $\tilde{\alpha}\left(n,2^{n R}, P_X,Q_Y,\gamma\right)$ in~\eqref{EqTAlpha}, of the DT and MC bounds, respectively, are~plotted in black diamonds. The~upper  bounds, i.e., $N\left(n,2^{n R}, P_X\right)$ in~\eqref{EqN} and $\tilde{N}\left(n,2^{n R}, P_X, Q_Y,\gamma\right)$ in~\eqref{EqTN}, are~plotted in blue squares. 
	The lower bounds of the DT and MC bounds, i.e., $D\left(n,M, P_X\right)$ in~\eqref{EqD} and $\tilde{D}\left(n,2^{n R}, P_X,Q_Y,\gamma\right)$ in~\eqref{EqTD}, are non-positive in these cases, and thus do not appear in the figures.
	The saddlepoint approximations of the DT and MC bounds, i.e., $\beta\left(n,2^{n R},\theta, P_X\right)$ in \eqref{EqBeta} and $\tilde{\beta}\left(n,\gamma,\theta, P_{X},Q_Y,2^{n R}\right)$ in \eqref{EqTBeta}, respectively, are plotted in black stars. 
	The upper bounds,  i.e., $S\left(n,2^{n R},\theta, P_X\right)$ in~\eqref{EqL} and $\tilde{S}\left(n,\gamma,\theta, P_{X},Q_Y,2^{n R}\right)$ in~\eqref{EqTL}, are plotted in blue upward-pointing triangles. The  lower bounds, i.e.,  $G\left(n,2^{n R},\theta, P_X\right)$ in~\eqref{EqU} and  $\tilde{G}\left(n,\gamma,\theta, P_{X},Q_Y,2^{n R}\right)$ in~\eqref{EqTU}, are plotted in  red downward-pointing  triangles.

	Figure \ref{FigBSCGB}  illustrates the case of a BSC with cross-over probability $\delta = 0.11$. The information rates are chosen to be $R = 0.32$ and $R = 0.42$ bits per channel use in Figure~\ref{FigBSCGB}a,b,  respectively. The functions $T$ and $C$ can be calculated exactly and thus they are plotted in magenta asterisks in Figure~\ref{FigBSCGB}a,b, respectively. 
	In these figures, it can be observed that the saddlepoint approximations of the DT and MC bounds, i.e., $\beta$ and  $\tilde{\beta}$, respectively, overlap with the functions $T$ and $C$. These observations are in line with those reported in~\citep{fJos2018}. Therein, the saddlepoint approximations of the RCU bound and the MC bound are both shown to be precise approximations. Alternatively, the normal approximations of the DT and MC bounds, i.e., $\alpha$ and $\tilde \alpha$, do not overlap with $T$ and $C$ respectively.

	In Figure \ref{FigBSCGB}, it can be observed that the new bounds on the DT and MC provided in Theorems~\ref{TheoNew} and \ref{TheoNewMC}, respectively,  are tighter than those in Theorems~\ref{Theo3} and \ref{Theo3MC}. Indeed, the upper-bounds $N$ and $\tilde N$ on the DT and MC bounds derived from the normal approximations $\alpha$ and $\tilde \alpha$, are several order of magnitude above $T$ and $C$, respectively. This observation remains valid for AWGN channels in Figure \ref{FigAWGNGB} and S$\alpha$S channels in Figure \ref{FigASGB}, respectively. Note that, in Figure~\ref{FigBSCGB}a, for $n > 1000$, the normal approximation $\alpha$ is below the lower bound $G$ showing that approximating $T$ by $\alpha$ is too optimistic. These results show that the use of the Berry--Esseen Theorem to approximate the DT and MC bounds may lead to erroneous conclusions due to the uncontrolled error made on the approximation.

	Figures~\ref{FigAWGNGB} and \ref{FigASGB} illustrate the cases of a real-valued AWGN channel and a S$\alpha$S channel, respectively.
	The signal-to-noise ratio (SNR) is $\mathrm{SNR} = 1$ for the AWGN channel. The information rate is $R = 0.425$ bits per channel use for the AWGN channel and $R = 0.38$ bits per channel use for the  S$\alpha$S channel with $(\alpha,\sigma)=(1.4,0.6)$.
	In both cases, the functions $T$ in~\eqref{EqT} and $C$ in~\eqref{EqMetConSym} can not be computed explicitly and hence does not appear in Figures~\ref{FigAWGNGB} and \ref{FigASGB}. In addition, the lower bounds $D\left(n,M, P_X\right)$ and $\tilde{D}\left(n,2^{n R}, P_X,Q_Y,\gamma\right)$ obtained from Theorems~\ref{Theo3} and \ref{Theo3MC} are non-positive in these cases, and thus, do~not appear on these figures.

	In Figure~\ref{FigAWGNGB}, note that the saddlepoint approximations, $\beta$ and $\tilde{\beta}$, are well bounded by  Theorems~\ref{TheoNew} and \ref{TheoNewMC} for a large range of blocklengths. Alternatively, the lower bounds $D$ and $\tilde D$ based on the normal approximation do not even exist in that case.

	In Figure~\ref{FigASGB}, note that the upper bounds $S$ and $\tilde S$ on the DT and MC respectively are relatively tight compared to those in AWGN channel case. This characteristic is of a particular importance in a channel such as S$\alpha$S channel, where the DT and MC bounds remain computable only by Monte Carlo simulations.

	\begin{figure}[H]
	\centering
			\begin{subfigure}{.495\textwidth}
				\centering
				\includegraphics[width=\linewidth]{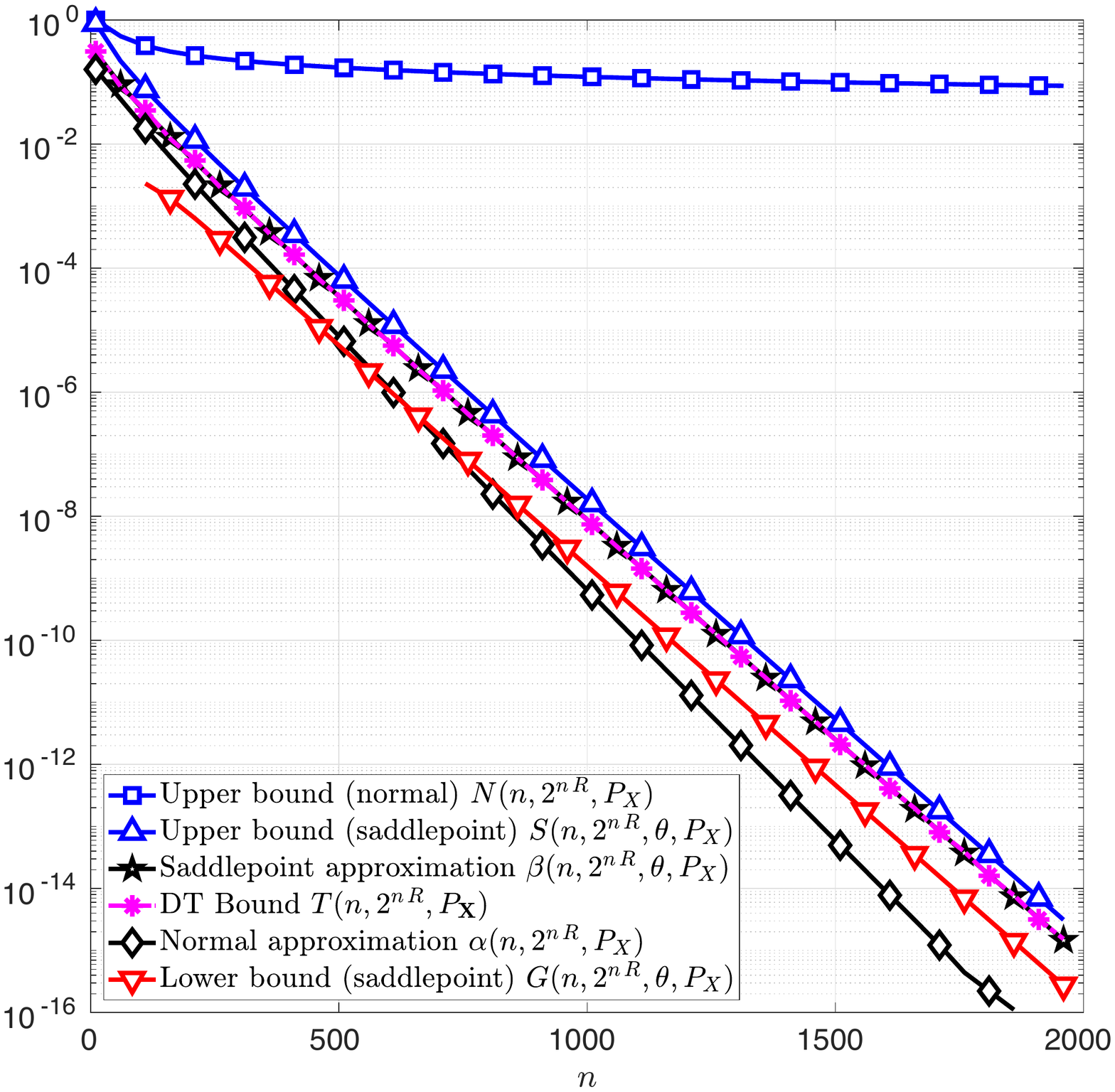}
				\caption{DT Bound}
				\label{FigBSC}
			\end{subfigure}
			\begin{subfigure}{.495\textwidth}
				\centering
				\includegraphics[width=\linewidth]{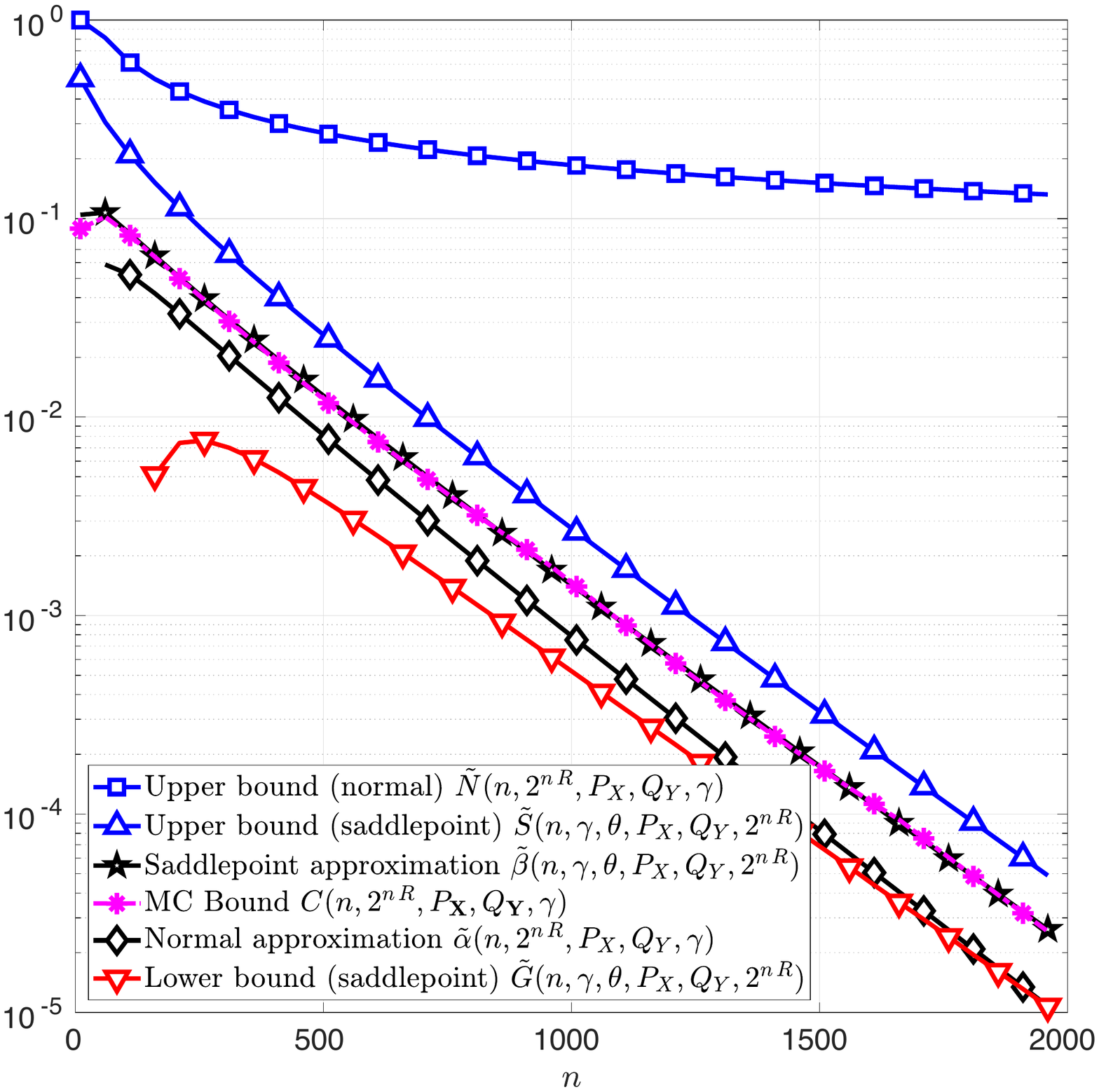}
				\caption{MC Bound}
				\label{FigBSCMC}
			\end{subfigure}
			\caption{Normal and saddlepoint approximations to the functions $T$  (Figure \ref{FigBSCGB}a) in~\eqref{EqT}  and $C$ (Figure~\ref{FigBSCGB}b) in~\eqref{EqMetConSym} as functions of the blocklength $n$ for the case of a BSC with cross-over probability $\delta = 0.11$. 
				The information rate is $R = 0.32$ and $R=0.42$ bits per channel use for Figure~\ref{FigBSCGB}a,b, respectively.
				The channel input distribution $P_X$ is chosen to be the uniform distribution, the output distribution $Q_Y$ is chosen to be the channel output distribution $P_Y$, and the parameter $\gamma$ is chosen to maximize $C$ in~\eqref{EqMetConSym}. 
				The parameter $\theta$ is chosen to be respectively the unique solution to $t$  in~\eqref{EqDTBoundThetaStarA} in Figure~\ref{FigBSCGB}a and in~\eqref{EqMCBoundThetaStarA}  in Figure~\ref{FigBSCGB}b.
			}
			\label{FigBSCGB}
	\end{figure}

	\begin{figure}[H]
	\centering
			\begin{subfigure}{.495\textwidth}
				\centering
				\includegraphics[width=\linewidth]{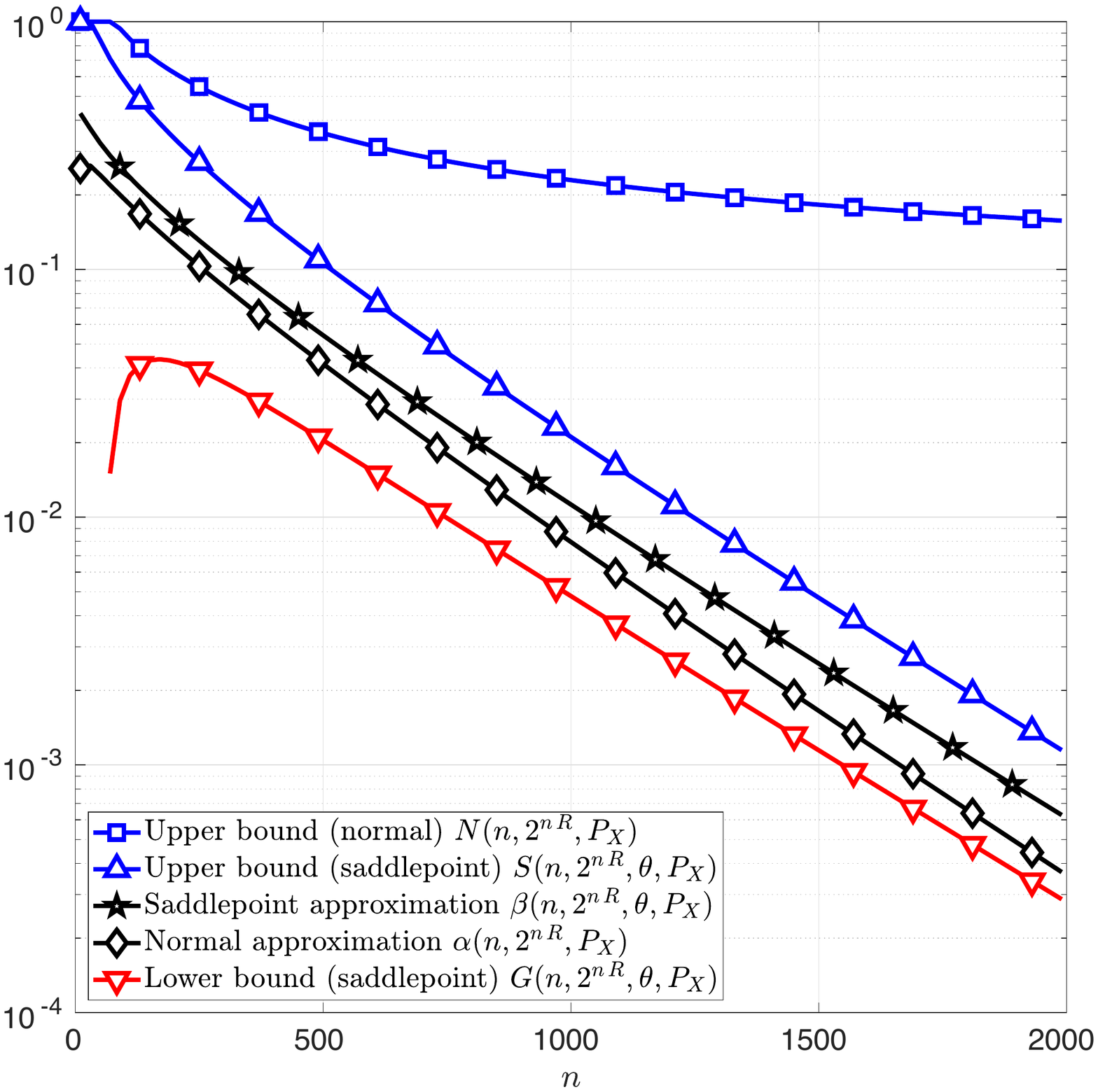}
				\caption{DT Bound}
				\label{FigAWGN}
			\end{subfigure}
			\begin{subfigure}{.495\textwidth}
				\centering
				\includegraphics[width=\linewidth]{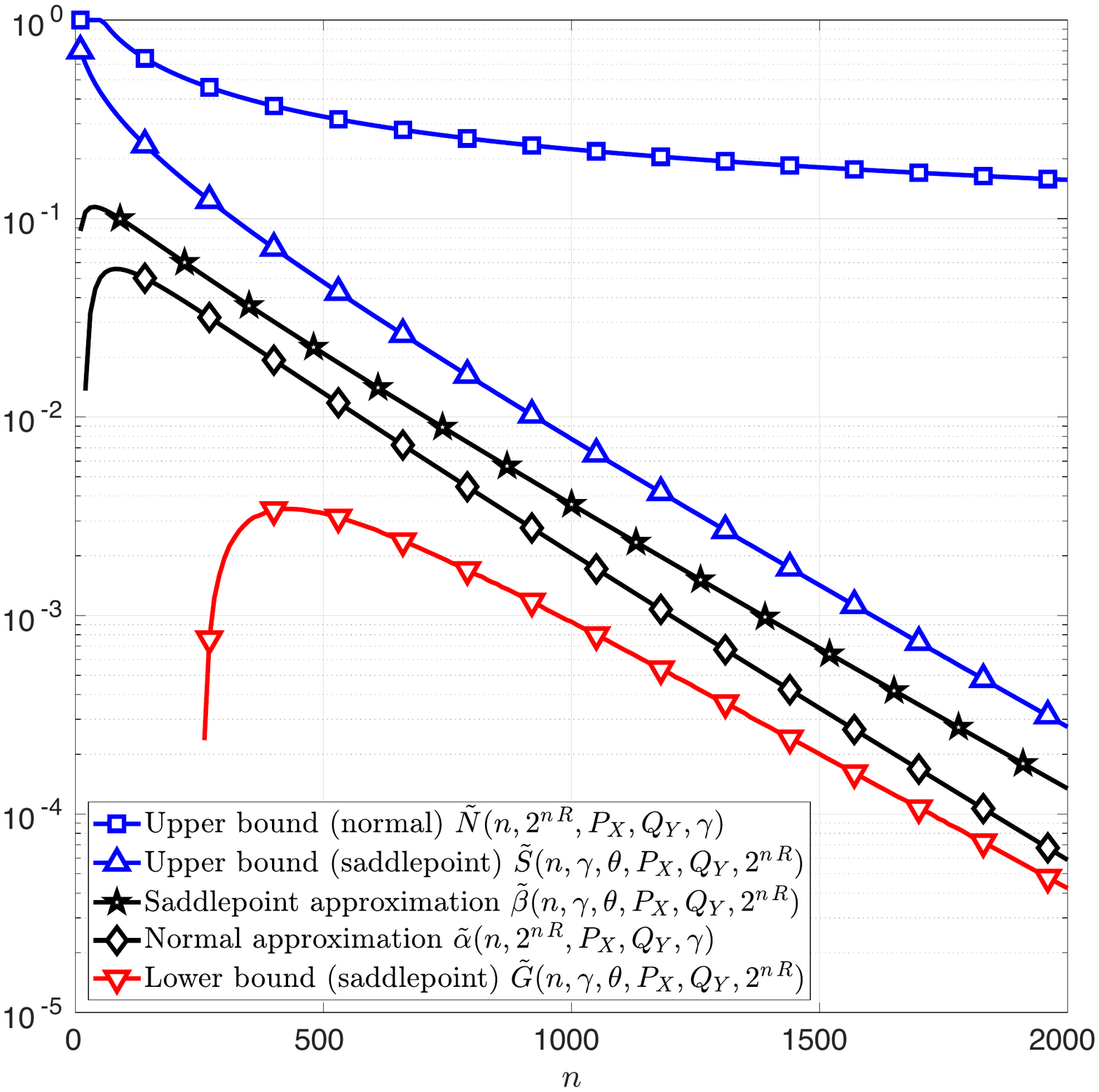}
				\caption{MC Bound}
				\label{FigAWGNMC}
			\end{subfigure}
			\caption{Normal and saddlepoint approximations to the functions $T$ (Figure~\ref{FigAWGNGB}a) in~\eqref{EqT} and $C$ (Figure~\ref{FigAWGNGB}b) in~\eqref{EqMetConSym} as  functions of the blocklength $n$ for the case of a real-valued AWGN channel with discrete channel inputs, $\mathcal{X} = \lbrace -1, 1 \rbrace$, signal to noise ratio $\mathrm{SNR} = 1$, and information rate $R = 0.425$ bits per channel use.
				The channel input distribution $P_X$ is chosen to be the uniform distribution, the~output distribution $Q_Y$ is chosen to be the channel output distribution $P_Y$, and  the parameter $\gamma$ is chosen to maximize $C$ in~\eqref{EqMetConSym}.
				The parameter $\theta$ is respectively chosen to be the unique solution to $t$  in~\eqref{EqDTBoundThetaStarA}  in Figure~\ref{FigAWGNGB}a and in~\eqref{EqMCBoundThetaStarA}  in Figure~\ref{FigAWGNGB}b.
			}
			\label{FigAWGNGB}
	\end{figure}
	\unskip
	\begin{figure}[H]
	\centering
			\begin{subfigure}{.495\textwidth}
				\centering
				\includegraphics[width=\linewidth]{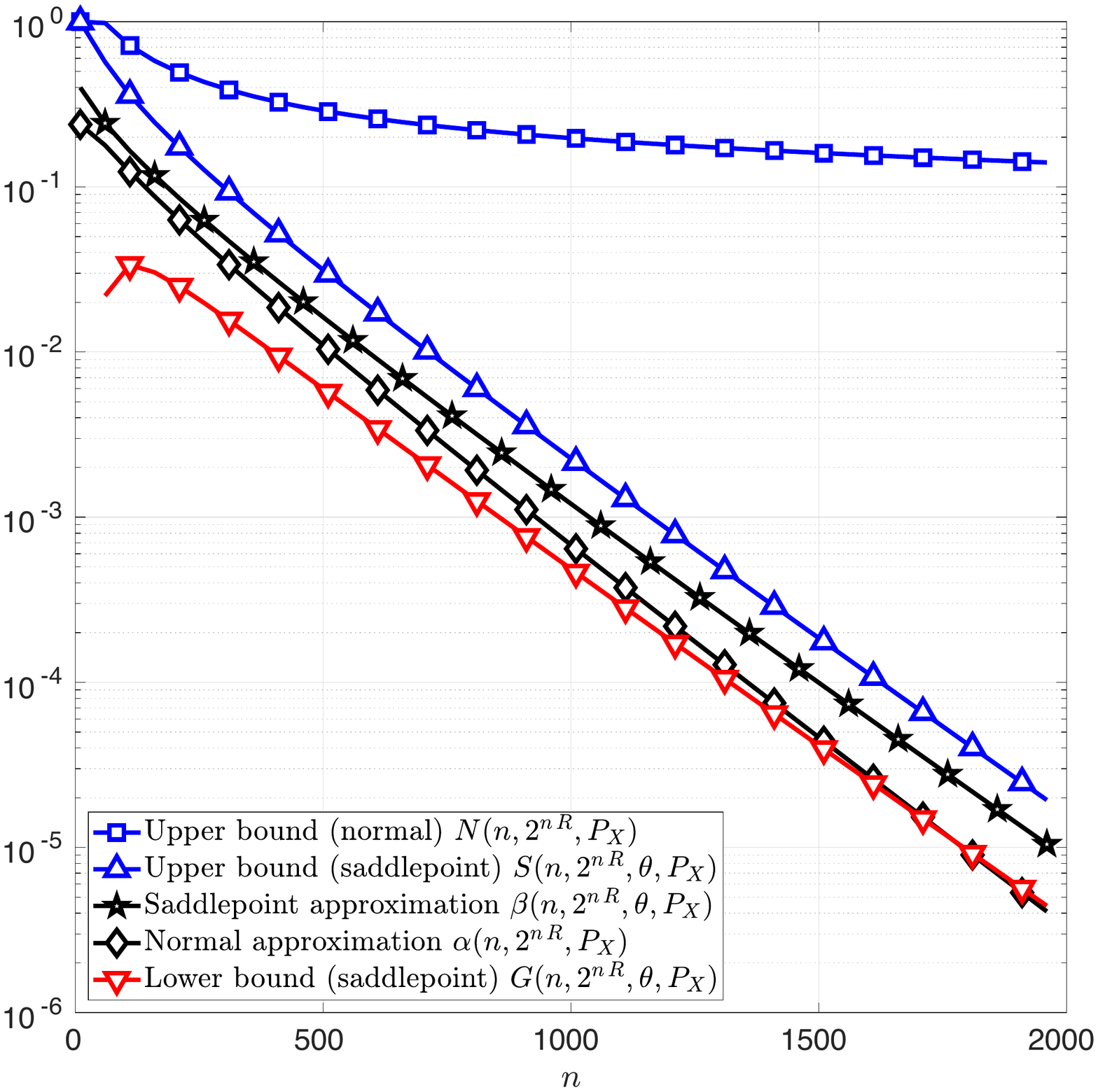}
				\caption{DT Bound}
				\label{FigAS}
			\end{subfigure}
			\begin{subfigure}{.495\textwidth}
				\centering
				\includegraphics[width=\linewidth]{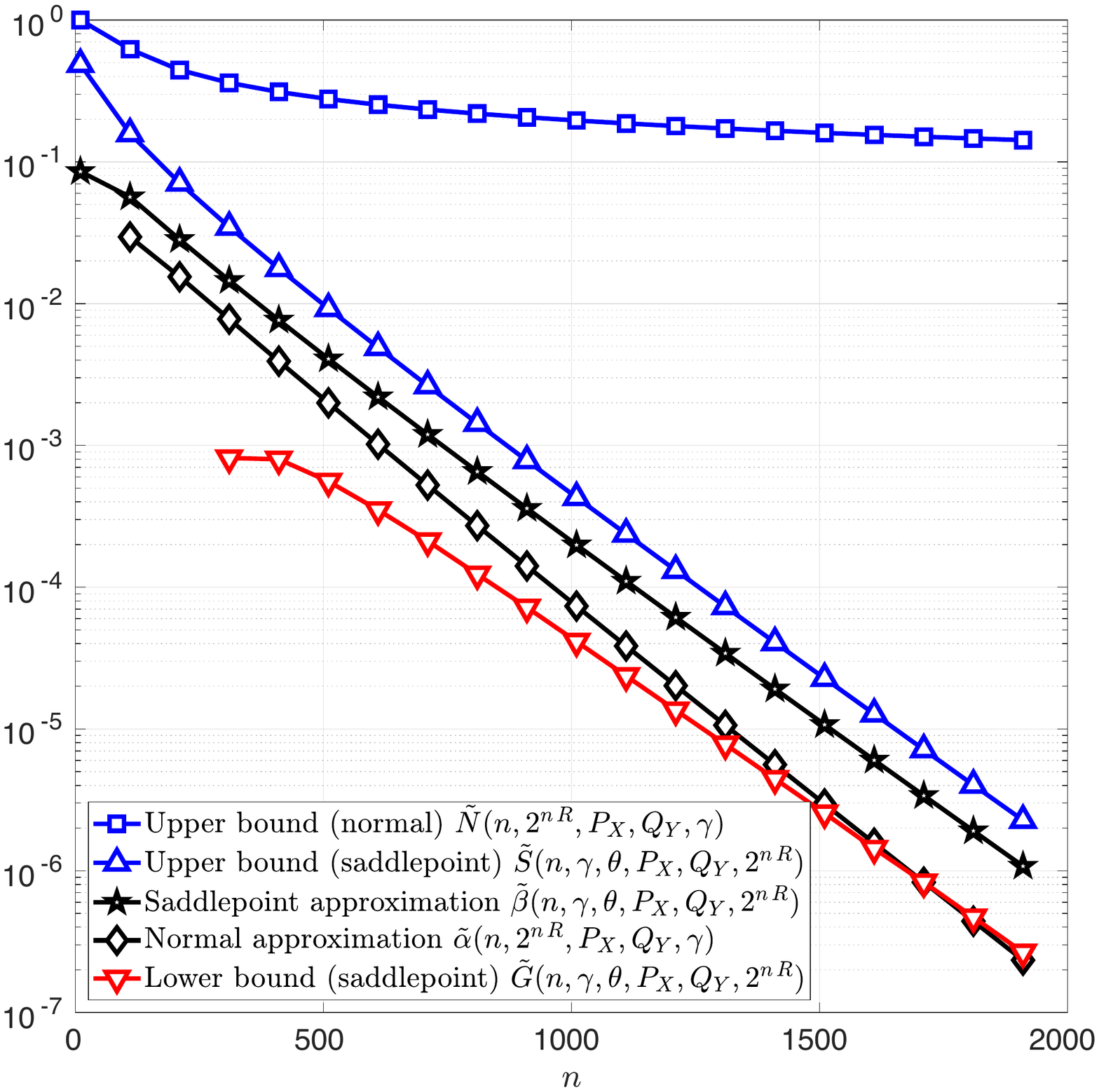}
				\caption{MC Bound}
				\label{FigASMC}
			\end{subfigure}
			\caption{Normal and saddlepoint approximation to the functions $T$ (Figure~\ref{FigASGB}a) in~\eqref{EqT} and $C$ (Figure~\ref{FigASGB}b) in~\eqref{EqMetConSym} as functions of the blocklength $n$ for the case of a real-valued symmetric $\alpha$-stable noise channel with discrete channel inputs $\mathcal{X} = \lbrace -1, 1 \rbrace$,  shape parameter $\alpha=1.4$,  dispersion parameter $\sigma=0.6$, and~information rate $R = 0.38$ bits per channel use.
				The channel input distribution $P_X$ is chosen to be the uniform distribution, the output distribution $Q_Y$ is chosen to be the channel output distribution $P_Y$, and the parameter $\gamma$ is chosen to maximize $C$ in~\eqref{EqMetConSym}.
				The parameter $\theta$ is respectively chosen to be the unique solution to $t$  in~\eqref{EqDTBoundThetaStarA} in Figure~\ref{FigASGB}a and in~\eqref{EqMCBoundThetaStarA} in  Figure~\ref{FigASGB}b.
			}
			\label{FigASGB}
	\end{figure} 
	
	\section{Discussion and Further Work}
	
	One of the main results of this work is Theorem~\ref{TheoSaddlePointBeryUni}, which gives an upper bound on the error induced by the saddlepoint approximation of the CDF of a sum of i.i.d. random variables. 
	This result paves the way to study channel coding problems at any finite blocklength and  any constraint on the DEP.
	In particular, Theorem~\ref{TheoSaddlePointBeryUni} is used to bound the DT and MC bounds in point-to-point memoryless channels. This leads to tighter bounds than those obtained from Berry--Esseen Theorem (Theorem~\ref{TheoBerry}), cf., examples in Section \ref{sec:numresults}, particularly for the small values of the DEP.

	The bound on the approximation error presented in Theorem~\ref{lem:saddlePointBeryUni} uses a triangle inequality in the proof of Lemma~\ref{lem:1ForLem2}, which is loose. This is essentially the reason why Theorem~\ref{lem:saddlePointBeryUni} is not reduced to the Berry--Esseen Theorem when the parameter $\theta$ is equal to zero. An interesting extension of this work is to tighten the inequality in Lemma~\ref{lem:1ForLem2} such that the Berry--Esseen Theorem can be obtained as a special case of Theorem~\ref{lem:saddlePointBeryUni}, i.e., when $\theta = 0$. 
	If such improvement on Theorem~\ref{lem:saddlePointBeryUni} is possible, Theorem~\ref{TheoSaddlePointBeryUni} will be {significantly} improved and it would be more precise everywhere and in particular  in the vicinity of the mean of the sum in~\eqref{EqXn}.
	\vspace{6pt}

	\authorcontributions{All authors have equally contributed to this work. All authors have read and agreed to the published version of the manuscript.}
	\funding{This work was partially funded by the French National Agency for Research (ANR) under grant ANR-16-CE25-0001.}
	\conflictsofinterest{The authors declare no conflict of interest.} 
	\appendixtitles{yes}
	
	\appendix
		\section{Proof of Theorem~\ref{lem:saddlePointBeryUni}}\label{plem:saddlePointBeryUni}
		
		The proof of Theorem \ref{lem:saddlePointBeryUni} relies on the notion of exponentially tilted distributions. 
		Let $\varphi_Y$ be the moment generating function of the distribution $P_Y$.
		Given  $\theta \in \Theta_{Y}$, let $Y^{(\theta)}_1$, $Y^{(\theta)}_2$, $\ldots$, $Y^{(\theta)}_n$ be random variables whose joint probability distribution, denoted by  $P_{Y^{(\theta)}_1 Y^{(\theta)}_2 \ldots Y^{(\theta)}_n}$, satisfies for all \mbox{$(y_1, y_2, \ldots, y_n)\in\mathbb{R}^n$,}
		\begin{IEEEeqnarray}{lcl}\label{EqRND1}
			\rndder{P_{Y^{(\theta)}_1 Y^{(\theta)}_2\ldots Y^{(\theta)}_n}}{P_{Y_1Y_2\ldots Y_n}}(y_1,y_2,\ldots,y_n) = \frac{\exp\left(\theta \sum_{j=1}^n y_j\right)}{\left( \varphi_{Y}(\theta) \right)^n}.
		\end{IEEEeqnarray} 
		That is, the distribution $P_{Y^{(\theta)}_1 Y^{(\theta)}_2\ldots Y^{(\theta)}_n}$ is an  exponentially tilted distribution with respect to $P_{Y_1 Y_2\ldots Y_n}$. 
		Using this notation, for all  $\set{A} \subseteq \mathbb{R}$ and for all $\theta\in\Theta_{Y}$,
		\begin{subequations}
			\begin{IEEEeqnarray}{lcl}
				P_{X_n}(\mathcal{A}) &=& \mathbb{E}_{P_{X_n}}[\mathds{1}_{\{X_n\in\set{A}\}}] \\
				&=& \mathbb{E}_{P_{Y_1Y_2\ldots Y_n}}\left[\mathds{1}_{\left\{\sum_{j=1}^n Y_j\in\set{A}\right\}}\right]\\
				&=& \mathbb{E}_{P_{Y^{(\theta)}_1 Y^{(\theta)}_2\ldots Y^{(\theta)}_n}}\left[\rndder{P_{Y_1Y_2\ldots Y_n}}{P_{Y^{(\theta)}_1 Y^{(\theta)}_2\ldots Y^{(\theta)}_n}}(Y^{(\theta)}_1,Y^{(\theta)}_2,\ldots,Y^{(\theta)}_n)\mathds{1}_{\left\{\sum_{j=1}^n Y^{(\theta)}_j\in\set{A}\right\}}\right]\label{EqAbsContEaO1}\\
				&=& \mathbb{E}_{P_{Y^{(\theta)}_1 Y^{(\theta)}_2\ldots Y^{(\theta)}_n}}\left[\hspace*{-0.8ex}\left(\rndder{P_{Y^{(\theta)}_1 Y^{(\theta)}_2\ldots Y^{(\theta)}_n}}{P_{Y_1Y_2\ldots Y_n}}(Y^{(\theta)}_1,Y^{(\theta)}_2,\ldots,Y^{(\theta)}_n)\right)^{-1}\mathds{1}_{\left\{\sum_{j=1}^n Y^{(\theta)}_j\in\set{A}\right\}}\hspace*{-0.8ex}\right]\label{EqAbsContEaO}\\
				&=& \mathbb{E}_{P_{Y^{(\theta)}_1 Y^{(\theta)}_2\ldots Y^{(\theta)}_n}}\left[\left(\frac{\exp{\left(\theta \sum_{j=1}^n Y^{(\theta)}_j\right)}}{\left(\varphi_{Y}(\theta)\right)^n}\right)^{-1}\mathds{1}_{\left\{\sum_{j=1}^n Y^{(\theta)}_j\in\set{A}\right\}}\right] \label{EqAbsContEaO2}\\
				\label{EqGosumiid}
				&=& \left(\varphi_{Y}(\theta)\right)^n \mathbb{E}_{P_{Y^{(\theta)}_1 Y^{(\theta)}_2\ldots Y^{(\theta)}_n}}\left[\exp{\left(-\theta \sum_{j=1}^n Y^{(\theta)}_j\right)}\mathds{1}_{\left\{\sum_{j=1}^n Y^{(\theta)}_j\in\set{A}\right\}}\right]
			\end{IEEEeqnarray}
		\end{subequations}
		For the ease of the notation, consider the random variable
		\begin{IEEEeqnarray}{lcl}\label{EqPsumiidExp}
			S_{n,\theta} = \sum_{j=1}^n Y^{(\theta)}_j,
		\end{IEEEeqnarray}
		whose probability distribution is denoted by $P_{S_{n,\theta}}$.
		Hence, plugging~\eqref{EqPsumiidExp} in~\eqref{EqGosumiid} yields,
		\begin{IEEEeqnarray}{lcl}
			P_{X_n}(\mathcal{A}) &=& \left(\varphi_{Y}(\theta)\right)^n \mathbb{E}_{P_{S_{n,\theta}}}\left[\exp{\left(-\theta S_{n,\theta}\right)}\mathds{1}_{\left\{S_{n,\theta}\in\set{A}\right\}}\right]. \label{EqwithsumVar}
		\end{IEEEeqnarray}
		The proof continues by upper bounding the following absolute difference
		\begin{IEEEeqnarray}{l}
			\label{EqGoDaddy}
			\left|P_{X_n}(\set{A}) - \left(\varphi_{Y}(\theta)\right)^n \mathbb{E}_{P_{Z_{n,\theta}}}\left[\exp{\left(-\theta Z_{n,\theta}\right)}\mathds{1}_{\left\{Z_{n,\theta}\in\set{A}\right\}}\right]\right|,
		\end{IEEEeqnarray}
		where   $Z_{n,\theta}$ is a Gaussian random variable with the same mean and variance as $S_{n,\theta}$, and probability distribution denoted by $P_{Z_{n, \theta}}$. 
		The relevance of the absolute difference in~\eqref{EqGoDaddy} is that it is equal to the error of calculating $P_{X_n}(\set{A})$ under the assumption that the resulting random variable $S_n$ follows a Gaussian distribution.
		The following lemma provides an upper bound on the absolute difference in~\eqref{EqGoDaddy}  in terms of the Kolmogorov--Smirnov distance between the distributions $P_{S_{n,\theta}}$ and $P_{Z_{n,\theta}}$, denoted~by
		\begin{IEEEeqnarray}{lcl}\label{EqSuppDistFuncDiff}
			\Delta \left(P_{S_{n,\theta}}, P_{Z_{n,\theta}} \right) \df \sup_{x   \in \mathbb{R}} \left| F_{S_{n,\theta}}(x) -  F_{Z_{n,\theta}}(x)\right|,
		\end{IEEEeqnarray} 
		where  $F_{S_{n,\theta}}$ and $F_{Z_{n,\theta}}$ are the CDFs of the random variables $S_{n,\theta}$ and $Z_{n,\theta}$, respectively.
		
		\begin{Lemma}\label{lem:1ForLem2}  
			Given $\theta \in \Theta_Y$ and $a \in \mathbb{R}$, consider the following conditions: \begin{enumerate}[leftmargin=2.3em,labelsep=4mm]
			\item[$(i)$] $\theta \leqslant 0$ and $\mathcal{A} = \left( -\infty, a \right]$, and  
			\item[$(ii)$] $\theta >  0$ and $\mathcal{A} = \left( a, \infty \right)$. 
			\end{enumerate}
			
			If at least one of the above conditions is satisfied, then the absolute difference in~\eqref{EqGoDaddy} satisfies
			\begin{eqnarray}
				\left|P_{X_n}(\set{A}) - \left(\varphi_{Y}(\theta)\right)^n \mathbb{E}_{P_{Z_{n,\theta}}}\left[\exp{\left(-\theta Z_{n,\theta}\right)}\mathds{1}_{\left\{Z_{n,\theta}\in\set{A}\right\}}\right]\right| 
				\label{EqOhYeah}
				\leqslant \frac{\left(\varphi_{Y}(\theta)\right)^n}{\exp(\theta a)} {\min\left\{1,2\,\Delta(P_{S_{n,\theta}},P_{Z_{n,\theta}})\right\}}.
			\end{eqnarray}
		\end{Lemma}
		\begin{proof}
			The proof of Lemma~\ref{lem:1ForLem2} is presented in Appendix~\ref{plem:1ForLem2}.
		\end{proof}

		The proof continues by providing an upper bound on $ \Delta \left(P_{S_{n,\theta}}, P_{Z_{n,\theta}} \right)$ in~\eqref{EqOhYeah} leveraging the observation that $S_{n,\theta}$ is the sum of $n$ independent and identically distributed random variables. This~follows immediately from the assumptions of  Theorem~\ref{lem:saddlePointBeryUni}, nonetheless, for the sake of completeness,  the following lemma provides a proof of this statement. 
		\begin{Lemma}\label{LemmaBeautiful}
			For all $\theta \in \Theta_{Y}$,  $Y^{(\theta)}_1$, $Y^{(\theta)}_2$, $\ldots$, $Y^{(\theta)}_n$ are mutually independent and identically distributed  random variables with probability distribution $P_{Y^{(\theta)}}$. Moreover, $P_{Y^{(\theta)}}$ is an exponential tilted distribution with respect to $P_{Y}$. That is, $P_{Y^{(\theta)}}$ satisfies for all $y\in\mathbb{R}$,
			\begin{eqnarray}\label{EqRND2}
				\rndder{P_{Y^{(\theta)}}}{P_{Y}}(y) = \frac{\exp\left(\theta y\right)}{ \varphi_{Y}(\theta) }.
			\end{eqnarray} 
		\end{Lemma}
		\begin{proof}
			The proof of Lemma \ref{LemmaBeautiful} is presented in Appendix \ref{ProofLemmaBeautiful}.
		\end{proof}

		Lemma~\ref{LemmaBeautiful}  paves the way for obtaining an upper bound on $ \Delta \left(P_{S_{n,\theta}}, P_{Z_{n,\theta}} \right)$ in~\eqref{EqOhYeah} via the Berry--Esseen Theorem (Theorem~\ref{TheoBerry}). 
		Let $\mu_{\theta}$, $V_{\theta}$, and {$T_{\theta}$} be the mean, the variance, and the third absolute central moment of the random variable $Y^{(\theta)}$, whose probability distribution is $P_{Y^{(\theta)}}$ in~\eqref{EqRND2}. More~specifically:
		\begin{IEEEeqnarray}{lcl}
			\mu_{\theta} & = &\mathbb{E}_{P_{Y^{(\theta)}}}[Y^{(\theta)}] 
			=  \mathbb{E}_{P_{Y}}\left[\frac{Y \exp{(\theta Y)}}{\varphi_{Y}(\theta)}\right],\\
			V_{\theta} & = &\mathbb{E}_{P_{Y^{(\theta)}}}[(Y^{(\theta)}-\mu_{\theta})^2] =\mathbb{E}_{P_{Y}}\left[\frac{(Y-\mu_{\theta})^2 \exp{(\theta Y)}}{\varphi_{Y}(\theta)}\right],
			\mbox{and}\\
			T_{\theta} & = &\mathbb{E}_{P_{Y^{(\theta)}}}[|Y^{(\theta)}-\mu_{\theta}|^3] 
			= \mathbb{E}_{P_{Y}}\left[\frac{|Y-\mu_{\theta}|^3 \exp{(\theta Y)}}{\varphi_{Y}(\theta)}\right].
		\end{IEEEeqnarray}
		Let also $\xi_{\theta}$ be
		\begin{IEEEeqnarray}{l}
			\xi_{\theta} = c_1\left(\frac{T_{\theta}}{V_{\theta}^{3/2}}+c_2\right),
		\end{IEEEeqnarray}
		with $c_1$ and $c_2$ defined in~\eqref{EqConstants}.

		From Theorem~\ref{TheoBerry}, it follows that $ \Delta \left(P_{S_{n,\theta}}, P_{Z_{n,\theta}} \right)$ in~\eqref{EqOhYeah} satisfies: 
		\begin{IEEEeqnarray}{l}
			\Delta(P_{S_{n,\theta}},P_{Z_{n,\theta}})  	\leqslant  \min\left\{1,\frac{\xi_{\theta}}{\sqrt{n } }\right\}
			\label{EqPfexpB}
			\leqslant  \frac{\xi_{\theta}}{\sqrt{n} }.
		\end{IEEEeqnarray}
		Plugging~\eqref{EqPfexpB} in~\eqref{EqOhYeah} yields
		\begin{IEEEeqnarray}{l}
			\left|\hspace*{-0.5ex}P_{X_n}\hspace*{-0.3ex}(\hspace*{-0.2ex}\set{A}\hspace*{-0.2ex}) \hspace*{-0.4ex} - \hspace*{-0.4ex} \frac{\left(\varphi_{Y}(\theta)\right)^n}{\exp(\theta b)} \mathbb{E}_{P_{Z_{n,\theta}}}\hspace*{-0.5ex}\left[\exp{\left(-\theta Z_{n,\theta}\right)}\mathds{1}\left\{Z_{n,\theta}\in\set{A}\right\}\right]\right|
			\label{EqOhYeahOh}
			\leqslant  \frac{\left(\varphi_{Y}(\theta)\right)^n}{\exp(\theta a)} \min\left\{\hspace*{-0.4ex}1,2 \frac{\xi_{\theta}}{\sqrt{n } } \hspace*{-0.4ex}\right\},\qquad
		\end{IEEEeqnarray}
		under the assumption that at least one of the conditions of Lemma~\ref{lem:1ForLem2} is met.

		The proof ends by obtaining a closed-form expression of the term $\mathbb{E}_{P_{Z_{n,\theta}}}\big[$ $\exp\left(-\theta Z_{n,\theta}\right)$ $\mathds{1}_{\{Z_{n,\theta}\in \set{A}\}}$~$\big]$ in~\eqref{EqOhYeahOh} under the assumption that at least one of the conditions of Lemma~\ref{lem:1ForLem2} is met.
		First, assuming that condition $(i)$ in Lemma~\ref{lem:1ForLem2} holds, it follows that:
		\begin{subequations}
			\begin{IEEEeqnarray}{l}
				\mathbb{E}_{P_{Z_{n,\theta}}}\left[\exp\left(-\theta Z_{n,\theta}\right)\mathds{1}_{\{Z_{n,\theta}\in \set{A}\}}\right]\nonumber \\
				= \int^a_{-\infty} \exp\left(-\theta z\right) \frac{1}{\sqrt{2\pi  n V_{\theta}}} \exp\left(-\frac{(z-   n \mu_{\theta} )^2}{2  n V_{\theta}}\right)\me{d}z\\
				= \int^a_{-\infty} \frac{1}{\sqrt{2\pi  n V_{\theta}}} \exp\left(-\frac{z^2 -2\, z\,  n \mu_{\theta}  +  n^2 \mu_{\theta} ^2 + 2 n \theta\, V_{\theta} \, z}{2  n V_{\theta}}\right)\me{d}z\\
				= \int^a_{-\infty} \frac{1}{\sqrt{2\pi  n V_{\theta}}} \exp\left(-\frac{(z- n \mu_{\theta}  +  n \theta V_{\theta})^2 -  n^2 \theta^2 V^2_{\theta}  + 2 n \mu_{\theta} \, n \theta V_{\theta}}{2  n V_{\theta}}\right)\me{d}z\\
				= \exp\left( -\theta n \mu_{\theta}  + \frac{1}{2} n V_{\theta}\theta^2\right)\int^a_{-\infty} \frac{1}{\sqrt{2\pi  n V_{\theta}}} \exp\left(-\frac{(z- n \mu_{\theta}  +  n \theta V_{\theta})^2}{2  n V_{\theta}}\right)\me{d}z \\
				= \exp\left( -\theta n \mu_{\theta}  + \frac{1}{2} n V_{\theta}\theta^2\right)\int^{\frac{a- n \mu_{\theta}  +  n \theta V_{\theta}}{\sqrt{ n V_{\theta}}}}_{-\infty} \frac{1}{\sqrt{2\pi}} \exp\left(-\frac{t^2}{2}\right)\me{d}t\\
				\label{EqpfGauExp1}
				= \exp\left( -\theta n \mu_{\theta}  + \frac{1}{2} n V_{\theta}\theta^2\right) Q\left(-\frac{a- n \mu_{\theta}  +  n \theta V_{\theta}}{\sqrt{n V_{\theta}}}\right).
			\end{IEEEeqnarray}
		\end{subequations}
		Second, assuming that condition $(ii)$ in Lemma~\ref{lem:1ForLem2} holds, it follows that:
		\begin{subequations}
			\begin{IEEEeqnarray}{lcl}
				\mathbb{E}_{P_{Z_{n,\theta}}}\left[\exp\left(-\theta Z_{n,\theta}\right)\mathds{1}_{\{Z_{n,\theta}\in \set{A}\}}\right]
				&=& \int_a^{\infty} \exp\left(-\theta z\right) \frac{1}{\sqrt{2\pi  n V_{\theta}}} \exp\left(-\frac{(z-   n \mu_{\theta} )^2}{2  n V_{\theta}}\right)\me{d}z\\
				&=& \exp\left( -\theta n \mu_{\theta}  + \frac{1}{2} n V_{\theta}\theta^2\right)\int_{\frac{a- n \mu_{\theta}  +  n \theta V_{\theta}}{\sqrt{ n V_{\theta}}}}^{\infty} \frac{1}{\sqrt{2\pi}} \exp\left(-\frac{t^2}{2}\right)\me{d}t\qquad\\
				\label{EqpfGauExp2}
				&=& \exp\left( -\theta n \mu_{\theta}  + \frac{1}{2} n V_{\theta}\theta^2\right) Q\left(\frac{a- n \mu_{\theta}  +  n \theta V_{\theta}}{\sqrt{n V_{\theta}}}\right),
			\end{IEEEeqnarray}
		\end{subequations}
		where $Q$ in~\eqref{EqpfGauExp1} and~\eqref{EqpfGauExp2} is the complementary CDF of the standard Gaussian distribution defined in~\eqref{EqGSCCDF}.

		The expressions in~\eqref{EqpfGauExp1} and~\eqref{EqpfGauExp2} can be jointly  written as follows:
		\begin{IEEEeqnarray}{l}
			\mathbb{E}_{P_{Z_{n,\theta}}}\left[\exp\left(-\theta Z_{n,\theta}\right)\mathds{1}_{\{Z_{n,\theta}\in \set{A}\}}\right]  =  \exp\left(\hspace*{-0.8ex}  -\theta n \mu_{\theta}  + \frac{1}{2} n V_{\theta}\theta^2\hspace*{-0.5ex} \right) 
			\label{EqGauFexp}
			Q\left(\hspace*{-0.8ex} (-1)^{\mathds{1}_{\{\theta \leqslant 0\}}}\frac{a- n \mu_{\theta}  +  n \theta V_{\theta}}{\sqrt{n V_{\theta}}}\hspace*{-0.5ex} \right), \qquad
		\end{IEEEeqnarray}
		under the assumption that at least one of the conditions $(i)$ or $(ii)$ in Lemma~\ref{lem:1ForLem2} holds.

		Finally, under the same assumption, plugging~\eqref{EqGauFexp} in~\eqref{EqOhYeahOh} yields
		\begin{IEEEeqnarray}{l}
			\label{Eqpfflbd}
			\Bigg|P_{X_n}(\set{A}) - \exp\left(n\ln{\varphi_{Y}(\theta)} - n \theta \mu_{\theta} + \frac{1}{2} n \theta^2 V_{\theta}  \right) \nonumber
			Q\left((-1)^{\mathds{1}_{\{\theta\leqslant 0\}}}\frac{a + n \theta V_{\theta}-  n \mu_{\theta} }{\sqrt{n V_{\theta}}} \right) \Bigg| \\
			\leqslant \exp\left(n\ln{\varphi_{Y}(\theta)}-\theta a \right) \min\left\{1, \frac{2\; \xi_{{\theta}}}{\sqrt{n}}\right\}.
		\end{IEEEeqnarray}
		Under condition $(i)$ in Lemma~\ref{lem:1ForLem2}, the inequality in~\eqref{Eqpfflbd} can be written as follows: 
		\begin{IEEEeqnarray}{l}
			\label{Eqpfflbd1}
			\Bigg|F_{X_n}(a) - \exp\left(n\ln{\varphi_{Y}(\theta)} - n \theta \mu_{\theta} + \frac{1}{2} n \theta^2 V_{\theta}  \right) \nonumber
			\cdot Q\left((-1)^{\mathds{1}_{\{\theta\leqslant 0\}}}\frac{a + n \theta V_{\theta}-  n \mu_{\theta} }{\sqrt{n V_{\theta}}} \right) \Bigg| \\
			\leqslant \exp\left(n\ln{\varphi_{Y}(\theta)}-\theta a \right) \min\left\{1, \frac{2\;  \xi_{{\theta}}}{\sqrt{n}}\right\}.
		\end{IEEEeqnarray}
		Alternatively, under condition $(ii)$ in Lemma~\ref{lem:1ForLem2}, it follows from~\eqref{Eqpfflbd} that 
		\begin{IEEEeqnarray}{l}
			\label{Eqpfflbd2}
			\Bigg|1-F_{X_n}(a) - \exp\left(n\ln{\varphi_{Y}(\theta)} - n \theta \mu_{\theta} + \frac{1}{2} n \theta^2 V_{\theta}  \right) \nonumber
			\cdot Q\left((-1)^{\mathds{1}_{\{\theta\leqslant 0\}}}\frac{a + n \theta V_{\theta}-  n \mu_{\theta} }{\sqrt{n V_{\theta}}} \right) \Bigg| \\
			\leqslant \exp\left(n\ln{\varphi_{Y}(\theta)}-\theta a \right) \min\left\{1, \frac{2\; \xi_{\theta}}{\sqrt{n}}\right\},
		\end{IEEEeqnarray}
		Then, jointly writing~\eqref{Eqpfflbd1} and~\eqref{Eqpfflbd2}, it follows that, for all $a \in \mathbb{R}$ and for all $\theta \in \Theta_Y$, 
		\begin{IEEEeqnarray}{l}
			\label{Eqpfflbdf}
			\left|F_{X_n}\hspace*{-0.5ex}(\hspace*{-0.3ex}a\hspace*{-0.3ex}) \hspace*{-0.5ex} - \hspace*{-0.5ex} \mathds{1}_{\{\hspace*{-0.3ex}\theta>0 \hspace*{-0.3ex}\}} \hspace*{-0.5ex}-\hspace*{-0.5ex} (\hspace*{-0.3ex}-1\hspace*{-0.2ex})^{\mathds{1}_{\{\hspace*{-0.2ex}\theta>0\hspace*{-0.2ex} \}}} \hspace*{-0.5ex}\exp\hspace*{-0.5ex}\left(\hspace*{-0.7ex}n\ln{\hspace*{-0.2ex}\varphi_{Y}(\theta)\hspace*{-0.2ex}}\hspace*{-0.5ex}-\hspace*{-0.5ex} n \theta \mu_{\theta}\hspace*{-0.5ex} +\hspace*{-0.5ex} \frac{1}{2} n \theta^2 V_{\theta}\hspace*{-0.7ex} \right) 
			\hspace*{-0.5ex}Q\hspace*{-0.5ex}\left(\hspace*{-0.5ex}(-1)^{\mathds{1}_{\{\theta\leqslant 0\}}}\frac{a + n \theta V_{\theta}-  n \mu_{\theta} }{\sqrt{n V_{\theta}}} \hspace*{-0.5ex}\right) \right| \nonumber\\
			\leqslant \exp\left(n\ln{\varphi_{Y}(\theta)}-\theta a \right) \min\left\{1, \frac{2\; \xi_{\theta}}{\sqrt{n}}\right\},
		\end{IEEEeqnarray}
		which can also be written as
		\begin{IEEEeqnarray}{l}
			\label{EqTilteapproxx}
			\left|F_{X_{n}}(a) -\eta_Y\left(\theta,a,n\right) \right| \leqslant  \exp\left(n K_{Y}(\theta)-\theta \, a \right)\min\left\{1,\frac{2\; \xi_Y(\theta)}{\sqrt{n}}\right\}.
		\end{IEEEeqnarray}
		This completes the proof.
		
		\section{Proof of Lemma~\ref{LemmaH}}\label{pLemmaH}
		Let $g: \mathbb{R}^2 \times \mathbb{N} \rightarrow \mathbb{R}$ be for all $(\theta,a,n) \in \mathbb{R}^2 \times \mathbb{N}$, 
		\begin{equation}
		g(\theta, a, n) = n K_Y(\theta) - \theta a = n  \me{ln}\left(\varphi_{Y}(\theta)\right) - \theta \, a.
		\end{equation}
		First, note that for all $\theta\in\Theta_Y$ and for all $n \in \mathbb{N}$, the function $g$ is a concave function of $a$. Hence, from~the definition of the function $h$ in~\eqref{EqFunctionH}, $h$  is concave. 
		
		Second, note that $0 \in \Theta_{Y}$ given that $\varphi_{Y}(0) = 1 < \infty$. Hence, 
		from~\eqref{EqFunctionH}, it holds that, for all $a\in\mathbb{R}$,
		\begin{subequations}
			\begin{IEEEeqnarray}{l}
				h(a) \leqslant n K_Y(0) = n\ln{\varphi_Y(0)} 
				= n\ln{1}
				= 0.
			\end{IEEEeqnarray}
		\end{subequations}
		This shows that the function $h$ in~\eqref{EqFunctionH} is not positive.

		Third, the next step of the proof consists of proving the equality in~\eqref{EqHRedef}.
		For doing so, 
		let~$\theta^{\star}:\mathbb{R} \times \mathbb{N} \rightarrow \mathbb{R}$ be for all $(a, n) \in \mathbb{R} \times \mathbb{N}$,
		\begin{IEEEeqnarray}{lcl}
			\label{eqthstdf}
			\theta^{\star}(a,n) = \underset{\theta \in \Theta_Y}{\mathrm{arg} \inf}\, g(\theta, a, n).
		\end{IEEEeqnarray} 
		
		Note that the function $g$ is a convex in $\theta$. 
		This follows by verifying that its second derivative with respect to $\theta$ is positive. That is,
		\begin{subequations}
			\begin{IEEEeqnarray}{lcl}
				\label{EqderG}
				\frac{\mathrm{d}}{\mathrm{d}\theta} g(\theta, a, n)  & =&  \frac{n}{\varphi_Y(\theta)}\frac{\mathrm{d}}{\mathrm{d}\theta} \varphi_Y(\theta) - a, \mbox{ and } \\ 
				\frac{\mathrm{d}^2}{\mathrm{d}\theta^2} g(\theta, a, n)& = &  \frac{n}{\left(\varphi_Y(\theta)\right)^2} \left( \varphi_Y(\theta) \frac{\mathrm{d}^2}{\mathrm{d}\theta^2} \varphi_Y(\theta)  - \left(\frac{\mathrm{d}}{\mathrm{d}\theta} \varphi_Y(\theta)\right)^2 \right)\\
				\label{EqPfgder2f1}
				& = & n \left(\frac{1}{\varphi_Y(\theta)}\frac{\mathrm{d}^2}{\mathrm{d}\theta^2} \varphi_Y(\theta)	 -  \left(\frac{1}{\varphi_Y(\theta)}\frac{\mathrm{d}}{\mathrm{d}\theta} \varphi_Y(\theta)\right)^2\right)\\
				& = & n \left(\frac{1}{\varphi_Y(\theta)}\frac{\mathrm{d}^2}{\mathrm{d}\theta^2} \mathbb{E}_{P_Y}[\exp(\theta Y)]	 -  \left(\frac{1}{\varphi_Y(\theta)}\frac{\mathrm{d}}{\mathrm{d}\theta} \mathbb{E}_{P_Y}[\exp(\theta Y)]\right)^2\right)\\
				\label{EqPfgder2f2}
				& = & n \left(\frac{\mathbb{E}_{P_Y}[Y^2 \exp(\theta Y)]}{\mathbb{E}_{P_Y}[\exp(\theta Y)]} -  \left(\frac{\mathbb{E}_{P_Y}[Y \exp(\theta Y)]}{\mathbb{E}_{P_Y}[ \exp(\theta Y)]}\right)^2\right)\\
				\label{EqPfgder2f3}
				& = &  n \left(\mathbb{E}_{P_Y}\left[\frac{Y^2 \exp(\theta Y)}{\mathbb{E}_{P_Y}[\exp(\theta Y)]}\right] -  \left(\mathbb{E}_{P_Y}\left[\frac{Y \exp(\theta Y)}{\mathbb{E}_{P_Y}[ \exp(\theta Y)]}\right]\right)^2\right)\\
				\label{EqPfgder2f4}
				& = &  n \Bigg(\mathbb{E}_{P_Y}\left[\frac{Y^2 \exp(\theta Y)}{\mathbb{E}_{P_Y}[\exp(\theta Y)]}\right] -  2\mathbb{E}_{P_Y}\left[\frac{Y \exp(\theta Y)}{\mathbb{E}_{P_Y}[ \exp(\theta Y)]}\right]K^{(1)}_Y(\theta)\nonumber +  \left(K^{(1)}_Y(\theta)\right)^2\Bigg) \\
				\label{EqPfgder2f5}
				& = & n \mathbb{E}_{P_Y}\left[\frac{\left(Y-K^{(1)}_Y(\theta)\right)^2 \exp(\theta Y)}{\mathbb{E}_{P_Y}[\exp(\theta Y)]}\right]
				\label{EqPfgder2f6}
				>  0.
			\end{IEEEeqnarray} 
		\end{subequations}
		
		Hence, if the first derivative of $g$ with respect to $\theta$ (see~\eqref{EqderG}) admits a zero in $\Theta_Y$, then $\theta^{\star}(a,n)$ is the unique solution in $\theta$ to the following equality:
		
		\begin{IEEEeqnarray}{lcl}
			\label{EqSolveTheta}
			\frac{\mathrm{d}}{\mathrm{d}\theta}g(\theta,a,n)& =&  \frac{n}{\varphi_Y(\theta)}\frac{\mathrm{d}}{\mathrm{d}\theta} \varphi_Y(\theta) - a = 0.
		\end{IEEEeqnarray}
		Equation~\eqref{EqSolveTheta} in $\theta$ can be rewritten as follows:
		\begin{subequations}
			\begin{IEEEeqnarray}{lcl}
				\frac{a}{n} &=& \frac{1}{\varphi_Y(\theta)}\frac{\mathrm{d}}{\mathrm{d}\theta} \varphi_Y(\theta) \\
				&=& \frac{1}{\mathbb{E}_{P_Y}[\exp(\theta Y)]}\frac{\mathrm{d}}{\mathrm{d}\theta} \mathbb{E}_{P_Y}[\exp(\theta Y)]\\
				&=& \frac{1}{\mathbb{E}_{P_Y}[\exp(\theta Y)]} \mathbb{E}_{P_Y}[Y\exp(\theta Y)]\\
				\label{EqSlvth1}
				&=& \mathbb{E}_{P_Y}\left[\frac{Y\exp(\theta Y)}{\mathbb{E}_{P_Y}[\exp(\theta Y)]}\right]\\
				\label{EqSlvth2}
				&=& K^{(1)}_Y(\theta).
			\end{IEEEeqnarray} 
		\end{subequations}
		From~\eqref{EqSlvth1}, it follows that $\frac{a}{n}$ is the mean of a random variable that follows an exponentially tilted distribution with respect to $P_Y$. Thus,  there exists a solution in $\theta$ for~\eqref{EqSlvth1} if and only if $\frac{a}{n}\in \mathrm{int}\set{C}_Y$---hence the equality in~\eqref{EqHRedef}.

		Finally, from~\eqref{EqSlvth1}, $a=n\mathbb{E}_{P_Y}[Y]$ implies that $\theta^{\star}(a,n)=0$. Hence, $h(n\mathbb{E}_{P_Y}[Y])=0$ from~\eqref{EqHRedef}. This completes the proof for $h(n\mathbb{E}_{P_Y}[Y])=0$.
		
		\section{Proof of Theorem~\ref{TheoSaddlePointBeryUni}}\label{pTheoSaddlePointBeryUni}
		From Lemma~\ref{LemmaH}, it holds that given $(a,n)\in\mathbb{R}\times\mathbb{N}$ such that $\frac{a}{n}\in\mathrm{int}\set{C}_Y$,
		\begin{eqnarray}\label{EqThePf}
		n K^{(1)}_Y(\theta^{\star}) = a.
		\end{eqnarray}

		Then, plugging~\eqref{EqThePf} in the expression of $\eta_Y(\theta^{\star},a,n)$, with function $\eta_Y$  defined in~\eqref{EqEtaApprox}, the~following holds:
		\begin{subequations}
			\begin{IEEEeqnarray}{l}
				\eta_Y(\theta^{\star},a,n) \nonumber\\
				= \hspace*{-0.7ex}\mathds{1}_{\{\theta^{\star} > 0 \}} \hspace*{-0.55ex}+\hspace*{-0.55ex} (\hspace*{-0.2ex}-1\hspace*{-0.2ex})^{\mathds{1}_{\{\theta^{\star} > 0 \}}}\exp\left( \frac{1}{2} n (\theta^{\star})^2  K^{(2)}_{Y}(\theta) \hspace*{-0.3ex}+\hspace*{-0.3ex} n K^{}_Y(\theta^{\star}) \hspace*{-0.4ex}-\hspace*{-0.4ex} \theta^{\star} a  \hspace*{-0.7ex}\right) Q\hspace*{-0.7ex}\left(\hspace*{-0.9ex}(\hspace*{-0.2ex}-1\hspace*{-0.2ex})^{\mathds{1}_{\{\hspace*{-0.2ex}\theta^{\star} \leqslant 0\hspace*{-0.2ex}\}}}\frac{a \hspace*{-0.3ex}+\hspace*{-0.3ex} n \theta^{\star} K^{(2)}_Y(\theta^{\star})\hspace*{-0.3ex}-\hspace*{-0.3ex} a }{\sqrt{n K^{(2)}_Y(\theta^{\star})}}\hspace*{-0.7ex} \right)\qquad\\
				= \hspace*{-0.7ex}\mathds{1}_{\{\hspace*{-0.3ex}\theta^{\star} > 0\hspace*{-0.3ex} \}} \hspace*{-0.55ex}+\hspace*{-0.55ex} (-1)^{\mathds{1}_{\{\theta^{\star} > 0 \}}}\exp\left( \frac{1}{2} n (\theta^{\star})^2  K^{(2)}_{Y}(\theta) \hspace*{-0.3ex}+\hspace*{-0.3ex} n K^{}_Y(\theta^{\star}) \hspace*{-0.4ex}-\hspace*{-0.4ex} \theta^{\star} a  \hspace*{-0.7ex}\right) Q\hspace*{-0.5ex}\left(\hspace*{-0.7ex}(-1)^{\mathds{1}_{\{\theta^{\star} \leqslant 0\}}}\theta^{\star}\sqrt{n K^{(2)}_Y(\theta^{\star})} \right)\\ 
				= \hspace*{-0.7ex}\mathds{1}_{\{\hspace*{-0.3ex}\theta^{\star} > 0\hspace*{-0.3ex} \}} \hspace*{-0.55ex}+\hspace*{-0.55ex} (-1)^{\mathds{1}_{\{\theta^{\star} > 0 \}}}\exp\left( \frac{1}{2} n (\theta^{\star})^2  K^{(2)}_{Y}(\theta) \hspace*{-0.3ex}+\hspace*{-0.3ex} n K^{}_Y(\theta^{\star}) \hspace*{-0.4ex}-\hspace*{-0.4ex} \theta^{\star} a  \hspace*{-0.7ex}\right) Q\hspace*{-0.5ex}\left(\hspace*{-0.7ex}|\theta^{\star}|\sqrt{n K^{(2)}_Y(\theta^{\star})} \right)\\ 
				\label{EqEtZeta}
				= \hat{F}_{X_n}(a),
			\end{IEEEeqnarray}
		\end{subequations}
		where equality in~\eqref{EqEtZeta} follows~\eqref{EqHatPDF}. Finally, plugging~\eqref{EqEtZeta} in~\eqref{EqTilteapprox} yields
		\begin{IEEEeqnarray}{lcl}
			\left|F_{X_n}(a)  -\hat{F}_{X_n}(a) \right|
			\leqslant \exp\left(n K^{}_Y(\theta^{\star})-\theta^{\star} a\right) {\min\left\{1, \frac{2\; \xi_Y(\theta^{\star}) }{\sqrt{n}}\right\}}. \label{EqPfCompApp1}
		\end{IEEEeqnarray}
		This completes the proof by observing that $\frac{a}{n}\in\mathrm{int}\set{C}_Y$ is equivalent to $a\in\mathrm{int}\set{C}_{X_n}$.
		
		\section{Proof of Lemma~\ref{lem:1ForLem2}}\label{plem:1ForLem2}
		The left-hand side of~\eqref{EqOhYeah} satisfies
		\begin{IEEEeqnarray}{lcl}
			& &\left|P_{X_n}(\set{A}) - \left(\varphi_{Y}(\theta)\right)^n \mathbb{E}_{P_{Z_{n,\theta}}}\left[\exp{\left(-\theta Z_{n,\theta}\right)}\mathds{1}_{\left\{Z_{n,\theta}\in\set{A}\right\}}\right]\right| \nonumber\\
			&=& \left(\varphi_{Y}(\theta)\right)^n \left| \mathbb{E}_{P_{S_{n,\theta}}}
			\left[\exp{\left(-\theta S_{n,\theta}\right)}\mathds{1}_{\left\{S_{n,\theta}\in\set{A}\right\}}\right] -  \mathbb{E}_{P_{Z_{n,\theta}}}\left[\exp{\left(-\theta Z_{n,\theta}\right)}\mathds{1}_{\left\{Z_{n,\theta}\in\set{A}\right\}}\right]\right|.
			\label{EqOhMyGod}
		\end{IEEEeqnarray}
		The focus is on obtaining explicit expressions for the terms $\mathbb{E}_{P_{S_{n,\theta}}}
		\left[\exp{\left(-\theta S_{n,\theta}\right)}\mathds{1}_{\left\{S_{n,\theta}\in\set{A}\right\}}\right]$ and $\mathbb{E}_{P_{Z_{n,\theta}}}\left[\exp{\left(-\theta Z_{n,\theta}\right)}\mathds{1}_{\left\{Z_{n,\theta}\in\set{A}\right\}}\right]$ in~\eqref{EqOhMyGod}.
		First, consider the case in which the random variable $S_{n,\theta}$ is absolutely continuous and denote its  probability density function by $f_{S_{n,\theta}}$ and its CDF by $F_{S_{n,\theta}}$. Then,
		\begin{IEEEeqnarray}{lcl}
			\label{EqPfAbRvE}
			\mathbb{E}_{P_{S_{n,\theta}}}\left[\exp\left(-\theta S_{n,\theta}\right)\mathds{1}_{\{S_{n,\theta}  \in \set{A}\}}\right] &=&  \int_{\set{A}} \exp\left(-\theta x\right) f_{S_{n,\theta}}(x) \mathrm{d}x.
		\end{IEEEeqnarray}
		Using integration by parts in~\eqref{EqPfAbRvE}, under the assumption $(i)$ or $(ii)$ in Lemma~\ref{lem:1ForLem2}, the following~holds:
		\begin{IEEEeqnarray}{lcl}
			&&\mathbb{E}_{P_{S_{n,\theta}}}\hspace*{-0.8ex}\left[\exp\left(-\theta S_{n,\theta}\right)\mathds{1}_{\{S_{n,\theta}  \in \set{A}\}}\right]
			\label{EqPfAbRvE2}
			\hspace*{-0.4ex}=\hspace*{-0.4ex} (-1)^{\mathds{1}_{\{\theta > 0\}}} \exp\left(-\theta a\right) F_{S_{n,\theta}}(a) - \hspace*{-0.8ex}\int_{\set{A}}\hspace*{-0.8ex} \theta \exp\left(-\theta x\right) F_{S_{n,\theta}}(x) \mathrm{d}x.\qquad
		\end{IEEEeqnarray}

		Second, consider the case in which the random variable $S_{n,\theta}$ is discrete and denote its  probability mass  function by $p_{S_{n,\theta}}$ and its CDF by $F_{S_{n,\theta}}$.  Let the support of $S_{n,\theta}$ be $\{s_0$, $s_1$, $\ldots$, $s_{\ell}\} \subset \mathbb{R}$, with $\ell  \in \mathbb{N}$.
		Assume that condition $(i)$ in Lemma~\ref{lem:1ForLem2} is satisfied. Then, 
		\begin{equation}
		\set{A}\cap \{s_0,s_1,\ldots,s_l\} = \{s_0,s_{1},\ldots,s_{u}\},
		\end{equation}
		with $u \leqslant \ell$, and
		\begin{subequations}
			\begin{IEEEeqnarray}{l}
				\mathbb{E}_{P_{S_{n,\theta}}}\left[\exp\left(-\theta S_{n,\theta}\right)\mathds{1}_{\{S_{n,\theta}  \in \set{A}\}}\right] \nonumber\\
				= \sum_{k=0}^{u}\exp\left(-\theta s_k\right) p_{S_{n,\theta}}(s_k)\\
				= F_{S_{n,\theta}}(s_{0})\exp\left(-\theta s_0\right) +  \sum_{k=1}^{u}\left(F_{S_{n,\theta}}(s_{k})-F_{S_{n,\theta}}(s_{k-1})\right)\exp\left(-\theta s_k\right)\\
				= \sum_{k=0}^{u}F_{S_{n,\theta}}(s_{k})\exp\left(-\theta s_k\right) -  \sum_{k=1}^{u} F_{S_{n,\theta}}(s_{k-1})\exp\left(-\theta s_{k}\right)\\
				= \sum_{k=0}^{u}F_{S_{n,\theta}}(s_{k})\exp\left(-\theta s_k\right) -  \sum_{k=0}^{u-1} F_{S_{n,\theta}}(s_{k})\exp\left(-\theta s_{k+1}\right)\\
				= F_{S_{n,\theta}}(s_{u})\exp\left(-\theta s_u\right) - \sum_{k=0}^{u-1}F_{S_{n,\theta}}(s_{k})\left(\exp\left(-\theta s_{k+1}\right) -  \exp\left(-\theta s_{k}\right)\right)\\
				= F_{S_{n,\theta}}(s_{u})\exp\left(-\theta s_u\right)- \sum_{k=0}^{u-1}\int_{s_{k}}^{s_{k+1}} \theta \exp\left(-\theta t\right) F_{S_{n,\theta}}(s_{k}) \mathrm{d}t\\
				= F_{S_{n,\theta}}(s_{u})\exp\left(-\theta s_u\right)- \int_{s_{0}}^{s_{u}}\theta \exp\left(-\theta t\right)F_{S_{n,\theta}}(t) \mathrm{d}t \\
				= F_{S_{n,\theta}}(a)\exp\left(\hspace*{-0.3ex} -\theta a\right)\hspace*{-0.5ex}-\hspace*{-0.5ex}  F_{S_{n,\theta}}(a)\exp\left(\hspace*{-0.3ex} -\theta a\right) \hspace*{-0.5ex} 
				+ \hspace*{-0.5ex} F_{S_{n,\theta}}(s_{u})\exp\left(\hspace*{-0.3ex} -\theta s_u\right)\hspace*{-0.5ex} - \hspace*{-0.9ex} \int_{s_{0}}^{s_{u}}\hspace*{-1.6ex} F_{S_{n,\theta}}(t)\theta \exp\left(\hspace*{-0.3ex} -\theta t\right) \mathrm{d}t\qquad\\
				=\hspace*{-0.5ex}  F_{S_{n,\theta}}(a)\exp\left(\hspace*{-0.3ex} -\theta a\right) \hspace*{-0.5ex}-\hspace*{-0.5ex}  F_{S_{n,\theta}}(s_u)\exp\left(\hspace*{-0.3ex} -\theta a\right)\hspace*{-0.5ex} 
				+ \hspace*{-0.5ex} F_{S_{n,\theta}}(s_{u})\exp\left(\hspace*{-0.3ex} -\theta s_u\right)\hspace*{-0.5ex} - \hspace*{-0.9ex} \int_{s_{0}}^{s_{u}}\hspace*{-1.6ex} \theta \exp\left(\hspace*{-0.3ex} -\theta t\right)F_{S_{n,\theta}}(t) \mathrm{d}t\\
				= F_{S_{n,\theta}}(a)\exp\left(-\theta a\right) - F_{S_{n,\theta}}(s_u)\left(\exp\left(-\theta a\right)-\exp\left(-\theta s_u\right)\right) 
				-\int_{s_{0}}^{s_{u}} \theta \exp\left(-\theta t\right) F_{S_{n,\theta}}(t) \mathrm{d}t\qquad\\
				= F_{S_{n,\theta}}(a)\exp\left(-\theta a\right) - \int_{s_u}^{a} \theta \exp\left(-\theta t\right) F_{S_{n,\theta}}(s_u) \mathrm{d}t 
				-\int_{s_{0}}^{s_{u}} \theta \exp\left(-\theta t\right) F_{S_{n,\theta}}(t) \mathrm{d}t\\
				= \exp\left(-\theta a\right) F_{S_{n,\theta}}(a) -\int_{s_{0}}^{a} \theta \exp\left(-\theta t\right) F_{S_{n,\theta}}(t) \mathrm{d}t \\
				\label{EqExD1}
				= \exp\left(-\theta a\right) F_{S_{n,\theta}}(a) -\int_{-\infty}^{a} \theta \exp\left(-\theta t\right) F_{S_{n,\theta}}(t) \mathrm{d}t,
			\end{IEEEeqnarray}
		\end{subequations}
		which is an expression of the same form as the one in~\eqref{EqPfAbRvE2}.
		Alternatively, assume that condition~$(ii)$ in Lemma~\ref{lem:1ForLem2} holds. Then, 
		\begin{equation}
		\set{A}\cap \{s_0,s_1,\ldots,s_l\} = \{s_u,s_{u+1},\ldots,s_{l}\},
		\end{equation}
		with $u \leqslant \ell$, and
		\begin{subequations}
			\begin{IEEEeqnarray}{l}
				\mathbb{E}_{P_{S_{n,\theta}}}\left[\exp\left(-\theta S_{n,\theta}\right)\mathds{1}_{\{S_{n,\theta}  \in \set{A}\}}\right] \nonumber\\
				= \sum_{k=u}^{l}\exp\left(-\theta s_k\right) p_{S_{n,\theta}}(s_k)\\
				= \left(F_{S_{n,\theta}}(s_{u})-F_{S_{n,\theta}}(a)\right)\exp\left(-\theta s_u\right) 
				+  \sum_{k=u+1}^{l}\left(F_{S_{n,\theta}}(s_{k})-F_{S_{n,\theta}}(s_{k-1})\right)\exp\left(-\theta s_k\right) \qquad\\
				= -F_{S_{n,\theta}}(a)\exp\left(-\theta s_u\right) +\sum_{k=u}^{l}F_{S_{n,\theta}}(s_{k})\exp\left(-\theta s_k\right) 
				-  \sum_{k=u+1}^{l} F_{S_{n,\theta}}(s_{k-1})\exp\left(-\theta s_{k}\right)\\
				= -F_{S_{n,\theta}}(a)\exp\left(-\theta s_u\right) + \sum_{k=u}^{l}F_{S_{n,\theta}}(s_{k})\exp\left(-\theta s_k\right) 
				- \sum_{k=u}^{l-1} F_{S_{n,\theta}}(s_{k})\exp\left(-\theta s_{k+1}\right)\\
				=F_{S_{n,\theta}}(s_{l})\exp\left(-\theta s_l\right)\hspace*{-0.5ex}  -\hspace*{-0.5ex} F_{S_{n,\theta}}(a)\exp\left(-\theta s_u\right)  
				\hspace*{-0.5ex} -\hspace*{-0.5ex}  \sum_{k=u}^{l-1}\hspace*{-0.5ex} F_{S_{n,\theta}}(s_{k})\left(\exp\left(-\theta s_{k+1}\right) -  \exp\left(-\theta s_{k}\right)\right)\qquad\\
				= -F_{S_{n,\theta}}(a)\exp\left(-\theta s_u\right)\hspace*{-0.5ex} -\hspace*{-0.8ex} \int_{s_l}^{\infty}\hspace*{-1.2ex} \theta \exp\left(-\theta s_t\right) F_{S_{n,\theta}}(s_{l})\mathrm{d}t
				- \sum_{k=u}^{l-1}\int_{s_{k}}^{s_{k+1}} \hspace*{-1.1ex} \theta \exp\left(-\theta t\right) F_{S_{n,\theta}}(s_{k}) \mathrm{d}t\\
				= -F_{S_{n,\theta}}(a)\exp\left(-\theta s_u\right) - \int_{s_{u}}^{\infty}\theta \exp\left(-\theta t\right)F_{S_{n,\theta}}(t) \mathrm{d}t \\
				= F_{S_{n,\theta}}(a)\exp\left(\hspace*{-0.3ex} -\theta a\right) \hspace*{-0.6ex} - \hspace*{-0.6ex} F_{S_{n,\theta}}(a)\exp\left(\hspace*{-0.3ex} -\theta a\right)
				\hspace*{-0.5ex} - \hspace*{-0.5ex} F_{S_{n,\theta}}(a)\exp\left(\hspace*{-0.3ex} -\theta s_u\right) \hspace*{-0.5ex} - \hspace*{-0.7ex} \int_{s_{u}}^{\infty}\hspace*{-1.2ex}  \theta \exp\left(\hspace*{-0.3ex} -\theta t\right)F_{S_{n,\theta}}(t) \mathrm{d}t\\
				= - F_{S_{n,\theta}}(a)\exp\left(-\theta a\right) \hspace*{-0.5ex}  - \hspace*{-0.5ex}  F_{S_{n,\theta}}(a)\, \left(\exp\left(-\theta s_u\right) -\exp\left(-\theta a\right)\right)
				-\hspace*{-0.5ex} \int_{s_{u}}^{\infty} \theta \exp\left(-\theta t\right)F_{S_{n,\theta}}(t) \mathrm{d}t\\
				= - F_{S_{n,\theta}}(a)\exp\left(-\theta a\right) - \int^{s_{u}}_{a} \theta \exp\left(-\theta t\right)F_{S_{n,\theta}}(a) \mathrm{d}t
				- \int_{s_{u}}^{\infty} \theta \exp\left(-\theta t\right)F_{S_{n,\theta}}(t) \mathrm{d}t\\
				\label{EqExD2}
				= - F_{S_{n,\theta}}(a)\exp\left(-\theta a\right) - \int_{a}^{\infty} \theta \exp\left(-\theta t\right)F_{S_{n,\theta}}(t) \mathrm{d}t,
			\end{IEEEeqnarray}
		\end{subequations}
		which is an expression of the same form as those in~\eqref{EqPfAbRvE2} and~\eqref{EqExD1}.

		Note that, under the assumption that at least one of the conditions in Lemma~\ref{lem:1ForLem2} holds, the~expressions in~\eqref{EqPfAbRvE2},~\eqref{EqExD1}, and~\eqref{EqExD2} can be jointly  written as follows:
		\begin{IEEEeqnarray}{l}
			\label{EqExpF}
			\mathbb{E}_{P_{S_{n,\theta}}}\hspace*{-0.5ex} \left[\exp\left(-\theta S_{n,\theta}\right)\mathds{1}_{\{S_{n,\theta}  \in \set{A}\}}\right]
			= (-1)^{\mathds{1}_{\{\theta > 0\}}} \exp\left(-\theta a\right) F_{S_{n,\theta}}(a) \hspace*{-0.5ex} - \hspace*{-0.8ex} \int_{\set{A}} \hspace*{-0.8ex} \theta \exp\left(-\theta x\right) F_{S_{n,\theta}}(x) \mathrm{d}x.\qquad
		\end{IEEEeqnarray}

		The expression in~\eqref{EqExpF} does not involve particular assumptions on the random variable $S_{n,\theta}$ other than being discrete or absolutely continuous. Hence, the same expression holds with respect to the random variable $Z_{n,\theta}$ in~\eqref{EqOhMyGod}. More specifically, 
		\begin{IEEEeqnarray}{l}
			\label{EqExpF2}
			\mathbb{E}_{P_{Z_{n,\theta}}}\hspace*{-0.5ex} \left[\exp\left(-\theta Z_{n,\theta}\right)\mathds{1}_{\{Z_{n,\theta}  \in \set{A}\}}\right]
			= (-1)^{\mathds{1}_{\{\theta > 0\}}} \exp\left(-\theta a\right) F_{Z_{n,\theta}}(a)\hspace*{-0.5ex}  - \hspace*{-0.8ex} \int_{\set{A}}\hspace*{-0.8ex}  \theta \exp\left(-\theta x\right) F_{Z_{n,\theta}}(x) \mathrm{d}x, \qquad 
		\end{IEEEeqnarray}
		where $F_{Z_{n,\theta}}$ is the CDF of the random variable $Z_{n,\theta}$.

		The proof ends by plugging~\eqref{EqExpF} and~\eqref{EqExpF2} into the right-hand side of~\eqref{EqOhMyGod}. This yields
		\begin{subequations}
			\begin{IEEEeqnarray}{l}
				\left|P_{X_n}(\set{A}) - \left(\varphi_{Y}(\theta)\right)^n \mathbb{E}_{P_{Z_{n,\theta}}}\left[\exp{\left(-\theta Z_{n,\theta}\right)}\mathds{1}_{\left\{Z_{n,\theta}\in\set{A}\right\}}\right]\right| \nonumber\\
				= \left(\varphi_{Y}(\theta)\right)^n \Big| (-1)^{\mathds{1}_{\{\theta > 0\}}}\exp\left(-\theta a\right)F_{S_{n,\theta}}(a) -
				\int_{\set{A}}\theta \exp\left(-\theta x\right) F_{S_{n,\theta}}(x) \me{d}x \nonumber \\
				- (-1)^{\mathds{1}_{\{\theta > 0\}}}\exp\left(-\theta a\right)F_{Z_{n,\theta}}(a) +
				\int_{\set{A}}\theta \exp\left(-\theta x\right) F_{Z_{n,\theta}}(x) \me{d}x \Big|\\
				= \hspace*{-0.5ex}  \left(\hspace*{-0.2ex} \varphi_{Y}(\theta)\hspace*{-0.2ex} \right)^n\left|\hspace*{-0.3ex} (\hspace*{-0.2ex} -1)^{\mathds{1}_{\{\theta > 0\}}} \exp\left(\hspace*{-0.3ex} -a\right)\left(\hspace*{-0.5ex} F_{S_{n,\theta}}(a)\hspace*{-0.6ex}-\hspace*{-0.6ex}  F_{Z_{n,\theta}}(a)\hspace*{-0.4ex} \right) 
				\hspace*{-0.9ex} -\hspace*{-0.5ex} \int_{\set{A}}\hspace*{-0.9ex} \theta \exp\left(\hspace*{-0.3ex} -\theta x\right) \left(\hspace*{-0.5ex} F_{S_{n,\theta}}(x) \hspace*{-0.7ex} -\hspace*{-0.7ex}  F_{Z_{n,\theta}}(x)\hspace*{-0.4ex} \right) \me{d}x \hspace*{-0.2ex}  \right| \qquad\\
				\label{EqtriangIneImp}
				\le \left(\varphi_{Y}(\theta)\right)^n\left( \left| \exp\left(-\theta a\right)\left(F_{S_{n,\theta}}(a) \hspace*{-0.5ex}-\hspace*{-0.5ex}  F_{Z_{n,\theta}}(a)\right)\right|
				\hspace*{-0.5ex} +\hspace*{-0.6ex} 
				\left|\int_{\set{A}}\theta \exp\left(-\theta x\right) \left(F_{S_{n,\theta}}(x) - F_{Z_{n,\theta}}(x)\right) \me{d}x  \right|\hspace*{-0.4ex} \right)\,\qquad \\
				\label{EqIntMon}
				\le \left(\varphi_{Y}(\theta)\right)^n\left( \exp\left(-\theta a\right)\Delta\left(P_{S_{n,\theta}}, P_{Z_{n,\theta}}\right)+
				\int_{\set{A}} \left|\theta \exp\left(-\theta x\right) \right| \Delta\left(P_{S_{n,\theta}}, P_{Z_{n,\theta}}\right) \me{d}x\right) \,\qquad \\
				\label{EqIntBackward}
				= \left(\varphi_{Y}(\theta)\right)^n\left( \exp\left(-\theta a\right)\Delta\left(P_{S_{n,\theta}}, P_{Z_{n,\theta}}\right)+
				\Delta\left(P_{S_{n,\theta}}, P_{Z_{n,\theta}}\right) \left|\int_{\set{A}} \theta \exp\left(-\theta x\right)   \me{d}x \right|\right) \,\qquad \\
				=  \left(\varphi_{Y}(\theta)\right)^n\left(\exp\left(-\theta a\right)\Delta\left(P_{S_{n,\theta}}, P_{Z_{n,\theta}}\right)+
				\Delta\left(P_{S_{n,\theta}}, P_{Z_{n,\theta}}\right) \exp\left(-\theta a\right)\right) \,\qquad \\
				\label{EqlemFuncExImp}
				= 2 \frac{\left(\varphi_{Y}(\theta)\right)^n}{\exp(\theta a)}\Delta\left(P_{S_{n,\theta}}, P_{Z_{n,\theta}}\right).
			\end{IEEEeqnarray} 
		\end{subequations}
		
		Finally,  under the assumption that at least one of the conditions in Lemma~\ref{lem:1ForLem2} holds, then
		\begin{subequations}
			\begin{IEEEeqnarray}{lcl}
				& &\left|P_{X_n}(\set{A}) - \left(\varphi_{Y}(\theta)\right)^n \mathbb{E}_{P_{Z_{n,\theta}}}\left[\exp{\left(-\theta Z_{n,\theta}\right)}\mathds{1}_{\left\{Z_{n,\theta}\in\set{A}\right\}}\right]\right| \nonumber\\
				&\le& \left(\varphi_{Y}(\theta)\right)^n \max\left( \mathbb{E}_{P_{S_{n,\theta}}}
				\left[\exp{\left(-\theta S_{n,\theta}\right)}\mathds{1}\left\{S_{n,\theta}\in\set{A}\right\}\right],  \mathbb{E}_{P_{Z_{n,\theta}}}\left[\exp{\left(-\theta Z_{n,\theta}\right)}\mathds{1}\left\{Z_{n,\theta}\in\set{A}\right\}\right]\right)\,\qquad \\
				&\le& \left(\varphi_{Y}(\theta)\right)^n \exp\left(-\theta a\right)
				\label{EqlemFuncExImp04}
				= \frac{\left(\varphi_{Y}(\theta)\right)^n}{\exp(\theta a)}.
			\end{IEEEeqnarray}
		\end{subequations}
		Under the same assumption, the expressions in~\eqref{EqlemFuncExImp} and~\eqref{EqlemFuncExImp04} can be jointly  written as follows:
		\begin{IEEEeqnarray}{l}
			\left|P_{X_n}(\set{A})\hspace*{-0.5ex}-\hspace*{-0.5ex} \left(\varphi_{Y}(\theta)\right)^n \mathbb{E}_{P_{Z_{n,\theta}}}\left[\exp{\left(-\theta Z_{n,\theta}\right)}\mathds{1}_{\left\{Z_{n,\theta}\in\set{A}\right\}}\right]\right| 
			\label{EqlemFuncExImp05}
			\leqslant \frac{\left(\varphi_{Y}(\theta)\right)^n}{\exp(\theta a)}{\min\left\{2\Delta\left(P_{S_{n,\theta}}, P_{Z_{n,\theta}}\right), 1\right\}}.\qquad
		\end{IEEEeqnarray}
		This concludes the proof of Lemma~\ref{lem:1ForLem2}.
		
		\section{Proof of Lemma~\ref{LemmaBeautiful}}\label{ProofLemmaBeautiful}	
		In the case in which $Y$  is discrete ($p_Y$, $p_{Y^{(\theta)}}$,   $p_{Y^{(\theta)}_1Y^{(\theta)}_2\ldots Y^{(\theta)}_n}$ denote probability mass functions) or absolutely continuous random variables ($p_Y$, $p_{Y^{(\theta)}}$,   $p_{Y^{(\theta)}_1Y^{(\theta)}_2\ldots Y^{(\theta)}_n}$ denote probability density functions), the following holds for all $(y_1, y_2, \ldots, y_n)\in\mathbb{R}^n$,
		\begin{IEEEeqnarray}{lcl}\label{EqpjointDensityCoM}
			\rndder{P_{Y^{(\theta)}_1 Y^{(\theta)}_2\ldots Y^{(\theta)}_n}}{P_{Y_1Y_2\ldots Y_n}}(y_1,y_2,\ldots,y_n) &=&\frac{p_{Y^{(\theta)}_1Y^{(\theta)}_2\ldots Y^{(\theta)}_n}(y_1,y_2,\ldots,y_n)}{\prod_{j=1}^n p_{Y}(y_j)}, 	
		\end{IEEEeqnarray}
		and for all $y \in\mathbb{R}$,
		\begin{IEEEeqnarray}{lcl}\label{EqpMargDensityCoM}
			\rndder{P_{Y^{(\theta)}}}{P_{Y}}(y) &=&\frac{p_{Y^{(\theta)}}(y)}{ p_{Y}(y)}.
		\end{IEEEeqnarray}
		Equating the right-hand side of both~\eqref{EqRND1} and~\eqref{EqpjointDensityCoM}, it yields for all $(y_1,y_2,\ldots,y_n)\in\mathbb{R}^n$
		\begin{IEEEeqnarray}{lcl}\label{EqEq1}
			p_{Y^{(\theta)}_1Y^{(\theta)}_2\ldots Y^{(\theta)}_n}(y_1,y_2,\ldots,y_n) &= \prod_{j=1}^n \frac{\exp{\left(\theta y_j\right)}}{\varphi_{ Y}(\theta)} p_{Y}(y_j).
		\end{IEEEeqnarray}
		Hence,  $Y^{(\theta)}_1$, $Y^{(\theta)}_2$, $\ldots$, $Y^{(\theta)}_n$ are mutually independent and identically distributed. Moreover, for all $y \in\mathbb{R}$,	
		\begin{IEEEeqnarray}{lcl}\label{EqEq2}
			p_{Y^{(\theta)}}(y) &=&  \frac{\exp{\left(\theta y\right)}}{\varphi_{ Y}(\theta)} p_{Y}(y). 
		\end{IEEEeqnarray}
		Finally, plugging~\eqref{EqEq2} in~\eqref{EqpMargDensityCoM} yields, for all $y \in\mathbb{R}$,
		\begin{IEEEeqnarray}{lcl}\label{EqptilDens}
			\rndder{P_{Y^{(\theta)}}}{P_{Y}}(y)  &=& \frac{\exp{\left(\theta y\right)}}{\varphi_{ Y}(\theta)},
		\end{IEEEeqnarray}
		which completes the proof.
		
		\section{Proof of Theorem \ref{TheoNew}}\label{pTheoNew}
		
		For a fixed product probability input distribution $P_X$ in \eqref{EqiidRandCod} and for the random transformation in~\eqref{EqMemChan}, the upper bound
		$T(n,M,P_{\vec{X}})$ in \eqref{EqT} can be written in the form of  a weighted sum of the CDF and the complementary CDF of the random variables variables $W_n$ and $V_n$ that are sums of i.i.d random variables, respectively. That is,
		\begin{IEEEeqnarray}{lcl}
			W_n &=& \sum_{t=1}^{n} \iota(X_t;Y_t), \mbox{ and }\\
			V_n &=& \sum_{t=1}^{n} \iota(\bar{X}_t;Y_t),
		\end{IEEEeqnarray}
		where $(X_t,Y_t)\sim P_X P_{Y|X}$ and $(\bar{X}_t,Y_t)\sim P_{\bar{X}} P_{Y}$ with $P_X=P_{\bar{X}}$. More specifically, the function $T$ in~\eqref{EqT} can be rewritten in the form
		\begin{IEEEeqnarray}{lcl}
			\label{EqLamUp}
			T(n,M,P_{\vec{X}}) &=& F_{W_n}\left(\ln{\frac{M-1}{2}}\right)  + \frac{M-1}{2} \left(1-F_{V_n}\left(\ln{\frac{M-1}{2}}\right) \right),
		\end{IEEEeqnarray}
		where $F_{W_n}$ and $F_{V_n}$ are the CDFs of $W_n$ and $V_n$, respectively.

		The next step derives the upper and lower bounds on $F_{W_n}\left(\ln{\frac{M-1}{2}}\right)$ and $1-F_{V_n}\left(\ln{\frac{M-1}{2}}\right)$ by using the result of Theorem \ref{TheoSaddlePointBeryUni}. That is,
		\begin{IEEEeqnarray}{l}
			F_{W_n}\left(\hspace*{-0.5ex}\ln{\hspace*{-0.5ex}\frac{M-1}{2}\hspace*{-0.4ex}}\hspace*{-0.5ex}\right) \nonumber\\
			\leqslant \hspace*{-0.5ex}\zeta_{\hspace*{-0.1ex}\iota(\hspace*{-0.1ex}X;\hspace*{-0.1ex}Y\hspace*{-0.1ex})\hspace*{-0.1ex}}\hspace*{-0.6ex}\left(\hspace*{-0.7ex}\theta,\hspace*{-0.2ex}\ln{\hspace*{-0.5ex}\hspace*{-0.5ex}\frac{M\hspace*{-0.2ex}-\hspace*{-0.2ex}1}{2}\hspace*{-0.4ex}\hspace*{-0.4ex}}\hspace*{-0.4ex},\hspace*{-0.3ex}n\hspace*{-0.5ex}\right) \hspace*{-0.8ex} + \hspace*{-0.5ex}\exp\hspace*{-0.5ex}\Bigg(\hspace*{-0.6ex}n \me{ln}\hspace*{-0.5ex}\left(\hspace*{-0.5ex}\varphi_{\iota(X;Y)}(\hspace*{-0.2ex}\theta\hspace*{-0.2ex})\hspace*{-0.5ex}\right)\hspace*{-0.5ex}-\hspace*{-0.5ex} \theta \hspace*{-0.1ex} \ln{\hspace*{-0.6ex}\frac{M-1}{2}\hspace*{-0.6ex}}\hspace*{-1.4ex}\Bigg)
			\label{EqFWUP}
			\min\left\{1,\frac{ 2\, \xi_{\iota(X;Y)}(\theta)}{\sqrt{n}}\hspace*{-0.7ex}\right\},\\
			F_{W_n}\left(\hspace*{-0.5ex}\ln{\hspace*{-0.5ex}\frac{M-1}{2}\hspace*{-0.4ex}}\hspace*{-0.5ex}\right) \nonumber\\
			\geqslant \hspace*{-0.5ex}\zeta_{\hspace*{-0.1ex}\iota(\hspace*{-0.1ex}X;\hspace*{-0.1ex}Y\hspace*{-0.1ex})\hspace*{-0.1ex}}\hspace*{-0.6ex}\left(\hspace*{-0.7ex}\theta,\hspace*{-0.2ex}\ln{\hspace*{-0.5ex}\hspace*{-0.5ex}\frac{M\hspace*{-0.2ex}-\hspace*{-0.2ex}1}{2}\hspace*{-0.4ex}\hspace*{-0.4ex}}\hspace*{-0.4ex},\hspace*{-0.3ex}n\hspace*{-0.5ex}\right) \hspace*{-0.8ex} - \hspace*{-0.5ex}\exp\hspace*{-0.5ex}\Bigg(\hspace*{-0.6ex}n \me{ln}\hspace*{-0.5ex}\left(\hspace*{-0.5ex}\varphi_{\iota(X;Y)}(\hspace*{-0.2ex}\theta\hspace*{-0.2ex})\hspace*{-0.5ex}\right)\hspace*{-0.5ex}-\hspace*{-0.5ex} \theta \hspace*{-0.1ex} \ln{\hspace*{-0.6ex}\frac{M-1}{2}\hspace*{-0.6ex}}\hspace*{-1.4ex}\Bigg)
			\label{EqFWLW}
			\hspace*{-0.5ex}\min\left\{1,\frac{ 2 \, \xi_{\iota(X;Y)}(\theta)}{\sqrt{n}}\hspace*{-0.7ex}\right\},\\
			1-F_{V_n}\left(\hspace*{-0.5ex}\ln{\hspace*{-0.5ex}\frac{M-1}{2}\hspace*{-0.4ex}}\hspace*{-0.5ex}\right) \nonumber\\
			\leqslant \hspace*{-0.5ex}1\hspace*{-0.5ex}-\hspace*{-0.5ex}\zeta_{\hspace*{-0.1ex}\iota(\hspace*{-0.1ex}\bar{X};\hspace*{-0.1ex}Y\hspace*{-0.1ex})\hspace*{-0.1ex}}\hspace*{-0.6ex}\left(\hspace*{-0.7ex}\theta,\hspace*{-0.2ex}\ln{\hspace*{-0.5ex}\hspace*{-0.5ex}\frac{M\hspace*{-0.2ex}-\hspace*{-0.2ex}1}{2}\hspace*{-0.4ex}\hspace*{-0.4ex}}\hspace*{-0.4ex},\hspace*{-0.3ex}n\hspace*{-0.5ex}\right) \hspace*{-0.8ex} + \hspace*{-0.5ex}\exp\hspace*{-0.5ex}\Bigg(\hspace*{-0.6ex}n \me{ln}\hspace*{-0.5ex}\left(\hspace*{-0.5ex}\varphi_{\iota(\bar{X};Y)}(\hspace*{-0.2ex}\theta\hspace*{-0.2ex})\hspace*{-0.5ex}\right)\hspace*{-0.5ex}-\hspace*{-0.5ex} \theta \hspace*{-0.1ex} \ln{\hspace*{-0.6ex}\frac{M-1}{2}\hspace*{-0.6ex}}\hspace*{-1.4ex}\Bigg)
			\label{EqFVUP}
			\hspace*{-0.5ex}\min\left\{1,\frac{ 2 \, \xi_{\iota(\bar{X};Y)}(\theta)}{\sqrt{n}}\hspace*{-0.7ex}\right\}\hspace*{-0.8ex},\mbox{ and }\\
			1-F_{V_n}\left(\hspace*{-0.5ex}\ln{\hspace*{-0.5ex}\frac{M-1}{2}\hspace*{-0.4ex}}\hspace*{-0.5ex}\right) \nonumber\\
			\geqslant \hspace*{-0.5ex}1\hspace*{-0.5ex}-\hspace*{-0.5ex}\zeta_{\hspace*{-0.1ex}\iota(\hspace*{-0.1ex}\bar{X};\hspace*{-0.1ex}Y\hspace*{-0.1ex})\hspace*{-0.1ex}}\hspace*{-0.6ex}\left(\hspace*{-0.7ex}\theta,\hspace*{-0.2ex}\ln{\hspace*{-0.5ex}\hspace*{-0.5ex}\frac{M\hspace*{-0.2ex}-\hspace*{-0.2ex}1}{2}\hspace*{-0.4ex}\hspace*{-0.4ex}}\hspace*{-0.4ex},\hspace*{-0.3ex}n\hspace*{-0.5ex}\right) \hspace*{-0.8ex} - \hspace*{-0.5ex}\exp\hspace*{-0.5ex}\Bigg(\hspace*{-0.6ex}n \me{ln}\hspace*{-0.5ex}\left(\hspace*{-0.5ex}\varphi_{\iota(\bar{X};Y)}(\hspace*{-0.2ex}\theta\hspace*{-0.2ex})\hspace*{-0.5ex}\right)\hspace*{-0.5ex}-\hspace*{-0.5ex} \theta \hspace*{-0.1ex} \ln{\hspace*{-0.6ex}\frac{M-1}{2}\hspace*{-0.6ex}}\hspace*{-1.4ex}\Bigg)
			\label{EqFVLW}
			\hspace*{-0.5ex}\min\left\{1,\frac{ 2 \, \xi_{\iota(\bar{X};Y)}(\theta)}{\sqrt{n}}\hspace*{-0.7ex}\right\}\hspace*{-0.8ex}, \qquad
		\end{IEEEeqnarray}
		where $\theta$ and $\tau$ satisfy
		\begin{IEEEeqnarray}{lcl}
			\label{EqChtt}
			n\mu_{\iota(X;Y)}(\theta) &=& \ln{\frac{M-1}{2}} = n\mu_{\iota(\bar{X};Y)}(\tau),
		\end{IEEEeqnarray}
		and for all $t\in \mathbb{R}$,
		\begin{IEEEeqnarray}{lcl}
			\label{EqMGFIotaD}
			\varphi_{\iota(X;Y)}(t) &=& \mathbb{E}_{P_{X}P_{Y|X}}\left[\exp(t\, \iota(X;Y))\right],\\
			\label{EqMGFIotaI}
			\varphi_{\iota(\bar{X};Y)}(t) &=& \mathbb{E}_{P_{\bar{X}}P_{Y}}\left[\exp\left(t\, \iota({\bar{X}};Y)\right)\right],\\
			\label{EqTIotaD}
			\xi_{\iota(X;Y)}(t) &=& c_1\left(\frac{\mathbb{E}_{P_XP_{Y|X}}\left[\left|\iota(X;Y) - \mu_{\iota(X;Y)}(t)\right|^3\frac{\exp(t\, \iota(X;Y))}{\varphi_{\iota(X;Y)}(t)}\right]}{\left(V_{\iota(X;Y)}(t)\right)^{3/2}}+c_2\right),\\
			\label{EqTIotaI}
			\xi_{\iota(\bar{X};Y)}(t) &=& c_1\left(\frac{\mathbb{E}_{P_{\bar{X}}P_{Y}}\left[\left|\iota({\bar{X}};Y) - \mu_{\iota(\bar{X};Y)}(t)\right|^3\frac{\exp\left(t\, \iota({\bar{X}};Y)\right)}{\varphi_{\iota(\bar{X};Y)}(t)}\right]}{\left(V_{\iota(\bar{X};Y)}(t)\right)^{3/2}}+c_2\right) ,\\
			\label{EqMuIotaD}
			\mu_{\iota(X;Y)}(t) &=& \mathbb{E}_{P_XP_{Y|X}}\left[\iota(X;Y)\frac{\exp(t\, \iota(X;Y))}{\varphi_{\iota(X;Y)}(t)}\right],\\
			\label{EqMuIotaI}
			\mu_{\iota(\bar{X};Y)}(t) &=&  \mathbb{E}_{P_{\bar{X}}P_{Y}}\left[\iota({\bar{X}};Y)\frac{\exp\left(t\, \iota({\bar{X}};Y)\right)}{\varphi_{\iota({\bar{X}};Y)}(t)} \right],\\
			\label{EqVIotaD}
			V_{\iota(X;Y)}(t) &=& \mathbb{E}_{P_XP_{Y|X}}\left[\left(\iota(X;Y) - \mu_{\iota(X;Y)}(t)\right)^2\frac{\exp(t\, \iota(X;Y))}{\varphi_{\iota(X;Y)}(t)}\right],\\
			\label{EqVIotaI}
			V_{\iota(\bar{X};Y)}(t) &=& \mathbb{E}_{P_{\bar{X}}P_{Y}}\left[\left(\iota({\bar{X}};Y) - \mu_{\iota(\bar{X};Y)}(t)\right)^2\frac{\exp\left(t\, \iota({\bar{X}};Y)\right)}{\varphi_{\iota(\bar{X};Y)}(t)} \right],
		\end{IEEEeqnarray}
		with $c_1$ and $c_2$ defined in~\eqref{EqConstants};
		and for all $(t,a,n)\in\mathbb{R}^2\times\mathbb{N}$
		\begin{IEEEeqnarray}{l}
			\nonumber
			\zeta_{\iota(X;Y)}(t, a, n)  \\
			\df\hspace*{-0.5ex} \mathds{1}_{\{t > 0 \}} 
			\hspace*{-0.5ex}+ \hspace*{-0.5ex} (-1)^{\mathds{1}_{\{t > 0 \}}} \exp\left( \frac{1}{2} n  t^2  V_{\iota(X;Y)}(t) + n \me{ln}\left(\varphi_{\iota(X;Y)}(t)\right) - t  a \right)
			\label{EqZetaApproxIotaD}
			Q\left( \abs{t} \,\sqrt{n \, V_{\iota(X;Y)}(t)}\right)\hspace*{-0.5ex},\qquad\\
			\nonumber
			\zeta_{\iota(\bar{X};Y)}(t, a, n) \\
			\df \hspace*{-0.5ex}  \mathds{1}_{\{t > 0 \}} \hspace*{-0.5ex}
			+ \hspace*{-0.5ex} (-1)^{\mathds{1}_{\{t > 0 \}}} \exp\left( \frac{1}{2} n  t^2  V_{\iota(\bar{X};Y)}(t) + n \me{ln}\left(\varphi_{\iota(\bar{X};Y)}(t)\right) - t  a \right)
			\label{EqZetaApproxIotaI}
			Q\left( \abs{t} \,\sqrt{n \, V_{\iota(\bar{X};Y)}(t)}\right).
		\end{IEEEeqnarray}

		The next step simplifies the expressions on the right hand-side of \eqref{EqFVUP} and \eqref{EqFVLW} by studying the relation between $\varphi_{\iota(X;Y)}$ and $\varphi_{\iota(\bar{X};Y)}$, $\theta$ and $\tau$, $V_{\iota(X;Y)}$ and $V_{\iota(\bar{X};Y)}$,
		$\xi_{\iota(X;Y)}$ and $\xi_{\iota(\bar{X};Y)}$.

		First, from \eqref{EqMGFIotaD}, using the change of measure from $P_XP_{Y|X}$ to $P_{\bar{X}}P_{Y}$ because $P_XP_{Y|X}$ is absolutely continuous with respect to $P_{\bar{X}}P_{Y}$, it holds that 
		\begin{IEEEeqnarray}{lcl}
			\label{EqMGFIotaDE}
			\varphi_{\iota(X;Y)}(t) &=& \mathbb{E}_{P_{\bar{X}}P_{Y}}\left[\rndder{P_{X}P_{Y|X}}{P_{\bar{X}}P_{Y}}\left({\bar{X}};Y\right)\exp(t\, \iota({\bar{X}};Y))\right]\\
			\label{EqMGFIotaDE1}
			&=& \mathbb{E}_{P_{\bar{X}}P_{Y}}\left[\exp\left((t+1)\, \iota({\bar{X}};Y)\right)\right].
		\end{IEEEeqnarray}
		Then, from~\eqref{EqMGFIotaD} and~\eqref{EqMGFIotaI}, it holds that
		\begin{IEEEeqnarray}{lcl}
			\label{EqMGFIota}
			\varphi_{\iota(X;Y)}(t)	&=& \varphi_{\iota(\bar{X};Y)}(t+1).
		\end{IEEEeqnarray}
		This concludes the relation between $\varphi_{\iota(X;Y)}$ and $\varphi_{\iota(\bar{X};Y)}$.

		Second, from \eqref{EqMuIotaD}, using the change of measure from $P_XP_{Y|X}$ to $P_{\bar{X}}P_{Y}$, it holds that 
		\begin{IEEEeqnarray}{lcl}
			\label{EqMuIotaDE}
			\mu_{\iota(X;Y)}(t) &=& \mathbb{E}_{P_{\bar{X}}P_{Y}}\left[\iota({\bar{X}};Y) \frac{\exp(t\, \iota({\bar{X}};Y))}{\varphi_{\iota(X;Y)}(t)} \rndder{P_XP_{Y|X}}{P_{\bar{X}}P_{Y}}\left({\bar{X}};Y\right)\right]\\
			\label{EqMuIotaDE1}
			&=& \mathbb{E}_{P_{\bar{X}}P_{Y}}\left[\iota({\bar{X}};Y)\frac{\exp\left((t+1)\, \iota({\bar{X}};Y)\right)}{\varphi_{\iota(X;Y)}(t)} \right].
		\end{IEEEeqnarray}
		Then, from \eqref{EqMGFIota} and \eqref{EqMuIotaDE1}, it holds that
		\begin{IEEEeqnarray}{lcl}
			\label{EqMuIotaDE2}
			\mu_{\iota(X;Y)}(t) &=& \mathbb{E}_{P_{\bar{X}}P_{Y}}\left[\iota({\bar{X}};Y)\frac{\exp\left((t+1)\, \iota({\bar{X}};Y)\right)}{\varphi_{\iota({\bar{X}};Y)}(t+1)} \right].
		\end{IEEEeqnarray}
		From \eqref{EqMuIotaI} and \eqref{EqMuIotaDE2}, it holds that
		\begin{IEEEeqnarray}{lcl}
			\label{EqMuIota}
			\mu_{\iota(X;Y)}(t)&=& \mu_{\iota(\bar{X};Y)}(t+1).
		\end{IEEEeqnarray}
		This concludes the relation between $\mu_{\iota(X;Y)}$ and $\mu_{\iota(\bar{X};Y)}$.

		Third, from \eqref{EqChtt} and \eqref{EqMuIota}, it holds that
		\begin{IEEEeqnarray}{lcl}
			\label{EqChttE}
			\tau = \theta + 1.
		\end{IEEEeqnarray}
		This concludes the relation between $\tau$ and $\theta$.

		Fourth, from \eqref{EqVIotaD}, using the change of measure from $P_XP_{Y|X}$ to $P_{\bar{X}}P_{Y}$, it holds that 
		\begin{IEEEeqnarray}{lcl}
			\label{EqVIotaDE}
			V_{\iota(X;Y)}(t) &=& \mathbb{E}_{P_{\bar{X}}P_{Y}}\left[\left(\iota({\bar{X}};Y) - \mu_{\iota(X;Y)}(t)\right)^2\frac{\exp(t\, \iota({\bar{X}};Y))}{\varphi_{\iota(X;Y)}(t)} \rndder{P_XP_{Y|X}}{P_{\bar{X}}P_{Y}}\left({\bar{X}};Y\right)\right]\\
			\label{EqVIotaDE1}
			&=& \mathbb{E}_{P_{\bar{X}}P_{Y}}\left[\left(\iota({\bar{X}};Y) - \mu_{\iota(X;Y)}(t)\right)^2\frac{\exp\left((t+1)\, \iota({\bar{X}};Y)\right)}{\varphi_{\iota(X;Y)}(t)} \right].
		\end{IEEEeqnarray}
		From \eqref{EqMGFIota}, \eqref{EqMuIota}, and \eqref{EqVIotaDE1}, it holds that
		\begin{IEEEeqnarray}{lcl}
			\label{EqVIotaDE2}
			V_{\iota(X;Y)}(t) &=& \mathbb{E}_{P_{\bar{X}}P_{Y}}\left[\left(\iota({\bar{X}};Y) - \mu_{\iota(\bar{X};Y)}(t+1)\right)^2\frac{\exp\left((t+1)\, \iota({\bar{X}};Y)\right)}{\varphi_{\iota(\bar{X};Y)}(t+1)} \right].
		\end{IEEEeqnarray}
		From \eqref{EqVIotaI} and \eqref{EqVIotaDE2}, it holds that
		\begin{IEEEeqnarray}{lcl}
			\label{EqVIota}
			V_{\iota(X;Y)}(t) &=& V_{\iota(\bar{X};Y)}(t+1).
		\end{IEEEeqnarray}
		This concludes the relation between $V_{\iota(X;Y)}$ and $V_{\iota(\bar{X};Y)}$.

		Fifth, from \eqref{EqTIotaD}, using the change of measure from $P_XP_{Y|X}$ to $P_{\bar{X}}P_{Y}$, it holds that 
		\begin{IEEEeqnarray}{lcl}
			\label{EqTIotaDE}
			\xi_{\iota(X;Y)}(t) &=& c_1\left(\frac{\mathbb{E}_{P_{\bar{X}}P_{Y}}\left[\left|\iota({\bar{X}};Y) - \mu_{\iota(X;Y)}(t)\right|^3\frac{\exp(t\, \iota({\bar{X}};Y))}{\varphi_{\iota(X;Y)}(t)} \rndder{P_XP_{Y|X}}{P_{\bar{X}}P_{Y}}\left({\bar{X}};Y\right)\right]}{\left(V_{\iota(X;Y)}(t)\right)^{3/2}}+c_2\right)\\
			\label{EqTIotaDE1}
			&=& c_1\left(\frac{\mathbb{E}_{P_{\bar{X}}P_{Y}}\left[\left|\iota({\bar{X}};Y) - \mu_{\iota(X;Y)}(t)\right|^3\frac{\exp\left((t+1)\, \iota({\bar{X}};Y)\right)}{\varphi_{\iota(X;Y)}(t)} \right]}{\left(V_{\iota(X;Y)}(t)\right)^{3/2}}+c_2\right).
		\end{IEEEeqnarray}
		From \eqref{EqMGFIota}, \eqref{EqMuIota}, \eqref{EqVIota}, and \eqref{EqTIotaDE1}, it holds that
		\begin{IEEEeqnarray}{lcl}
			\label{EqTIotaDE2}
			\xi_{\iota(X;Y)}(t) &=& c_1\left(\frac{\mathbb{E}_{P_{\bar{X}}P_{Y}}\left[\left|\iota({\bar{X}};Y) - \mu_{\iota(\bar{X};Y)}(t+1)\right|^3\frac{\exp\left((t+1)\, \iota({\bar{X}};Y)\right)}{\varphi_{\iota(\bar{X};Y)}(t+1)} \right]}{\left(V_{\iota(\bar{X};Y)}(t+1)\right)^{3/2}}+c_2\right).
		\end{IEEEeqnarray}
		From \eqref{EqTIotaI} and \eqref{EqTIotaDE2}, it holds that
		\begin{IEEEeqnarray}{lcl}
			\label{EqTIota}
			\xi_{\iota(X;Y)}(t) &=& \xi_{\iota(\bar{X};Y)}(t+1).
		\end{IEEEeqnarray}
		This concludes the relation between $\xi_{\iota(X;Y)}$ and $\xi_{\iota(\bar{X};Y)}$.

		Sixth, plugging \eqref{EqMGFIota}, \eqref{EqMuIota}, and \eqref{EqVIota} into \eqref{EqZetaApproxIotaD}, for all $t\in \mathbb{R}$, it holds that
		\begin{IEEEeqnarray}{l}
			\nonumber
			\zeta_{\iota(\bar{X};Y)}(t, a, n)  \\
			\df \hspace*{-0.5ex} \mathds{1}_{\{t > 0 \}} \hspace*{-0.5ex}
			+ \hspace*{-0.5ex} (-1)^{\mathds{1}_{\{t > 0 \}}} \hspace*{-0.5ex}\exp\hspace*{-0.5ex}\left(\hspace*{-0.5ex} \frac{1}{2} n  t^2  V_{\iota(X;Y)}(\hspace*{-0.2ex}t\hspace*{-0.2ex}-\hspace*{-0.2ex}1\hspace*{-0.2ex}) \hspace*{-0.5ex}+\hspace*{-0.5ex} n \me{ln}\hspace*{-0.5ex}\left(\hspace*{-0.5ex}\varphi_{\iota(X;Y)}(t\hspace*{-0.2ex}-\hspace*{-0.2ex}1)\hspace*{-0.4ex}\right)\hspace*{-0.5ex} -\hspace*{-0.5ex} t  a \hspace*{-0.7ex}\right)\hspace*{-0.5ex}
			\label{EqZetaApproxIotaIE}
			Q\hspace*{-0.5ex}\left(\hspace*{-0.5ex} \abs{t} \hspace*{-0.2ex}\sqrt{n \hspace*{-0.2ex} V_{\iota(X;Y)}(\hspace*{-0.2ex}t\hspace*{-0.2ex}-\hspace*{-0.2ex}1\hspace*{-0.2ex})}\hspace*{-0.5ex}\right)\hspace*{-0.5ex}.\qquad
		\end{IEEEeqnarray}
		Then, from \eqref{EqBeta2} and \eqref{EqZetaApproxIotaIE}, it holds that
		\begin{IEEEeqnarray}{lcl}
			\label{EqZetaApproxIotaIE2}
			\zeta_{\iota(\bar{X};Y)}\left(t, \ln{\frac{M-1}{2}}, n\right) = 1-\beta_2(n,M,t-1,P_X).
		\end{IEEEeqnarray}

		Then, plugging \eqref{EqMGFIota}, \eqref{EqMuIota}, \eqref{EqChttE}, \eqref{EqVIota}, \eqref{EqTIota}, and \eqref{EqZetaApproxIotaIE2} into the right hand-side of \eqref{EqFVUP}, it~holds that
		\begin{IEEEeqnarray}{l}
			1-F_{V_n}\left(\ln{\frac{M-1}{2}}\right) \nonumber\\
			\leqslant \hspace*{-0.5ex}\beta_2(n,M,\theta,P_X) \hspace*{-0.5ex}+\hspace*{-0.5ex} \exp\hspace*{-0.5ex}\Bigg(\hspace*{-0.8ex} n \me{ln}\left(\varphi_{\iota(X;Y)}(\theta)\right)\hspace*{-0.5ex} -\hspace*{-0.5ex} \left(\theta+1\right)  \ln{\hspace*{-0.5ex}\frac{M-1}{2}\hspace*{-0.5ex}}\hspace*{-1.3ex}\Bigg)\hspace*{-0.5ex} \min\left\{1,\frac{ 2 \, \xi_{\iota(X;Y)}(\theta)}{\sqrt{n}}\hspace*{-0.7ex}\right\}\qquad\\
			\leqslant \beta_2(n,M,\theta,P_X) + \exp\Bigg( n \me{ln}\left(\varphi_{\iota(X;Y)}(\theta)\right) - \left(\theta+1\right) \, \ln{\frac{M-1}{2}}\Bigg) 
			\label{EqFVUPE}
			\frac{ 2 \, \xi_{\iota(X;Y)}(\theta)}{\sqrt{n}}.
		\end{IEEEeqnarray}
		Alternatively, plugging \eqref{EqMGFIota}, \eqref{EqMuIota}, \eqref{EqChttE}, \eqref{EqVIota}, \eqref{EqTIota}, and \eqref{EqZetaApproxIotaIE2} into the right hand-side of \eqref{EqFVLW}, it holds that
		\begin{IEEEeqnarray}{l}
			1-F_{V_n}\left(\ln{\frac{M-1}{2}}\right) \nonumber\\
			\geqslant \hspace*{-0.5ex}\beta_2(n,M,\theta,P_X)\hspace*{-0.5ex} -\hspace*{-0.5ex} \exp\hspace*{-0.5ex}\Bigg(\hspace*{-0.5ex} n \me{ln}\left(\varphi_{\iota(X;Y)}(\theta)\right) \hspace*{-0.5ex}- \hspace*{-0.5ex}\left(\theta+1\right)  \ln{\hspace*{-0.5ex}\frac{M-1}{2}\hspace*{-0.5ex}}\hspace*{-1.2ex}\Bigg) \hspace*{-0.5ex}\min\left\{1,\frac{ 2 \, \xi_{\iota(X;Y)}(\theta)}{\sqrt{n}}\hspace*{-0.5ex}\right\}\qquad\\
			\geqslant \beta_2(n,M,\theta,P_X) - \exp\Bigg( n \me{ln}\left(\varphi_{\iota(X;Y)}(\theta)\right)- \left(\theta+1\right) \ln{\frac{M-1}{2}}\Bigg) \,
			\label{EqFVLPE1}
			\frac{ 2 \, \xi_{\iota(X;Y)}(\theta)}{\sqrt{n}}\\
			\label{EqFVLPE}
			= G_2(n,M,\theta,P_X),
		\end{IEEEeqnarray}
		where the equality in~\eqref{EqFVLPE} follows from \eqref{EqG2}.
		Observing that $1-F_{V_n}$ is a positive function, then~from~\eqref{EqFVLPE1}, it holds that
		
		\begin{IEEEeqnarray}{lcl}
			1\hspace*{-0.5ex}-\hspace*{-0.5ex}F_{V_n}\left(\hspace*{-0.5ex}\ln{\hspace*{-0.5ex}\frac{M-1}{2}\hspace*{-0.5ex}}\hspace*{-0.5ex}\right) &\geqslant& \max\left\{0,G_2(n,M,\theta,P_X)\right\}.
		\end{IEEEeqnarray}

		Seventh, from \eqref{EqBeta1} and \eqref{EqZetaApproxIotaD}, it holds that 
		\begin{IEEEeqnarray}{lcl}
			\label{EqZetaApproxIotaDE}
			\zeta_{\iota(X;Y)}\left(t, \ln{\frac{M-1}{2}}, n\right) = \beta_1(n,M,t,P_X).
		\end{IEEEeqnarray}
		
		Then, plugging \eqref{EqMGFIota}, \eqref{EqMuIota}, \eqref{EqChttE}, \eqref{EqVIota}, \eqref{EqTIota}, and \eqref{EqZetaApproxIotaDE} into the right hand-side of \eqref{EqFWUP}, it~holds that
		\begin{IEEEeqnarray}{l}
			F_{W_n}\left(\ln{\frac{M-1}{2}}\right)\nonumber \\
			\leqslant \beta_1(n,M,\theta,P_X) + \exp\Bigg( n \me{ln}\left(\varphi_{\iota(X;Y)}(\theta)\right) - \theta \, \ln{\frac{M-1}{2}}\Bigg)
			\min\left\{1,\frac{ 2 \, \xi_{\iota(X;Y)}(\theta)}{\sqrt{n}}\right\}\qquad\\
			\leqslant \beta_1(n,M,\theta,P_X) + \exp\Bigg( n \me{ln}\left(\varphi_{\iota(X;Y)}(\theta)\right) - \theta \, \ln{\frac{M-1}{2}}\Bigg)
			\label{EqFWUPE}
			\frac{ 2 \,\xi_{\iota(X;Y)}(\theta)}{\sqrt{n}}.
		\end{IEEEeqnarray}
		Alternatively, plugging \eqref{EqMGFIota}, \eqref{EqMuIota}, \eqref{EqChttE}, \eqref{EqVIota}, \eqref{EqTIota}, and \eqref{EqZetaApproxIotaIE2} into the right hand-side of \eqref{EqFWLW}, it holds that
		\begin{IEEEeqnarray}{l}
			F_{W_n}\left(\ln{\frac{M-1}{2}}\right)\nonumber \\
			\geqslant \beta_1(n,M,\theta,P_X) - \exp\Bigg( n \me{ln}\left(\varphi_{\iota(X;Y)}(\theta)\right) - \theta \, \ln{\frac{M-1}{2}}\Bigg)
			\min\left\{1,\frac{ 2 \, \xi_{\iota(X;Y)}(\theta)}{\sqrt{n}}\right\}\qquad\\
			\geqslant \beta_1(n,M,\theta,P_X) - \exp\Bigg( n \me{ln}\left(\varphi_{\iota(X;Y)}(\theta)\right) - \theta \, \ln{\frac{M-1}{2}}\Bigg)
			\label{EqFWLPE1}
			\frac{ 2 \,  \xi_{\iota(X;Y)}(\theta)}{\sqrt{n}}\\
			\label{EqFWLPE}
			= G_1(n,M,\theta,P_X),
		\end{IEEEeqnarray}
		where the equality in~\eqref{EqFWLPE} follows from \eqref{EqG1}.
		Observing that $F_{W_n}$ is a positive function, \mbox{then, from \eqref{EqFWLPE1}}, it holds that
		\begin{IEEEeqnarray}{lcl}
			F_{W_n}\left(\ln{\frac{M-1}{2}}\right) \geqslant \max\left\{0,G_1(n,M,\theta,P_X)\right\}.
		\end{IEEEeqnarray}

		Finally, plugging \eqref{EqFVUPE} and \eqref{EqFWUPE} in \eqref{EqLamUp}, it holds that
		\begin{IEEEeqnarray}{l}
			T(n,M,P_{\vec{X}})\nonumber \\
			\leqslant\hspace*{-0.5ex} \beta_1\hspace*{-0.1ex}(\hspace*{-0.1ex}n\hspace*{-0.1ex},\hspace*{-0.1ex}M\hspace*{-0.1ex},\hspace*{-0.1ex}\theta\hspace*{-0.1ex},\hspace*{-0.1ex}P_X\hspace*{-0.1ex})\hspace*{-0.5ex}+\hspace*{-0.5ex} \frac{M\hspace*{-0.5ex}-\hspace*{-0.4ex}1}{2}\beta_2\hspace*{-0.1ex}(\hspace*{-0.1ex}n\hspace*{-0.1ex},\hspace*{-0.1ex}M\hspace*{-0.1ex},\hspace*{-0.1ex}\theta\hspace*{-0.1ex},\hspace*{-0.1ex}P_X\hspace*{-0.1ex})\hspace*{-0.5ex}+\hspace*{-0.5ex} \exp\hspace*{-0.8ex}\Bigg(\hspace*{-0.8ex} n \me{ln}\left(\hspace*{-0.3ex}\varphi_{\iota(\hspace*{-0.1ex}X;Y\hspace*{-0.1ex})\hspace*{-0.3ex}}(\hspace*{-0.1ex}\theta\hspace*{-0.1ex})\hspace*{-0.3ex}\right)\hspace*{-0.4ex}-\hspace*{-0.4ex}\theta \hspace*{-0.1ex} \ln{\hspace*{-0.7ex}\frac{\hspace*{-0.1ex}M\hspace*{-0.1ex}-\hspace*{-0.1ex}1\hspace*{-0.1ex}}{2}\hspace*{-0.7ex}}\hspace*{-1.4ex}\Bigg)\hspace*{-0.4ex}
			\frac{ 4 \xi_{\iota(X;Y)}(\theta)}{\sqrt{n}}\qquad\\
			= \beta(n,M,\theta,P_X)+\exp\hspace*{-0.5ex}\Bigg(\hspace*{-0.6ex}n \ln{\hspace*{-0.5ex}\varphi_{\iota(X;Y)}(\theta)\hspace*{-0.3ex}} 
			\hspace*{-0.3ex}-\hspace*{-0.3ex}\theta \, \ln{\hspace*{-0.5ex}\frac{M-1}{2}\hspace*{-0.4ex}}\hspace*{-0.8ex}\Bigg)
			\label{EqTUPE1}
			\frac{ 4 \,  \xi_{\iota(X;Y)}(\theta)}{\sqrt{n}},
		\end{IEEEeqnarray}
		where the equality in~\eqref{EqTUPE1} follows from \eqref{EqBeta}.
		Observing that $T(n,M,P_{\vec{X}}) \leqslant 1$, from \eqref{EqTUPE1}, \mbox{it holds that}
		\begin{IEEEeqnarray}{lcl}
			T(\hspace*{-0.3ex}n,M\hspace*{-0.3ex},\hspace*{-0.3ex}P_{\vec{X}}) &\leqslant& \min\left\{1,\beta(n,M,\theta,P_X)\hspace*{-0.4ex}+\hspace*{-0.4ex}\exp\hspace*{-0.5ex}\Bigg(\hspace*{-0.6ex}n \ln{\hspace*{-0.5ex}\varphi_{\iota(X;Y)}(\theta)\hspace*{-0.3ex}} 
			\hspace*{-0.4ex}-\hspace*{-0.4ex}\theta \, \ln{\hspace*{-0.5ex}\frac{M\hspace*{-0.4ex}-\hspace*{-0.4ex}1}{2}\hspace*{-0.4ex}}\hspace*{-1.5ex}\Bigg)\hspace*{-0.65ex}
			\frac{ 4 \, \xi_{\iota(X;Y)}(\theta)}{\sqrt{n}}\right\}\qquad
			\\
			\label{EqTUPE}
			&=&  S(n,M,\theta,P_X),
		\end{IEEEeqnarray}
		where the equality in~\eqref{EqFWLPE} follows from \eqref{EqL}.

		Alternatively, plugging~\eqref{EqFVLPE} and~\eqref{EqFWLPE} in~\eqref{EqLamUp}, it holds that
		\begin{IEEEeqnarray}{lcl}
			T(n,M,P_{\vec{X}}) &\geqslant& \max\left\{0,G_1(n,M,\theta,P_X)\right\} + \frac{M-1}{2}\max\left\{0,G_2(n,M,\theta,P_X)\right\} \\
			\label{EqTLE}
			&=&  G(n,M,\theta,P_X),
		\end{IEEEeqnarray}
		where the equality in~\eqref{EqFWLPE} follows from~\eqref{EqL}.
		Combining~\eqref{EqTUPE} and~\eqref{EqTLE} concludes the proof.

		\section{Proof of Theorem \ref{TheoNewMC}}\label{pTheoNewMC}
		
		Note that, for given distributions $P_{\vec{X}}$ subject~\eqref{EqiidRandCod},  $Q_{\vec{Y}}$ subject to~\eqref{EqGenIndDensMut}, and for a random transformation in~\eqref{EqGeneralRamdonTransfomation} subject to~\eqref{EqMemChan}, the lower bound $C(n$,$M$,$P_{\vec{X}}$,$Q_{\vec{Y}}$,$\gamma)$ in~\eqref{EqMetConSym} can be written in the form of a weighted sum of the CDF and the complementary CDF of the random variables variables $W_n$ and $V_n$ that are sums of i.i.d random variables, respectively. That is,
		\begin{IEEEeqnarray}{lcl}
			W_n &=& \sum_{t=1}^{n} \tilde{\iota}(X_t;Y_t|Q_Y),\\
			V_n &=& \sum_{t=1}^{n} \tilde{\iota}(\bar{X}_t;Y_t|Q_{Y}),
		\end{IEEEeqnarray}
		where $(X_t,Y_t)\sim P_X P_{Y|X}$ and $(\bar{X}_t,Y_t)\sim P_{\bar{X}} Q_{Y}$ with $P_X=P_{\bar{X}}$. More specifically, the function $C$ in~\eqref{EqMetConSym} can be written in the form
		\begin{IEEEeqnarray}{lcl}
			\label{EqLamLw}
			C(n,M,P_{\vec{X}},Q_{\vec{Y}},\gamma) &=& F_{W_n}\left(\ln{\gamma}\right)  + \gamma \left(1-F_{V_n}\left(\ln{\gamma}\right) \right) - \frac{\gamma}{M},
		\end{IEEEeqnarray}
		where $F_{W_n}$ and $F_{V_n}$ are the CDFs of the random variables $W_n$ and $V_n$, respectively.

		The next step derives the upper and lower bounds on $F_{W_n}\left(\ln{\gamma}\right)$ and $1-F_{V_n}\left(\ln{\gamma}\right)$ by using the result of Theorem~\ref{TheoSaddlePointBeryUni}. That is,
		\begin{IEEEeqnarray}{l}
			F_{W_n}\left(\ln{\gamma}\right)
			\label{EqFWMCUP}
			\leqslant \hspace*{-0.5ex}\zeta_{\tilde{\iota}(X;Y|Q_Y)}(\theta,\ln{\gamma},n)\hspace*{-0.5ex} +\hspace*{-0.5ex} \exp\hspace*{-0.5ex}\left(\hspace*{-0.5ex} n \me{ln}\left(\hspace*{-0.5ex}\varphi_{\tilde{\iota}(X;Y|Q_Y)}(\theta)\hspace*{-0.5ex}\right)\hspace*{-0.5ex}-\hspace*{-0.5ex} \theta \ln{\gamma}\hspace*{-0.3ex}\right)\hspace*{-0.5ex}
			\min\hspace*{-0.5ex}\left\{\hspace*{-0.5ex}1,\frac{ 2 \, \xi_{\tilde{\iota}(X;Y|Q_Y)}(\theta)}{\sqrt{n}}\hspace*{-0.5ex}\right\},\\
			F_{W_n}\left(\ln{\gamma}\right)
			\label{EqFWMCLW}
			\geqslant  \hspace*{-0.5ex}\zeta_{\tilde{\iota}(X;Y|Q_Y)}(\theta,\ln{\gamma},n)\hspace*{-0.5ex} -\hspace*{-0.5ex} \exp\hspace*{-0.5ex}\left(\hspace*{-0.5ex} n \me{ln}\left(\hspace*{-0.5ex}\varphi_{\tilde{\iota}(X;Y|Q_Y)}(\theta)\hspace*{-0.5ex}\right)\hspace*{-0.5ex}-\hspace*{-0.5ex} \theta \ln{\gamma}\hspace*{-0.3ex}\right)\hspace*{-0.5ex}
			\min\hspace*{-0.5ex}\left\{\hspace*{-0.5ex}1,\frac{ 2 \, \xi_{\tilde{\iota}(X;Y|Q_Y)}(\theta)}{\sqrt{n}}\hspace*{-0.5ex}\right\},\\
			1-F_{V_n}\left(\ln{\gamma}\right) 
			\leqslant \hspace*{-0.5ex}1\hspace*{-0.5ex}- \hspace*{-0.5ex}\zeta_{\tilde{\iota}(\bar{X};Y|Q_Y)}(\theta,\ln{\gamma},n)\hspace*{-0.5ex} +\hspace*{-0.5ex} \exp\hspace*{-0.5ex}\left(\hspace*{-0.5ex} n \me{ln}\left(\hspace*{-0.5ex}\varphi_{\tilde{\iota}(\bar{X};Y|Q_Y)}(\theta)\hspace*{-0.5ex}\right)\hspace*{-0.5ex}-\hspace*{-0.5ex} \theta \ln{\gamma}\hspace*{-0.3ex}\right)\hspace*{-0.5ex}
			\min\hspace*{-0.5ex}\left\{\hspace*{-0.5ex}1,\frac{ 2 \, \xi_{\tilde{\iota}(\bar{X};Y|Q_Y)}(\theta)}{\sqrt{n}\hspace*{-0.5ex}}\hspace*{-0.5ex}\right\}\hspace*{-0.5ex},
			\label{EqFVMCUP}\nonumber\\
		\end{IEEEeqnarray}
		and
		\begin{IEEEeqnarray}{l}
			1-F_{V_n}\left(\ln{\gamma}\right)
			\geqslant \hspace*{-0.5ex}1\hspace*{-0.5ex}-\hspace*{-0.5ex}\zeta_{\tilde{\iota}(\bar{X};Y|Q_Y)}(\theta,\ln{\gamma},n)\hspace*{-0.5ex} -\hspace*{-0.5ex} \exp\hspace*{-0.5ex}\left(\hspace*{-0.5ex} n \me{ln}\left(\hspace*{-0.5ex}\varphi_{\tilde{\iota}(\bar{X};Y|Q_Y)}(\theta)\hspace*{-0.5ex}\right)\hspace*{-0.5ex}-\hspace*{-0.5ex} \theta \ln{\gamma}\hspace*{-0.3ex}\right)\hspace*{-0.5ex}
			\min\hspace*{-0.5ex}\left\{\hspace*{-0.5ex}1,\frac{ 2 \, \xi_{\tilde{\iota}(\bar{X};Y|Q_Y)}(\theta)}{\sqrt{n}\hspace*{-0.5ex}}\hspace*{-0.5ex}\right\}\hspace*{-0.5ex},
			\label{EqFVMCLW}\nonumber\\
		\end{IEEEeqnarray}
		where $\theta$ and $\tau$ satisfy
		\begin{IEEEeqnarray}{lcl}
			\label{EqChttMC}
			n\mu_{\tilde{\iota}(X;Y|Q_Y)}(\theta) &=& \ln{\gamma} = n\mu_{\tilde{\iota}(\bar{X};Y|Q_Y)}(\tau),
		\end{IEEEeqnarray}
		{and} for all $t\in \mathbb{R}$
		\begin{IEEEeqnarray}{lcl}
			\label{EqMGFIotaDMC}
			\varphi_{\tilde{\iota}(X;Y|Q_Y)}(t) &=& \mathbb{E}_{P_{X}P_{Y|X}}\left[\exp\left(t\, \tilde{\iota}(X;Y|Q_Y)\right)\right],\\
			\label{EqMGFIotaIMC}
			\varphi_{\tilde{\iota}(\bar{X};Y|Q_Y)}(t) &=& \mathbb{E}_{P_{\bar{X}}Q_{Y}}\left[\exp\left(t\, \tilde{\iota}({\bar{X}};Y|Q_Y)\right)\right],\\
			\label{EqTIotaDMC}
			\xi_{\tilde{\iota}(X;Y|Q_Y)}(t) &=& c_1\left(\frac{\mathbb{E}_{P_XP_{Y|X}}\left[\left|\tilde{\iota}(X;Y|Q_Y) - \mu_{\tilde{\iota}(X;Y|Q_Y)}(t)\right|^3\frac{\exp(t\, \tilde{\iota}(X;Y|Q_Y))}{\varphi_{\tilde{\iota}(X;Y|Q_Y)}(t)}\right]}{\left(V_{\tilde{\iota}(X;Y|Q_Y)}(t)\right)^{3/2}} + c_2\right),\\
			\label{EqTIotaIMC}
			\xi_{\tilde{\iota}(\bar{X};Y|Q_Y)}(t) &=& c_1\left(\frac{\mathbb{E}_{P_{\bar{X}}Q_{Y}}\left[\left|\tilde{\iota}({\bar{X}};Y|Q_Y) - \mu_{\tilde{\iota}(\bar{X};Y|Q_Y)}(t)\right|^3\frac{\exp\left(t \tilde{\iota}({\bar{X}};Y|Q_Y)\right)}{\varphi_{\tilde{\iota}(\bar{X};Y|Q_Y)}(t)}\right]}{\left(V_{\tilde{\iota}(\bar{X};Y|Q_Y)}(t)\right)^{3/2}}+ c_2\right),\\
			\label{EqMuIotaDMC}
			\mu_{\tilde{\iota}(X;Y|Q_Y)}(t) &=& \mathbb{E}_{P_XP_{Y|X}}\left[\tilde{\iota}(X;Y|Q_Y)\frac{\exp(t\, \tilde{\iota}(X;Y|Q_Y))}{\varphi_{\tilde{\iota}(X;Y|Q_Y)}(t)}\right],\\
			\label{EqMuIotaIMC}
			\mu_{\tilde{\iota}(\bar{X};Y|Q_Y)}(t) &=&  \mathbb{E}_{P_{\bar{X}}Q_{Y}}\left[\tilde{\iota}({\bar{X}};Y|Q_Y)\frac{\exp\left(t \tilde{\iota}({\bar{X}};Y|Q_Y)\right)}{\varphi_{\tilde{\iota}({\bar{X}};Y|Q_Y)}(t)} \right],\\
			\label{EqVIotaDMC}
			V_{\tilde{\iota}(X;Y|Q_Y)}(t) &=& \mathbb{E}_{P_XP_{Y|X}}\left[\left(\tilde{\iota}(X;Y|Q_Y) - \mu_{\tilde{\iota}(X;Y|Q_Y)}(t)\right)^2\frac{\exp(t\, \tilde{\iota}(X;Y|Q_Y))}{\varphi_{\tilde{\iota}(X;Y|Q_Y)}(t)}\right],\\
			\label{EqVIotaIMC}
			V_{\tilde{\iota}(\bar{X};Y|Q_Y)}(t) &=& \mathbb{E}_{P_{\bar{X}}Q_{Y}}\left[\left(\tilde{\iota}({\bar{X}};Y|Q_Y) - \mu_{\tilde{\iota}(\bar{X};Y|Q_Y)}(t)\right)^2\frac{\exp\left(t \tilde{\iota}({\bar{X}};Y|Q_Y)\right)}{\varphi_{\tilde{\iota}(\bar{X};Y|Q_Y)}(t)} \right],
		\end{IEEEeqnarray}
		with $c_1$ and $c_2$ defined in~\eqref{EqConstants};
		and for all $(t,a,n)\in\mathbb{R}^2\times\mathbb{N}$
		\begin{IEEEeqnarray}{l}
			\nonumber
			\label{EqZetaApproxIotaDMC}
			\zeta_{\tilde{\iota}(X;Y|Q_Y)}(t, a, n)   \\
			\df \hspace*{-0.5ex}\mathds{1}_{\{\hspace*{-0.1ex}t > 0 \hspace*{-0.1ex}\}} \hspace*{-0.55ex}
			+ \hspace*{-0.55ex} (\hspace*{-0.2ex}-\hspace*{-0.2ex}1)^{\mathds{1}_{\{t > 0 \}}} \hspace*{-0.5ex}\exp\hspace*{-0.5ex}\left(\hspace*{-0.5ex} \frac{1}{2} n\hspace*{-0.1ex}  t^2 \hspace*{-0.3ex} V_{\tilde{\iota}(X;Y|Q_Y)}(\hspace*{-0.1ex}t\hspace*{-0.1ex}) \hspace*{-0.5ex}+\hspace*{-0.5ex} n \me{ln}\hspace*{-0.3ex}\left(\hspace*{-0.3ex}\varphi_{\tilde{\iota}(X;Y|Q_Y)}(t)\hspace*{-0.3ex}\right) \hspace*{-0.5ex}-\hspace*{-0.5ex} t  a \hspace*{-0.8ex}\right)
			\hspace*{-0.5ex}Q\hspace*{-0.5ex}\left(\hspace*{-0.5ex} \abs{t} \hspace*{-0.2ex}\sqrt{n  V_{\tilde{\iota}(X;Y|Q_Y)}(t)}\hspace*{-0.5ex}\right),\\
			\nonumber
			\label{EqZetaApproxIotaIMC}
			\zeta_{\tilde{\iota}(\bar{X};Y|Q_Y)}(t, a, n) \\
			\df \hspace*{-0.5ex}\mathds{1}_{\{\hspace*{-0.1ex}t > 0 \hspace*{-0.1ex}\}} \hspace*{-0.55ex}
			+ \hspace*{-0.55ex} (\hspace*{-0.2ex}-\hspace*{-0.2ex}1)^{\mathds{1}_{\{t > 0 \}}} \hspace*{-0.5ex}\exp\hspace*{-0.5ex}\left(\hspace*{-0.5ex} \frac{1}{2} n\hspace*{-0.1ex}  t^2 \hspace*{-0.3ex} V_{\tilde{\iota}(\bar{X};Y|Q_Y)}(\hspace*{-0.1ex}t\hspace*{-0.1ex}) \hspace*{-0.5ex}+\hspace*{-0.5ex} n \me{ln}\hspace*{-0.3ex}\left(\hspace*{-0.3ex}\varphi_{\tilde{\iota}(\bar{X};Y|Q_Y)}(t)\hspace*{-0.3ex}\right) \hspace*{-0.5ex}-\hspace*{-0.5ex} t  a \hspace*{-0.8ex}\right)
			\hspace*{-0.5ex}Q\hspace*{-0.5ex}\left(\hspace*{-0.5ex} \abs{t} \hspace*{-0.2ex}\sqrt{n  V_{\tilde{\iota}(\bar{X};Y|Q_Y)}(t)}\hspace*{-0.5ex}\right).\qquad
		\end{IEEEeqnarray}

		The next step simplifies the expressions on the right hand-side of \eqref{EqFVMCUP} and \eqref{EqFVMCLW} by studying the relation between $\varphi_{\tilde{\iota}(X;Y|Q_Y)}$ and $\varphi_{\tilde{\iota}(\bar{X};Y|Q_Y)}$, $\theta$ and $\tau$, $V_{\tilde{\iota}(X;Y|Q_Y)}$ and $V_{\tilde{\iota}(\bar{X};Y|Q_Y)}$,
		$\xi_{\tilde{\iota}(X;Y|Q_Y)}$ and $\xi_{\tilde{\iota}(\bar{X};Y|Q_Y)}$ when the $P_{Y|X}$ is absolutely continuous with respect to $Q_Y$.

		First, from \eqref{EqMGFIotaDMC}, using the change of measure from $P_XP_{Y|X}$ to $P_{\bar{X}}Q_{Y}$ because $P_XP_{Y|X}$ is absolutely continuous with respect to $P_{\bar{X}}Q_{Y}$, it holds that 
		\begin{IEEEeqnarray}{lcl}
			\label{EqMGFIotaDMCE}
			\varphi_{\tilde{\iota}(X;Y|Q_Y)}(t) &=& \mathbb{E}_{P_{\bar{X}}Q_{Y}}\left[\rndder{P_{X}P_{Y|X}}{P_{\bar{X}}Q_{Y}}\left({\bar{X}};Y\right)\exp(t\, \tilde{\iota}({\bar{X}};Y|Q_Y)\right]\\
			\label{EqMGFIotaDMCE1}
			&=& \mathbb{E}_{P_{\bar{X}}Q_{Y}}\left[\exp\left((t+1)\, \tilde{\iota}({\bar{X}};Y|Q_Y)\right)\right].
		\end{IEEEeqnarray}
		Then, from \eqref{EqMGFIotaDMC} and \eqref{EqMGFIotaIMC}, it holds that
		\begin{IEEEeqnarray}{lcl}
			\label{EqMGFIotaMC}
			\varphi_{\tilde{\iota}(X;Y|Q_Y)}(t)	&=& \varphi_{\tilde{\iota}(\bar{X};Y|Q_Y)}(t+1).
		\end{IEEEeqnarray}
		This concludes the relation between $\varphi_{\tilde{\iota}(X;Y|Q_Y)}$ and $\varphi_{\tilde{\iota}(\bar{X};Y|Q_Y)}$.

		Second, from \eqref{EqMuIotaDMC}, using the change of measure from $P_XP_{Y|X}$ to $P_{\bar{X}}Q_{Y}$, it holds that 
		\begin{IEEEeqnarray}{lcl}
			\label{EqMuIotaDMCE}
			\mu_{\tilde{\iota}(X;Y|Q_Y)}(t) &=& \mathbb{E}_{P_{\bar{X}}Q_{Y}}\left[\tilde{\iota}({\bar{X}};Y|Q_Y) \frac{\exp(t\, \tilde{\iota}({\bar{X}};Y|Q_Y)}{\varphi_{\tilde{\iota}(X;Y|Q_Y)}(t)} \rndder{P_XP_{Y|X}}{P_{\bar{X}}Q_{Y}}\left({\bar{X}};Y\right)\right]\\
			\label{EqMuIotaDMCE1}
			&=& \mathbb{E}_{P_{\bar{X}}Q_{Y}}\left[\tilde{\iota}({\bar{X}};Y|Q_Y)\frac{\exp\left((t+1)\, \tilde{\iota}({\bar{X}};Y|Q_Y)\right)}{\varphi_{\tilde{\iota}(X;Y|Q_Y)}(t)} \right].
		\end{IEEEeqnarray}
		Then, from \eqref{EqMGFIotaMC} and \eqref{EqMuIotaDMCE1}, it holds that
		\begin{IEEEeqnarray}{lcl}
			\label{EqMuIotaDMCE2}
			\mu_{\tilde{\iota}(X;Y|Q_Y)}(t) &=& \mathbb{E}_{P_{\bar{X}}Q_{Y}}\left[\tilde{\iota}({\bar{X}};Y|Q_Y)\frac{\exp\left((t+1)\, \tilde{\iota}({\bar{X}};Y|Q_Y)\right)}{\varphi_{\tilde{\iota}({\bar{X}};Y|Q_Y)}(t+1)} \right].
		\end{IEEEeqnarray}
		From \eqref{EqMuIotaIMC} and \eqref{EqMuIotaDMCE2}, it holds that
		\begin{IEEEeqnarray}{lcl}
			\label{EqMuIotaMC}
			\mu_{\tilde{\iota}(X;Y|Q_Y)}(t)&=& \mu_{\tilde{\iota}(\bar{X};Y|Q_Y)}(t+1).
		\end{IEEEeqnarray}
		This concludes the relation between $\mu_{\tilde{\iota}(X;Y|Q_Y)}$ and $\mu_{\tilde{\iota}(\bar{X};Y|Q_Y)}$.

		Third, from \eqref{EqChttMC} and \eqref{EqMuIotaMC}, it holds that
		\begin{IEEEeqnarray}{lcl}
			\label{EqChttMCE}
			\tau = \theta + 1.
		\end{IEEEeqnarray}
		This concludes the relation between $\tau$ and $\theta$.

		Fourth, from \eqref{EqVIotaDMC}, using the change of measure from $P_XP_{Y|X}$ to $P_{\bar{X}}Q_{Y}$, it holds that 
		\begin{IEEEeqnarray}{lcl}
			\label{EqVIotaDMCE}
			V_{\tilde{\iota}(X;Y|Q_Y)}(t) &=& \mathbb{E}_{P_{\bar{X}}Q_{Y}}\left[\left(\tilde{\iota}({\bar{X}};Y|Q_Y) - \mu_{\tilde{\iota}(X;Y|Q_Y)}(t)\right)^2\frac{\exp(t\, \tilde{\iota}({\bar{X}};Y|Q_Y)}{\varphi_{\tilde{\iota}(X;Y|Q_Y)}(t)} \rndder{P_XP_{Y|X}}{P_{\bar{X}}Q_{Y}}\left({\bar{X}};Y\right)\right]\qquad\\
			\label{EqVIotaDMCE1}
			&=& \mathbb{E}_{P_{\bar{X}}Q_{Y}}\left[\left(\tilde{\iota}({\bar{X}};Y|Q_Y) - \mu_{\tilde{\iota}(X;Y|Q_Y)}(t)\right)^2\frac{\exp\left((t+1)\, \tilde{\iota}({\bar{X}};Y|Q_Y)\right)}{\varphi_{\tilde{\iota}(X;Y|Q_Y)}(t)} \right].
		\end{IEEEeqnarray}
		From \eqref{EqMGFIotaMC}, \eqref{EqMuIotaMC}, and \eqref{EqVIotaDMCE1}, it holds that
		\begin{IEEEeqnarray}{lcl}
			\label{EqVIotaDMCE2}
			V_{\tilde{\iota}(X;Y|Q_Y)}(t) &=& \mathbb{E}_{P_{\bar{X}}Q_{Y}}\left[\left(\tilde{\iota}({\bar{X}};Y|Q_Y) - \mu_{\tilde{\iota}(\bar{X};Y|Q_Y)}(t+1)\right)^2\frac{\exp\left((t+1)\, \tilde{\iota}({\bar{X}};Y|Q_Y)\right)}{\varphi_{\tilde{\iota}(\bar{X};Y|Q_Y)}(t+1)} \right].\qquad
		\end{IEEEeqnarray}
		From \eqref{EqVIotaIMC} and \eqref{EqVIotaDMCE2}, it holds that
		\begin{IEEEeqnarray}{lcl}
			\label{EqVIotaMC}
			V_{\tilde{\iota}(X;Y|Q_Y)}(t) &=& V_{\tilde{\iota}(\bar{X};Y|Q_Y)}(t+1).
		\end{IEEEeqnarray}
		This concludes the relation between $V_{\tilde{\iota}(X;Y|Q_Y)}$ and $V_{\tilde{\iota}(\bar{X};Y|Q_Y)}$.

		Fifth, from \eqref{EqTIotaDMC}, using the change of measure from $P_XP_{Y|X}$ to $P_{\bar{X}}Q_{Y}$, it holds that 
		\begin{IEEEeqnarray}{lcl}
			\label{EqTIotaDMCE}
			\xi_{\tilde{\iota}(X;Y|Q_Y)}(t) &=& c_1\left(\frac{\mathbb{E}_{P_{\bar{X}}Q_{Y}}\left[\left|\tilde{\iota}({\bar{X}};Y|Q_Y) - \mu_{\tilde{\iota}(X;Y|Q_Y)}(t)\right|^3\frac{\exp(t\, \tilde{\iota}({\bar{X}};Y|Q_Y)}{\varphi_{\tilde{\iota}(X;Y|Q_Y)}(t)} \rndder{P_XP_{Y|X}}{P_{\bar{X}}Q_{Y}}\left({\bar{X}};Y\right)\right]}{\left(V_{\tilde{\iota}(X;Y|Q_Y)}(t)\right)^{3/2}} + c_2\right)\nonumber\\
			\\
			\label{EqTIotaDMCE1}
			&=& c_1\left(\frac{\mathbb{E}_{P_{\bar{X}}Q_{Y}}\left[\left|\tilde{\iota}({\bar{X}};Y|Q_Y) - \mu_{\tilde{\iota}(X;Y|Q_Y)}(t)\right|^3\frac{\exp\left((t+1)\, \tilde{\iota}({\bar{X}};Y|Q_Y)\right)}{\varphi_{\tilde{\iota}(X;Y|Q_Y)}(t)} \right]}{\left(V_{\tilde{\iota}(X;Y|Q_Y)}(t)\right)^{3/2}} + c_2\right).
		\end{IEEEeqnarray}
		From \eqref{EqMGFIotaMC}, \eqref{EqMuIotaMC}, \eqref{EqVIotaMC}, and \eqref{EqTIotaDMCE1}, it holds that
		\begin{IEEEeqnarray}{lcl}
			\label{EqTIotaDMCE2}
			\xi_{\tilde{\iota}(X;Y|Q_Y)}(t) &=& c_1\left(\frac{ \mathbb{E}_{P_{\bar{X}}Q_{Y}}\left[\left|\tilde{\iota}({\bar{X}};Y|Q_Y) - \mu_{\tilde{\iota}(\bar{X};Y|Q_Y)}(t+1)\right|^3\frac{\exp\left((t+1)\, \tilde{\iota}({\bar{X}};Y|Q_Y)\right)}{\varphi_{\tilde{\iota}(\bar{X};Y|Q_Y)}(t+1)} \right]}{\left(V_{\tilde{\iota}(\bar{X};Y|Q_Y)}(t+1)\right)^{3/2}} + c_2\right).\qquad
		\end{IEEEeqnarray}
		From \eqref{EqTIotaIMC} and \eqref{EqTIotaDMCE2}, it holds that
		\begin{IEEEeqnarray}{lcl}
			\label{EqTIotaMC}
			\xi_{\tilde{\iota}(X;Y|Q_Y)}(t) &=& \xi_{\tilde{\iota}(\bar{X};Y|Q_Y)}(t+1).
		\end{IEEEeqnarray}
		This concludes the relation between $\xi_{\tilde{\iota}(X;Y|Q_Y)}$ and $\xi_{\tilde{\iota}(\bar{X};Y|Q_Y)}$.

		Sixth, plugging \eqref{EqMGFIotaMC}, \eqref{EqMuIotaMC}, and \eqref{EqVIotaMC} into \eqref{EqZetaApproxIotaDMC}, for all $t\in \mathbb{R}$, it holds that
		\begin{IEEEeqnarray}{l}
			\nonumber
			\zeta_{\tilde{\iota}(\bar{X};Y|Q_Y)}(t, a, n) \\
			\df  \hspace*{-0.7ex}\mathds{1}_{\{t > 0 \}} \hspace*{-0.6ex}
			+ \hspace*{-0.5ex} (\hspace*{-0.3ex}-\hspace*{-0.2ex}1\hspace*{-0.2ex})^{\mathds{1}_{\{\hspace*{-0.2ex}t > 0 \hspace*{-0.2ex}\}}} \hspace*{-0.5ex}\exp\hspace*{-0.5ex}\left(\hspace*{-0.5ex} \frac{1}{2} n  t^2  V_{\tilde{\iota}(X;Y|Q_Y)}(\hspace*{-0.2ex}t\hspace*{-0.2ex}-\hspace*{-0.2ex}1\hspace*{-0.2ex})\hspace*{-0.5ex}+\hspace*{-0.5ex} n \me{ln}\hspace*{-0.4ex}\left(\hspace*{-0.4ex}\varphi_{\tilde{\iota}(X;Y|Q_Y)\hspace*{-0.2ex}}(\hspace*{-0.2ex}t\hspace*{-0.3ex}-\hspace*{-0.3ex}1\hspace*{-0.2ex})\hspace*{-0.5ex}\right)\hspace*{-0.7ex} - \hspace*{-0.5ex} t \hspace*{-0.2ex} a \hspace*{-0.8ex}\right)
			\label{EqZetaApproxIotaIMCE}
			\hspace*{-0.3ex}Q\hspace*{-0.5ex}\left(\hspace*{-0.5ex} \abs{t} \hspace*{-0.4ex}\sqrt{\hspace*{-0.2ex}n \hspace*{-0.2ex}V_{\tilde{\iota}(X;Y|Q_Y)}(\hspace*{-0.2ex}t\hspace*{-0.2ex}-\hspace*{-0.2ex}1\hspace*{-0.2ex})\hspace*{-0.2ex}}\hspace*{-0.5ex}\right)\hspace*{-0.8ex}.\qquad
		\end{IEEEeqnarray}
		Then, from \eqref{EqTBeta2} and \eqref{EqZetaApproxIotaIMCE}, it holds that
		\begin{IEEEeqnarray}{lcl}
			\label{EqZetaApproxIotaIMCE2}
			\zeta_{\tilde{\iota}(\bar{X};Y|Q_Y)}(t, \ln{\gamma}, n) = 1-\tilde{\beta}_2(n,\gamma,t-1,P_{X},Q_Y).
		\end{IEEEeqnarray}

		Then, plugging \eqref{EqMGFIotaMC}, \eqref{EqMuIotaMC}, \eqref{EqChttMCE}, \eqref{EqVIotaMC}, \eqref{EqTIotaMC}, and \eqref{EqZetaApproxIotaIMCE2} into the right hand-side of~\eqref{EqFVMCUP}, it holds that
		\begin{IEEEeqnarray}{l}
			1-F_{V_n}\left(\ln{\gamma}\right) \nonumber \\
			\leqslant \hspace*{-0.5ex}\tilde{\beta}_2(n,\gamma,\theta,P_{X},Q_Y)\hspace*{-0.5ex} +\hspace*{-0.5ex} \exp\hspace*{-0.5ex}\Bigg(\hspace*{-0.9ex} n \me{ln}\left(\varphi_{\tilde{\iota}(X;Y|Q_Y)}(\theta)\right) - \left(\theta+1\right)  \ln{\gamma}\hspace*{-1.2ex}\Bigg)
			\hspace*{-0.5ex}\min\left\{1,\frac{ 2 \xi_{\tilde{\iota}(X;Y|Q_Y)}(\theta)}{\sqrt{n}}\hspace*{-0.8ex}\right\}\qquad\\
			\leqslant \tilde{\beta}_2(n,\gamma,\theta,P_{X},Q_Y) + \exp\Bigg( n \me{ln}\left(\hspace*{-0.5ex}\varphi_{\tilde{\iota}(X;Y|Q_Y)}(\theta)\hspace*{-0.5ex}\right) - \left(\theta+1\right) \, \ln{\gamma}\Bigg) 
			\label{EqFVMCUPE}
			\frac{ 2 \, \xi_{\tilde{\iota}(X;Y|Q_Y)}(\theta)}{\sqrt{n}}.
		\end{IEEEeqnarray}
		Alternatively, plugging \eqref{EqMGFIotaMC}, \eqref{EqMuIotaMC}, \eqref{EqChttMCE}, \eqref{EqVIotaMC}, \eqref{EqTIotaMC}, and \eqref{EqZetaApproxIotaIMCE2} into the right hand-side of~\eqref{EqFVMCLW}, it holds that
		\begin{IEEEeqnarray}{l}
			1-F_{V_n}\left(\ln{\gamma}\right) \nonumber \\
			\geqslant \hspace*{-0.5ex}\tilde{\beta}_2(n,\gamma,\theta,P_{X},Q_Y)\hspace*{-0.5ex} -\hspace*{-0.5ex} \exp\hspace*{-0.5ex}\Bigg(\hspace*{-0.9ex} n \me{ln}\left(\varphi_{\tilde{\iota}(X;Y|Q_Y)}(\theta)\right) - \left(\theta+1\right)  \ln{\gamma}\hspace*{-1.2ex}\Bigg)
			\hspace*{-0.5ex}\min\left\{1,\frac{ 2 \xi_{\tilde{\iota}(X;Y|Q_Y)}(\theta)}{\sqrt{n}}\hspace*{-0.8ex}\right\}\qquad\\
			\geqslant \tilde{\beta}_2(n,\gamma,\theta,P_{X},Q_Y) - \exp\Bigg( n \me{ln}\left(\varphi_{\tilde{\iota}(X;Y|Q_Y)}(\theta)\right)- \left(\theta+1\right) \ln{\gamma}\Bigg) \, 
			\label{EqFVMCLPE1}
			\frac{ 2 \, \xi_{\tilde{\iota}(X;Y|Q_Y)}(\theta)}{\sqrt{n}}\\
			\label{EqFVMCLPE}
			= \tilde{G}_2(n,\gamma,\theta,P_{X},Q_Y),
		\end{IEEEeqnarray}
		where the equality in~\eqref{EqFVMCLPE} follows from \eqref{EqTG2}.
		Observing that $1-F_{V_n}$ is a positive function, \mbox{then, from \eqref{EqFVMCLPE1},} it holds that
		\begin{IEEEeqnarray}{l}
			1-F_{V_n}\left(\ln{\gamma}\right)\geqslant \max\left\{0,\tilde{G}_2(n,\gamma,\theta,P_{X},Q_Y)\right\}.
		\end{IEEEeqnarray}

		Seventh, from \eqref{EqTBeta1} and \eqref{EqZetaApproxIotaDMC}, it holds that 
		\begin{IEEEeqnarray}{lcl}
			\label{EqZetaApproxIotaDMCE}
			\zeta_{\tilde{\iota}(X;Y|Q_Y)}(t, \ln{\gamma}, n) = \tilde{\beta}_1(n,\gamma,t,P_{X},Q_Y).
		\end{IEEEeqnarray}
		
		Then, plugging \eqref{EqMGFIotaMC}, \eqref{EqMuIotaMC}, \eqref{EqChttMCE}, \eqref{EqVIotaMC}, \eqref{EqTIotaMC}, and \eqref{EqZetaApproxIotaDMCE} into the right hand-side of~\eqref{EqFWMCUP}, it holds that
		\begin{IEEEeqnarray}{lcl}
			F_{W_n}\left(\ln{\gamma}\right) \nonumber\\
			\leqslant \hspace*{-0.5ex}\tilde{\beta}_1(n,\gamma,\theta,P_{X},Q_Y) \hspace*{-0.5ex}+\hspace*{-0.5ex} \exp\Bigg( n \me{ln}\left(\varphi_{\tilde{\iota}(X;Y|Q_Y)}(\theta)\right) - \theta \, \ln{\gamma}\Bigg) \min\left\{1,\frac{ 2 \, \xi_{\tilde{\iota}(X;Y|Q_Y)}(\theta)}{\sqrt{n}}\hspace*{-0.5ex}\right\}\qquad\\
			\leqslant \tilde{\beta}_1(n,\gamma,\theta,P_{X},Q_Y) + \exp\Bigg( n \me{ln}\left(\varphi_{\tilde{\iota}(X;Y|Q_Y)}(\theta)\right) - \theta \, \ln{\gamma}\Bigg)
			\label{EqFWMCUPE}
			\frac{ 2 \, \xi_{\tilde{\iota}(X;Y|Q_Y)}(\theta)}{\sqrt{n}}.
		\end{IEEEeqnarray}
		Alternatively, plugging \eqref{EqMGFIotaMC}, \eqref{EqMuIotaMC}, \eqref{EqChttMCE}, \eqref{EqVIotaMC}, \eqref{EqTIotaMC}, and \eqref{EqZetaApproxIotaIMCE2} into the right hand-side of~\eqref{EqFWMCLW}, it holds that
		\begin{IEEEeqnarray}{lcl}
			F_{W_n}\left(\ln{\gamma}\right) \nonumber\\
			\geqslant \hspace*{-0.5ex}\tilde{\beta}_1(n,\gamma,\theta,P_{X},Q_Y) \hspace*{-0.5ex}-\hspace*{-0.5ex} \exp\Bigg( n \me{ln}\left(\varphi_{\tilde{\iota}(X;Y|Q_Y)}(\theta)\right) - \theta \, \ln{\gamma}\Bigg) \min\left\{1,\frac{ 2 \, \xi_{\tilde{\iota}(X;Y|Q_Y)}(\theta)}{\sqrt{n}}\hspace*{-0.5ex}\right\}\qquad\\
			\geqslant \tilde{\beta}_1(n,\gamma,\theta,P_{X},Q_Y) - \exp\Bigg( n \me{ln}\left(\varphi_{\tilde{\iota}(X;Y|Q_Y)}(\theta)\right) - \theta \, \ln{\gamma}\Bigg)
			\label{EqFWMCLPE1}
			\frac{ 2 \, \xi_{\tilde{\iota}(X;Y|Q_Y)}(\theta)}{\sqrt{n}}\\
			\label{EqFWMCLPE}
			= \tilde{G}_1(n,\gamma,\theta,P_{X},Q_Y),
		\end{IEEEeqnarray}
		where the equality in~\eqref{EqFWMCLPE} follows from \eqref{EqTG1}.
		Observing that $F_{W_n}$ is a positive function, \mbox{then from \eqref{EqFWMCLPE1}}, it holds that
		\begin{IEEEeqnarray}{l}
			F_{W_n}\left(\ln{\gamma}\right) \geqslant \max\left\{0,\tilde{G}_1(n,\gamma,\theta,P_{X},Q_Y)\right\}.
		\end{IEEEeqnarray}

		Finally, plugging \eqref{EqFVMCUPE} and \eqref{EqFWMCUPE} in \eqref{EqLamLw}, it holds that
		\begin{IEEEeqnarray}{l}
			C(n,M,P_{\vec{X}},Q_{\vec{Y}},\gamma) \nonumber \\
			\leqslant \hspace*{-0.5ex}\tilde{\beta}_1(n,\gamma,\theta,P_{X},Q_Y)\hspace*{-0.3ex} +\hspace*{-0.3ex} \gamma\tilde{\beta}_2(n,\gamma,\theta,P_{X},Q_Y)\hspace*{-0.3ex} + \hspace*{-0.3ex} \exp\hspace*{-0.7ex}\Bigg(\hspace*{-0.8ex} n\hspace*{-0.1ex} \me{ln}\hspace*{-0.3ex}\left(\hspace*{-0.4ex}\varphi_{\tilde{\iota}(X;Y|Q_Y)}(\theta)\hspace*{-0.3ex}\right) \hspace*{-0.5ex}
			-\hspace*{-0.5ex} \theta \hspace*{-0.1ex} \ln{\gamma}\hspace*{-1.2ex}\Bigg)\hspace*{-0.4ex}\frac{ 4 \xi_{\tilde{\iota}(X;Y|Q_Y)}(\theta)}{\sqrt{n}} \hspace*{-0.5ex}-\hspace*{-0.5ex}\frac{\gamma}{M}\nonumber\\
			\\
			=  \tilde{\beta}(n,\gamma,\theta,P_{X},Q_Y,M) + \exp\Bigg( n \me{ln}\left(\varphi_{\tilde{\iota}(X;Y|Q_Y)}(\theta)\right) 
			- \theta \, \ln{\gamma}\Bigg)
			\label{EqCUPE1}
			\frac{ 4  \xi_{\tilde{\iota}(X;Y|Q_Y)}(\theta)}{\sqrt{n}},
		\end{IEEEeqnarray}
		where the equality in~\eqref{EqFWMCLPE1} follows from \eqref{EqTBeta}.
		Observing that $C(n,M,P_{\vec{X}},Q_{\vec{Y}},\gamma)+\frac{\gamma}{M} \leqslant 1$, from~\eqref{EqCUPE1}, it holds that
		\begin{IEEEeqnarray}{l}
			C(n,M,P_{\vec{X}},Q_{\vec{Y}},\gamma)\nonumber \\
			\leqslant \hspace*{-0.5ex}\min\left\{1, \tilde{\beta}(n,\gamma,\theta,P_{X},Q_Y) \hspace*{-0.5ex} +\hspace*{-0.5ex} \exp\Bigg( n \me{ln}\left(\varphi_{\tilde{\iota}(X;Y|Q_Y)}(\theta)\right)  - \theta \, \ln{\gamma}\Bigg)\frac{ 4 \, \xi_{\tilde{\iota}(X;Y|Q_Y)}(\theta)}{\sqrt{n}}\right\}\qquad\\
			\label{EqCUPE}
			=  \tilde{S}(n,\gamma,\theta,P_{X},Q_Y,M),
		\end{IEEEeqnarray}
		where \eqref{EqCUPE} follows from \eqref{EqTL}.

		Alternatively, plugging \eqref{EqFVMCLPE} and \eqref{EqFWMCLPE} in \eqref{EqLamLw}, it holds that
		\begin{IEEEeqnarray}{lcl}
			C(n,M,P_{\vec{X}},Q_{\vec{Y}},\gamma) &\geqslant& \max\left\{0,\tilde{G}_1(n,\gamma,\theta,P_{X},Q_Y)\right\} \hspace*{-0.5ex}+\hspace*{-0.5ex} \gamma \max\left\{0,\tilde{G}_2(n,\gamma,\theta,P_{X},Q_Y)\right\}\hspace*{-0.7ex}-\hspace*{-0.7ex} \frac{\gamma}{M} \qquad \\
			\label{EqCLE}
			&=&  \tilde{G}(n,\gamma,\theta,P_{X},Q_Y,M),
		\end{IEEEeqnarray}
		where the equality in~\eqref{EqCLE} follows from \eqref{EqTU}.
		Combining \eqref{EqCUPE} and \eqref{EqCLE} concludes the proof.
	
	\reftitle{References}

\end{document}